\documentclass[11pt]{article}

\usepackage[margin=1in]{geometry}
\usepackage[T1]{fontenc}

\usepackage{amsmath, amssymb, amsthm, mathtools}
\usepackage{bm}
\usepackage{gensymb}
\usepackage{booktabs}
\usepackage{array}
\usepackage{tabularx}
\usepackage{multirow}
\usepackage{longtable}
\usepackage{graphicx}
\usepackage{float}
\usepackage{caption}
\usepackage{subcaption}
\usepackage{rotating}

\usepackage{enumitem}
\usepackage[hidelinks]{hyperref}
\usepackage{cleveref}
\usepackage{url}
\usepackage{authblk}

\usepackage[numbers,sort&compress]{natbib}
\newcommand{\Ls}{\ensuremath{L_s}}
\newcommand{\MY}{\ensuremath{\mathrm{MY}}}
\newcommand{\CO}{\ensuremath{\mathrm{CO_2}}}
\newcommand{\HO}{\ensuremath{\mathrm{H_2O}}}
\newcommand{\vu}{\ensuremath{\mathbf{u}}}
\newcommand{\vh}[1]{\ensuremath{\hat{\mathbf{e}}_{#1}}}
\newcommand{\taud}{\ensuremath{\tau_d}}

\title{Towards a Foundation Model for the Martian Atmosphere}

\author[1,2]{Sujit Roy}
\author[1]{Udayshankar Nair}
\author[1]{Yuling Wu}
\author[1]{Georgios Priftis}
\author[3]{Liping Wang}
\author[4*]{Anastasia Georgiou}
\author[4*]{Anne Jones}
\author[4*]{Bj\"orn L\"utjens}
\author[4*]{Johannes Schmude}
\author[4*]{Campbell Watson}
\author[5]{Rachel A. Slank}
\author[1]{Ankur Kumar}
\author[6]{Anirbit Mukherjee}
\author[7]{Procheta Sen}
\author[8,9,10]{Ramin Lolachi}
\author[3]{Haonan Chen}
\author[2]{Manil Maskey}
\author[4]{Juan Bernab\'e-Moreno}
\author[2]{Rahul Ramachandran}
{\footnotesize
\affil[1]{Earth System Science Center, University of Alabama in Huntsville, AL, USA}
\affil[2]{NASA Marshall Space Flight Center, Huntsville, AL, USA}
\affil[3]{Department of Electrical \& Computer Engineering, Colorado State University, Fort Collins, CO, USA}
\affil[4]{IBM Research}
\affil[5]{Science and Technology Institute/Universities Space Research Association (USRA), Huntsville, AL, USA}
\affil[6]{Department of Computer Science, The University of Manchester, UK}
\affil[7]{School of Computer Science, University of Liverpool, Liverpool, UK}
\affil[8]{Center for Space Sciences and Technology, University of Maryland, Baltimore County, Baltimore, MD, USA}
\affil[9]{NASA Goddard Space Flight Center, Greenbelt, MD, USA}
\affil[10]{Center for Research and Exploration in Space Science and Technology, NASA/GSFC, Greenbelt, MD, USA}
\affil[*]{Equal contribution}
}

\date{}
  
\begin{document}
\maketitle

\begin{abstract}
\noindent 
The martian atmosphere hosts dynamical phenomena ranging from planet-encircling dust storms to mesoscale orographic clouds and nocturnal low-level jets. General circulation model show capability to simulate these phenomena, but is computationally expensive at resolution needed to resolve mesoscale features. While assimilation of satellite remote sensing observation enable forecasting capabilities using such models, observation record is often sparse, short and fragmented across instrument generators.
These constraints motivate the development of a data-driven foundation model for the Martian atmosphere.

Foundation models live in a complex design landscape. There is an interplay between the available data, the physics of the underlying processes and corresponding developments in AI. Even though the idea of a foundation model is to address multiple use cases in a data- and compute-efficient manner, it is important to have a clear picture what applications can sensibly addressed by a single model. 

The purpose of this paper is to elucidate this design landscape. We discuss available data ranging from atmospheric retrievals to reanalysis datasets as well as existing physical models. Moreover, we identify a wide range of candidate downstream applications. Finally, we consider relevant recent developments in artificial intelligence (AI) that can be leveraged in this context. Here, we put a particular emphasis on AI models for atmospheric physics, data-driven approaches to data assimilation as well as methods to work in a limited data setting.
\end{abstract}
\clearpage
\tableofcontents
\clearpage

\section{Introduction and Overview}
\label{sec:intro}

The atmosphere of Mars is thin, cold, and compositionally distinct from that of Earth. It exhibits diverse dynamical behavior that bears directly on both fundamental planetary science as well as the practical requirements of robotic and human exploration. Surface pressure averages less than 1\% of Earth's value (6 mbar, \cite{haberle2001possibility}). The atmosphere is composed almost entirely of carbon dioxide (95\% by volume), with nitrogen (2.8\%) and argon (2\%) making up most of the remainder, and trace amounts of water vapor whose presence and availability are of outsized scientific interest despite their negligible thermodynamic contribution \cite{mahaffy2013abundance,sanchezlavega2024dynamical}. The low pressure and temperature confine CO$_2$ and H$_2$O primarily to the gaseous and solid phases: stable liquid water is not present on the surface, and the CO$_2$ that constitutes the bulk of the atmosphere can freeze out onto the polar caps in winter, removing roughly a quarter of the total atmospheric mass and driving a seasonal surface-pressure cycle with a peak-to-peak amplitude of 25--30\% \cite{hess1979seasonal,hourdin1993meteorological,kelly2006seasonal,leighton1966behavior,michaels2017timothy}. These cycles are further shaped by Mars's large topographic contrasts, hemispheric asymmetry, and eccentric orbit, which together produce strong seasonal and regional variations in circulation, temperature, and volatile exchange \cite{richardson2002topographically,hansen2024comparison}. This large magnitude of mass exchange between atmosphere and surface has no analog on Earth and is a first-order constraint on any model that attempts to simulate martian weather beyond a few sols.

Two defining properties distinguish the martian atmosphere from its terrestrial counterpart. First, dust is ubiquitous and serves as the primary driver of atmospheric heating and circulation. Through absorption of solar radiation, dust warms the convective boundary layer on diurnal timescales, while at regional to global scales, suspended dust can substantially modify the thermal structure of the entire atmosphere \cite{fernandez1997martian,kulowski2017seasonal,wang2015origin,wang2023martian,sanchezlavega2024dynamical,kahre2006modeling,smith2004tes,kass2020mars}. Dust activity spans a wide range of spatial and temporal scales, from short-lived, microscale dust devils to planet-encircling storms that persist for months, with global events recurring on average every three Mars years \cite{cantor2001martian,cantor2006mars,wang2015duststorm}. Second, water vapor, although thermodynamically subdominant, plays an active role in atmospheric processes through radiative effects, cloud formation, and seasonal transport between low and high latitudes. This includes phenomena such as the Aphelion Cloud Belt encircling the tropics, orographic clouds over major volcanoes, and the Arsia Mons Elongated Cloud \citep{campbell2020estimating,hernandezbernal2021elongated,clancy1996water,smith2004tes,larsen2025mid,montmessin2004cloud}. Taken together, the martian atmosphere is governed by strong coupling among the CO$_{2}$, dust, and water cycles, and models that treat these components independently are unlikely to capture the cross-cycle feedbacks that drive the most scientifically and operationally relevant phenomena.

Numerical modeling of the martian atmosphere has a long history, spanning from the early general circulation model (GCM) studies of Leovy and Mintz to modern systems such as the NASA Ames Mars Global Climate Model (MGCM) \citep{haberle2019ames,harris2021gfdl} and the Mars configuration of the Weather Research and Forecasting model (MarsWRF) \citep{richardson2007planetwrf,skamarock2008wrf}. These models produce physically consistent representations of the global atmosphere and incorporate comprehensive physical parameterizations, including radiatively active dust, CO$_{2}$ condensation, water-ice microphysics, and gravity-wave processes. They form the foundation of both operational forecasting for landed missions and the reanalysis products that provide the most complete gridded representation of the martian atmospheric state.

A widely used reanalysis dataset is OpenMARS \cite{holmes2018reanalysis,holmes2020openmars}, which assimilates spacecraft retrievals from the Thermal Emission Spectrometer (TES) \citep{christensen1992tes,christensen2001tes,smith2004tes}, the Mars Climate Sounder (MCS) \cite{mccleese2007mcs}, the Atmospheric Chemistry Suite (ACS) \cite{korablev2018acs}, and the Nadir and Occultation for Mars Discovery (NOMAD) \cite{vandaele2018nomad} into an underlying GCM. OpenMARS provides a global record at $5^\circ \times 5^\circ$ horizontal resolution across Mars Years (MY) 24--35, with the most recent version, v5, spanning MY 34--35 and offering hourly cadence, doubled vertical resolution, and resolved water vapor. An independent reanalysis product, the Ensemble Mars Atmosphere Reanalysis System (EMARS) \cite{greybush2019emars}, offers a complementary estimate based on ensemble data assimilation methods.

The current modeling infrastructure remains subject to significant constraints. GCM integrations at resolutions matching the $\sim 5$--10~km footprint of the best orbital retrievals are computationally prohibitive for multi-Mars-year archives: a single sol at 54~km resolution in MarsWRF costs $\sim 34$ wall-hours, and the cost scales steeply with further refinement. The reanalysis record is short by Earth standards --- eight Mars years of continuous coverage, roughly fifteen Earth years --- and is fragmented across instrument generations with differing biases and orbital geometries. Many atmospheric phenomena of interest, including dust storm life cycles, frontal-system dynamics, and the diurnal signatures of water-ice clouds, require temporal resolution and spatial detail that the reanalysis only partially provides.

On Earth, the past three years have seen a transformation in weather prediction. Data-driven models trained on the ERA5 reanalysis now match or exceed operational numerical weather prediction at medium-range lead times \citep{pathak2022fourcastnet,keisler2022forecasting,bi2023pangu,lam2023graphcast,chen2023fengwu,chen2023fuxi,bonev2023sfno,lang2024aifs}. The paradigm is straightforward: a deep neural network $f_\theta$ takes two consecutive atmospheric states as input and predicts the next, and autoregressive application of $f_\theta$ extends the forecast arbitrarily far forward. Refinements including rollout tuning, latitude-weighted losses, and diffusion-based ensemble generation have pushed these models to operational status at ECMWF \citep{lang2024aifs,lang2026aifs}. Separately, self-supervised foundation models trained with masked-autoencoder and related objectives have demonstrated that a single pretrained backbone can be fine-tuned for multiple downstream tasks such as forecasting and downscaling \cite{nguyen2023climax,lessig2023atmorep,Schmude2024,bodnar2025foundation}.

Transferring this paradigm to Mars is well motivated but not straightforward. The single dominant reanalysis product, OpenMARS, has a coarser grid ($5^\circ$ versus $0.25^\circ$ for ERA5), fewer vertical levels (35 versus 137), and a shorter temporal record. The atmosphere itself is structurally different: the CO$_2$ mass cycle, the radiative dominance of dust, and the absence of liquid water mean that the loss functions, normalization strategies, and physical inductive biases that work on Earth do not transfer without modification. And the observational network is sparse, heterogeneous, and evolving: each Mars orbiter has a distinct orbital geometry, a distinct retrieval suite, and a finite mission lifetime, so any model that cannot fuse multiple instruments with varying resolutions and coverages will underuse the available data.

We present a scope, design, and preliminary implementation study for a Mars Atmospheric Foundation Model (MAFM): a single multi-scale model intended to represent, forecast, and detect the principal phenomena of the martian atmosphere. The paper has three goals. First, we characterize the data landscape in detail, including the OpenMARS reanalysis, five orbital retrieval suites (TES, MCS, ACS, NOMAD, and EMIRS), and the two operational GCMs (MGCM and MarsWRF), and identify the scope tradeoffs that the data structure imposes on the model. Second, we systematically define the downstream applications that the MAFM should serve including dust storms, frontal systems, double annular cyclones, low-level jets, orographic and elongated clouds, the Aphelion Cloud Belt, frost cycles, global mass-budget forecasting, spatial downscaling, and reanalysis--observation fusion, and link each application to the datasets that support it. Third, we present preliminary results from three architecture families:  a Mars-adapted GraphCast, based on a graph neural network; a Mars-adapted Prithvi-WxC, based on a vision transformer; and a Spherical Neural Operator trained on OpenMARS surface variables and evaluated on short-to-medium-range wind forecasting. We use this comparison to motivate a hybrid architecture direction.

The remainder of this paper is organized as follows. Section~\ref{sec:data} describes the data landscape: the OpenMARS reanalysis, the orbital retrieval suites, the two GCMs, the downstream-task--dataset linkage, the scope tradeoffs, and the candidate data-fusion strategies. Section~\ref{sec:applications} defines the downstream applications, organized by atmospheric phenomenon. Section~\ref{sec:ai_developments} discusses relevant AI developments. There are two major subsections. Section~\ref{sec:ai_recent_developments} focuses on recent developments in data-driven atmospheric modeling, data assimilation, and data efficiency relevant to MAFM design. This is followed by initial results towards an MAFM in Section~\ref{sec:ai_initial_results}. The section focuses on surface variable forecasting as well as possible ways to address the data-limited setting. Section~\ref{sec:discussion} synthesizes the findings and identifies the open questions that will guide the next phase of MAFM development.

\section{The Data Perspective}
\label{sec:data}

The martian atmospheric data record can be broadly divided into two categories: gridded reanalysis products, which provide a physically consistent and dynamically complete estimate of the global atmospheric state, and atmospheric retrievals from orbiting spacecraft, which offer finer spatial resolution, additional chemical species, and partial temporal redundancy through multi-instrument coverage. Technically speaking we also have observations made by the rovers. By their very nature they are of extremely limited coverage and thus of little use for training a planet-wide model beyond validation.

Reanalysis products such as OpenMARS offer the regularity, global coverage, and dynamical completeness that makes supervised training of a predictive model rather straightforward: every grid point at every timestep carries a full set of state variables, and the fields are mutually consistent by construction because they are produced by a general circulation model constrained by observations. The cost is coarse resolution ($5^\circ \times 5^\circ$ for OpenMARS), dependence on the underlying GCM where observations are sparse, and a limited variable set that excludes, for example, water vapor in all but the most recent release. In addition, one has to keep in mind that a model trained on a reanalysis dataset learns the dynamics of the reanalysis. Not that of the underlying system. Any imperfection of the dataset is a ceiling for the performance of the data-driven model.

Atmospheric retrievals, by contrast, resolve the atmosphere at scales an order of magnitude finer than the reanalysis and provide species and vertical detail that the reanalysis lacks, but they are spatially and temporally irregular, instrument-specific, and carry retrieval biases that vary across missions and viewing geometries.

Whether a MAFM should be pretrained on reanalysis, retrievals or some combination thereof is not a decision that can be made in the abstract. It depends on the selected downstream task, the target resolution, and the tolerance for bias. A model aimed at synoptic-scale forecasting may do well with reanalysis-only pretraining; a model aimed at resolving mesoscale cloud features or at ingesting new observations in real time may require direct exposure to retrieval data during training. In addition, one needs to consider the performance tradeoffs of corresponding AI approaches. We therefore begin by describing each data source on its own terms: its resolution, coverage, variable set, and known limitations.

\subsection{Reanalysis data}

Atmospheric data assimilation on Mars is particularly challenging compared to Earth. The main limitation comes from the observing system, which is sparse and irregular, relying almost entirely on orbital measurements with limited spatial and temporal coverage and limited in situ observations \cite{navarro2017challenge}. Because of this, the analysis tends to depend strongly on the underlying general circulation model (GCM), rather than being tightly constrained by observations. In addition, dust plays a central role in the martian atmosphere, but it is highly variable and not well observed, while also being strongly coupled to radiation and dynamics. This adds another layer of uncertainty and nonlinearity to the problem. Overall, the combination of limited observations, strong model dependence, and the lack of independent validation makes it difficult to assess the quality of the assimilated fields.

Despite these limitations, several reanalysis products have been developed to provide a consistent picture of the martian atmosphere. Examples include Ensemble Mars Atmosphere Reanalysis System (EMARS) Version 1.0 \cite{greybush2019ensemble}, OpenMARS dataset, and Mars Analysis Correction Data Assimilation (MACDA). These systems combine Mars GCMs with available orbital retrievals using either variational or ensemble-based data assimilation methods. While they provide the best available estimates of the atmospheric state, differences in model configuration, assimilation approach, and dust representation lead to noticeable differences between datasets. Therefore, comparing multiple reanalysis datasets is often necessary to quantify uncertainties in the representation of processes in the martian Atmosphere.

\subsubsection{The OpenMARS Reanalysis}
\label{sec:openmars}

OpenMARS \cite{holmes2018openmarsOzone,holmes2020openmars} is a gridded, time-dependent reanalysis of the martian atmosphere generated by combining the Mars Global Circulation Model (MGCM) with the assimilation of spacecraft observations. Five versions of OpenMARS have been published, of which the latter four are currently publicly accessible. These versions differ primarily in the assimilated instrument data, temporal coverage across martian years, and spatial and temporal resolution of the reanalysis output. We first summarize the shared structure of the OpenMARS products before describing the version-specific differences among the four accessible versions. See section \ref{sec:mgcm} for a discussion of the MGCM.

\paragraph{Assimilation scheme.}
The reanalysis makes use of the ``AC scheme'' \cite{lorenc1991meteorological} in which analysis steps are interleaved with each model dynamical time step. At any given time step, corrections are made using
\begin{equation}\label{eq:open_mars_assimilation_scheme}
    \mathbf{x} = \mathbf{x}_b + \mathbf{B}\mathbf{H}^T\mathbf{R}^{-1}\mathbf{\tilde{Q}}(\mathbf{y}_o - H(\mathbf{x}_b)).
\end{equation}
Here, $\mathbf{x}$ is the analysis, $\mathbf{x}_b$ the background, $B^{-1}$ the background error covariance and $R^{-1}$ the covariance of the combined observation and observation operator error. $H$ is the observation operator and $\mathbf{H}$ its linearization. $\mathbf{\tilde{Q}}$ finally contains normalizing factors.\footnote{For readers whose background is primarily in AI, note that update schemes like \eqref{eq:open_mars_assimilation_scheme} can be derived from variational objectives like
\begin{equation*}
  J(\mathbf{x}, \mathbf{x}_b) = \frac{1}{2} \left(\mathbf{x} - \mathbf{x}_b\right)^T \mathbf{B}^{-1} \left(\mathbf{x}-\mathbf{x}_b\right) + \frac{1}{2} \left(\mathbf{y}_o - H(\mathbf{x})\right)^T R^{-1} \left(\mathbf{y}_o - H(\mathbf{x})\right).
\end{equation*}
See section \ref{sec:ai_data_assimilation} for further remarks on data assimilation. In particular in the context of AI.}

\paragraph{Dataset and coordinates.}
OpenMARS is defined on a structured latitude--longitude grid over the sphere $\mathbb{S}^2$, parameterized by
\begin{equation}
(\lambda, \phi) \in [-\pi, \pi) \times \left[-\frac{\pi}{2}, \frac{\pi}{2}\right],
\end{equation}
with $N_\lambda = 72$ longitude points and $N_\phi = 36$ latitude points, giving a horizontal resolution of $5^\circ \times 5^\circ$.

\paragraph{Time coordinate.}
The temporal coordinate is a sequence of $N_t$ discrete samples,
\begin{equation}
t \in \mathcal{T} = \{t_1, t_2, \dots, t_{N_t}\},
\end{equation}
where each $t_i$ is expressed in martian sols (a sol is 24 hours, 39 minutes, and 35 seconds) since a reference epoch. The standard versions have $N_t = 360$ per Mars year at a cadence of $\Delta t \approx 2$~h ($\approx 0.0833$~sol); version 5 doubles the cadence to $\Delta t \approx 1$~h and therefore has $N_t = 720$ per Mars year. Two auxiliary temporal variables accompany each timestep: the solar longitude $\Ls : \mathcal{T} \to [0, 360^\circ)$, measured from the northern hemisphere spring equinox, and the Mars year $\MY : \mathcal{T} \to \mathbb{Z}$, with MY~1 beginning on 11~April 1955.

\paragraph{Vertical coordinate.}
The vertical coordinate is a dimensionless terrain-following sigma coordinate,
\begin{equation}
\sigma = \frac{p}{p_s}, \qquad \sigma \in (0,1],
\end{equation}
where $p$ is atmospheric pressure and $p_s(\lambda,\phi,t)$ is surface pressure. The standard versions use $N_\sigma = 35$ monotonically decreasing levels with
\begin{equation}
\sigma_1 \approx 0.9995, \qquad \sigma_{N_\sigma} \approx 5.08 \times 10^{-5},
\end{equation}
corresponding to pressure levels from the surface to $\sim 0.01$~Pa ($\sim 105$~km altitude). Version 5 doubles the vertical resolution to $N_\sigma = 70$. The sigma coordinate is terrain-following by construction: $\sigma = 1$ is the surface everywhere, $\sigma \to 0$ is the upper atmosphere. We emphasize that the definition $\sigma = p/p_s$ is a coordinate transformation. It does not imply that the pressure field is determined by the surface pressure, which evolves under the governing dynamical and thermodynamical equations.

\paragraph{Variables.}
Surface fields $f_s : \mathbb{S}^2 \times \mathcal{T} \to \mathbb{R}$ are summarized in Table~\ref{tab:openmars-surface}; atmospheric fields on the sigma levels are summarized in Table~\ref{tab:openmars-atmos}. The horizontal wind is provided as its eastward ($u$) and northward ($v$) components on each sigma level,
\begin{equation}
\vu(\lambda,\phi,k,t) = u(\lambda,\phi,k,t)\,\vh{\lambda} + v(\lambda,\phi,k,t)\,\vh{\phi},
\end{equation}
with $\vh{\lambda}$ and $\vh{\phi}$ the unit vectors in the zonal and meridional directions and k indexing the ${\sigma}$ levels. No vertical velocity is provided in any OpenMARS version.

\begin{table}[t]
\centering
\caption{Surface variables in OpenMARS. Starred entries appear only in v5 (MY~34--35); daggered entries appear only in the non-standard MY~24--25 water-vapor release.}
\label{tab:openmars-surface}
\begin{tabular}{@{}llll@{}}
\toprule
Symbol & Field & Dimensions & Units \\
\midrule
$p_s$ & Surface pressure & $(\lambda, \phi, t)$ & Pa \\
$T_\text{surf}$ & Surface temperature & $(\lambda, \phi, t)$ & K \\
$\CO^\text{ice}$ & Surface \CO{} ice mass & $(\lambda, \phi, t)$ & kg m$^{-2}$ \\
$\taud$ & Visible column dust optical depth & $(\lambda, \phi, t)$ & -- \\
$\mathrm{vapcol}^{*\dagger}$ & Water vapor column abundance & $(\lambda, \phi, t)$ & kg m$^{-2}$ \\
\bottomrule
\end{tabular}
\end{table}

\begin{table}[t]
\centering
\caption{Atmospheric variables in OpenMARS, defined on sigma levels. Starred entries appear only in v5.}
\label{tab:openmars-atmos}
\begin{tabular}{@{}llll@{}}
\toprule
Symbol & Field & Dimensions & Units \\
\midrule
$T$ & Atmospheric temperature & $(\lambda, \phi, k, t)$ & K \\
$u$ & Zonal (eastward) wind & $(\lambda, \phi, k, t)$ & m s$^{-1}$ \\
$v$ & Meridional (northward) wind & $(\lambda, \phi, k, t)$ & m s$^{-1}$ \\
$\mathrm{vmr\_h2ovap}^*$ & Water vapor volume mixing ratio & $(\lambda, \phi, k, t)$ & ppmv \\
\bottomrule
\end{tabular}
\end{table}

\paragraph{Versions.}
Four OpenMARS releases are publicly available and differ primarily in the spacecraft observations assimilated into the reanalysis. Table~\ref{tab:openmars-versions} summarizes these releases. The shortest span, MY~24--25\_WV, is a non-standard release designed for water vapor studies and is the only pre-v5 product that provides vapor column data. The longest continuous span, v4, spans MY~28--35, and is therefore the natural choice for multi-year pretraining focused on atmospheric of dynamics, dust, and the \CO{} cycle. Version 5 spans MY 34--35 and is the only release that provides both hourly cadence and resolved water vapor.

\begin{table}[t]
\centering
\caption{OpenMARS versions, assimilated spacecraft instruments, and temporal coverage. $\tau$ denotes dust opacity. The MY~24--25\_WV release is non-standard and water-vapor-specific.}
\label{tab:openmars-versions}
\small
\begin{tabular}{@{}lllll@{}}
\toprule
Version & Instruments assimilated & Mars-year span & $N_\sigma$ & $\Delta t$ \\
\midrule
MY~24--25\_WV (non-std.) & TES & $\Ls{=}173^\circ$ MY24 -- $\Ls{=}186^\circ$ MY25 & 35 & 2 h \\
MY~24--27 (v2) & TES ($T$-profiles, $\tau$) & MY 24--27 & 35 & 2 h \\
MY~28--32 (v3) & MCS ($T$-profiles, $\tau$) & MY 28--32 & 35 & 2 h \\
MY~28--35 (v4) & MCS ($T$-profiles, $\tau$) & MY 28--35 & 35 & 2 h \\
MY~34--35 (v5) & MCS, NOMAD, ACS & $\Ls{=}159^\circ$ MY34 -- $\Ls{=}359^\circ$ MY35 & 70 & 1 h \\
\bottomrule
\end{tabular}
\end{table}

\subsection{Atmospheric retrievals}
\label{sec:spacecraft}

The $5^\circ \times 5^\circ$ spatial resolution of OpenMARS is insufficient to resolve many of the atmospheric processes targeted in Section 3. Higher-resolution observational constraints from atmospheric retrievals are therefore required for both validation and refinement at finer spatial scales. We thus discuss retrieval products spanning five mission generations. Each instrument contributes a distinct observational capability while also introducing specific sampling characteristics and retrieval limitations. MGS/TES provides historical context as it is the only instrument covering MY 24-27 at sufficient fidelity. MRO/MCS establishes a long-term baseline. TGO/ACS and TGO/NOMAD offer high-vertical-resolution constraints on trace gases and aerosols. Finally, Hope/EMIRS is unique in that it provides full-disk diurnal coverage.

\paragraph{Thermal Emission Spectrometer (TES).}
The Thermal Emission Spectrometer was a nadir-sounding instrument on NASA's Mars Global Surveyor (MGS) \citep{Christensen1992tesMission,Christensen2001tesRes}. TES operated from March 1999 to August 2004 in a sun-synchronous orbit, with equator-crossing times near 02:00 and 14:00 local solar time \citep{smith2004tes}. It acquired 12 narrow orbital strips per sol and retrieved atmospheric temperature profiles, visible column dust opacity,and water vapor column abundances. Temperature retrievals extended to $\sim 40$~km altitude, with a vertical resolution of $\sim 10$~km, or roughly one pressure scale height. Water vapor column retrievals had a horizontal sampling of $\sim 10$~km. TES is the earliest spacecraft remote sensing dataset in our atmospheric retrieval pool and the only dataset that covers MY~24--27 at this level of fidelity. It is therefore essential for cross-year analyses that extend to the pre-MCS era.

\paragraph{Mars Climate Sounder (MCS).}
The Mars Climate Sounder (MCS) is a passive nine-channel radiometer on NASA's Mars Reconnaissance Orbiter (MRO) that acquires limb and nadir measurements from visible through far infrared wavelengths \citep{mccleese2007mcs}. MCS has operated since September 2006 in a sun-synchronous orbit, with equator-crossing times near 03:00 and 15:00 local solar time \citep{kleinbohl2009mcsThermaltide}. Its limb viewing strategy provides substantially greater sensitivity than nadir viewing for vertical atmospheric profiling. In-track scans retrieve vertical profiles of temperature, water-ice opacity, and dust opacity with a vertical resolution of $\sim 5$~km, extending to $\sim 85$~km altitude. Observations are acquired in narrow swaths with $\sim 8.9$~km field of view (FOV) width, although, the projected sample length of along-track limb soundings is $\sim 200$~km. Observations are made every 1.86 degrees along the track (110km) with  12 orbits per sol providing longitudinal coverage. Interspersed side looking cross-track, and off-track scans broaden local time sampling and have been used to investigate atmospheric phenomena such as martian thermal tides \citep{kleinbohl2009mcsThermaltide}.  MCS provides the longest continuous record of any instrument in our retrieval pool, spanning 2006 to the present, and the along-track observations form the backbone of OpenMARS versions 3--5.

\paragraph{Atmospheric Chemistry Suite (ACS).}
The Atmospheric Chemistry Suite (ACS) on ESA's ExoMars Trace Gas Orbiter (TGO) entered Mars orbit in October 2016 and began routine science observations in April 2018 at an altitude of $\sim 400$~km \citep{korablev2018acs}. ACS consists of three spectrometers covering near(0.7-1.7 $\mu$m), mid (2.3-4.6 $\mu$m), and far (1.7-17 $\mu$m) infrared wavelengths. The instrument operates in solar occultation and limb/nadir geometries, and retrieves profiles of temperature, \CO, dust opacity, and water-ice mixing ratios. ACS products have a vertical resolution of $\sim 2$~km, horizontal resolution of 12--15~km, and extend to $\sim 60$--65~km altitude. Unlike MGS and MRO, TGO is not in a sun-synchronous orbit but completes roughly 12 orbits per sol. As discussed below, this orbital geometry leads to an approximate 108 sol revist time for any given geographic location. 

\paragraph{Nadir and Occultation for Mars Discovery (NOMAD).}
The Nadir and Occultation for Mars Discovery (NOMAD) instrument, also on TGO, operates in solar occultation and limb/nadir geometries across ultraviolet to mid-infrared wavelengths \citep{vandaele2018nomad}. NOMAD retrieves vertical profiles of temperature, atmospheric composition, clouds, and dust with a vertical resolution of $\sim 1$~km and horizontal resolutions of $\sim 5$~km (nadir viewing) and $\sim 0.5$~km x $17$~km (limb), covering altitudes from surface to $\sim 200$~km. Because ACS and NOMAD have different boresights and alternating occultation times, the two instruments produce distinct, alternating water vapor profiles rather than a single unified record. NOMAD provides the highest vertical resolution of any instrument listed here and is the only pre-Hope dataset that samples high thermosphere.

\paragraph{Emirates Mars Infrared Spectrometer (EMIRS).}
The Emirates Mars Infrared Spectrometer (EMIRS) on the UAE Hope mission has operated in martian orbit since February 2021 in a highly elliptical orbit ranging from $\sim 20\,000$~km to $\sim 43\,000$~km altitude \citep{edwards2021emirs}. Hope's orbital geometry is unusual among Mars assets. The spacecraft completes $\sim 0.44$ orbits per sol, but each orbit provides full-disk coverage of Mars. Repeated observations of a given region reconstruct the full diurnal cycle. EMIRS retrieves column dust opacity, column water-ice opacity, water-vapor column abundance, and atmospheric temperature profiles. Temperature profiles extend to 50 km altitude with a vertical resolution of 10 km, and retrievals have a horizontal resolution of 100--300~km. Although EMIRS is spatially coarser than MCS, ACS, or NOMAD, its diurnal coverage is unique and provides the only dataset in our pool capable of characterizing a given region across all local times with a single Mars year. EMIRS is not currently assimilated into any OpenMARS release.

\begin{table}[t]
\centering
\caption{Salient atmospheric retrievals. Stars indicate data not currently assimilated into OpenMARS. Daggers denote the along-track sample length of the limb observation. ($\tau_d$ is column dust optical depth and $\tau_i$ is column water-ice optical depth)}
\label{tab:satellite}
\small
\begin{tabular}{@{}p{2.3cm}p{3.2cm}p{1.9cm}p{1.9cm}p{1.7cm}p{3.5cm}@{}}
\toprule
Instrument & Retrievals & scanning method & Horizontal res. & Vertical res. & Temporal coverage \\
\midrule
TES (MGS) & $T_\text{surf}$, $p_s$, $T$ profiles, $\taud$, $\tau_i$, $\HO$ vapor column & nadir & $\sim 10$ km (vapor) & $\sim 10$ km, up to $\sim 40$ km & $\Ls{=}90^\circ$ MY 24 -- $\Ls{=}90^\circ$ MY 27 \\
MCS (MRO) & $T_\text{surf}$, $p_s$, $T$, dust, $\taud$, $\tau_i$, $\HO$ vapor, $\HO$ ice, \CO{} ice columns & nadir, limb & $\sim 8.9$ km, 200 km$^{\dagger}$ & $\sim 5$ km, up to $\sim 85$ km & 2006-09-24 -- present \\
ACS (TGO) & $T$, \CO, \CO{} ice, $\HO$ vapor, $\HO$ ice, dust & nadir, limb, occultation & 12--15 km & $\sim 2$ km, up to $\sim 65$ km & 2018-04 -- present, $\sim 108$ sol revisit \\
NOMAD (TGO) & $T$, composition, \CO, \CO{} ice, $\HO$ vapor, dust, $\taud$, $\tau_i$ & nadir, limb, occultation & $\sim 5$ km,  0.5 x 17 km & $\sim 1$ km, up to $\sim 200$ km & MY 34--36 (2019--2022) \\
EMIRS$^*$ (Hope) & $T$, $\taud$, $\tau_i$, $\HO$ vapor, full-disk diurnal coverage & nadir & 100--300 km & 5--10 km & 2021-02 -- present \\
\bottomrule
\end{tabular}
\end{table}

\subsection{Mars Global Climate Models as Dual-Role Data Sources}
\label{sec:gcms}

Having discussed data sources -- reanalysis and spacecraft retrievals -- it is time to turn to models. These can serve a dual purpose for an MAFM. To start, one can use them to generate additional data to leverage during training, thus infusing the trained FM with model physics. Second, they provide baselines for forecasting tasks against which the FM can be evaluated. We describe each model in turn.

\subsubsection{MarsWRF}
\label{sec:marswrf}

MarsWRF is the Mars-specific global configuration of the planetWRF model \citep{richardson2007planetwrf}, which is built upon the terrestrial Weather Research and Forecasting (WRF) model developed at the National Center for Atmospheric Research \citep{skamarock2008wrf}. WRF is one of the most widely used atmospheric modeling systems for Earth, with applications spanning regional weather forecasting, climate downscaling, and large-eddy simulations (LES) of boundary-layer turbulence. Its dynamical core is fully compressible and non-hydrostatic, uses a terrain-following hydrostatic pressure vertical coordinate on an Arakawa C-grid and supports one- and two-way nesting for high resolution applications. 

MarsWRF was developed by adapting WRF for global planetary use \cite{richardson2007planetwrf}, including generalized grid support for non-conformal projections, abstracted planetary parameters, implementation of the Ls calendar, and replacement of the terrestrial physics suite with Mars-specific parameterizations. MarsWRF has since been incorporated into the official NCAR WRF releases (\url{www.wrf-model.org}), while Mars-specific physics parameterizations are distributed through the planetWRF project (\url{www.planetwrf.com}).

In this context, MarsWRF effectively bridges the gap between global and limited-area models. The same framework can be run as a coarse global GCM, as a global model with embedded high-resolution nests, or, in principle, as a stand-alone limited-area model. In practice, global integrations and global simulations with nested domains are the most common configurations, distinguishing MarsWRF from mesoscale-only Mars models such as MRAMS \citep{rafkin2001mrams}. Published global MarsWRF configurations span horizontal grid spacings from $\sim$500 km to $\sim$30 km \citep{toigo2012resolution} while nested configurations have reached horizontal grid spacing of $\sim$490 m \citep{newman2017bagnold}. Large eddy simulation (LES)-coupled runs extend to still finer scales \citep{wu2021les, temel2021les}. Non-hydrostatic effects become important on Mars primarily at sub 10 km scales, where they are needed to represent slope flows on Tharsis, deep dry convection in the 8-10 km martian planetary boundary layer (PBL) and gravity waves over steep topography.

The Mars physics suite includes the fast radiative transfer scheme of \citet{mischna2012radiation}, handling the 15 $\mu$m CO2 band, near-IR CO2 absorption and radiatively active dust and water ice; a fully coupled dust cycle with interactive lifting via a moment method (or prescribed scenarios); CO2 condensation and sublimation tied to the polar caps; water ice microphysics; and Mars-tuned PBL schemes. Data assimilation uses the Data Assimilation Research Testbed (DART), an Ensemble Kalman Filter (EnKF) framework demonstrated with MGS TES radiances \citep{lee2011dart} and extensible to other observations including MRO and Curiosity. This is one of three distinct DA approaches in current Mars practice, alongside the LMD nudging scheme and the LETKF used in EMARS, an ensemble Mars reanalysis.

For an MAFM, MarsWRF provides a configurable simulation environment in which initial conditions can be drawn from EMARS reanalysis fields and integrated forward at horizontal resolutions substantially finer than OpenMARS. In practice, the workflow constructs a grid-consistent input template at the target resolutions and inserts the EMARS atmospheric state into the template prior to integration. The experiments are run as global integrations rather than as nested limited-area forecasts driven by time-varying lateral boundary conditions.

\paragraph{Computational cost.}
The primary constraint on using MarsWRF as a data source is computational cost, which also motivates the FM approach. We performed short, one-hour integrations with $dt$ equal to 4 times the horizontal grid spacing (in km) to characterize how computational expense scales with horizontal resolution and vertical-level count. All experiments were conducted using 64 MPI tasks with approximately 8~GB memory allocated per CPU. Table~\ref{tab:marswrf_cost} summarizes the results. For 21 number of vertical levels, walltime per martian hour increases sharply with increasing resolution, from $\sim$9.3~s at 216~km to $\sim$3.0~min at 108~km and $\sim$27.6~min at 54~km. Accounting for the $\sim$2.75\% difference between a martian sol and an Earth day (1 martian hour = $24.6597/24 \approx 1.0275$ Earth hours), the extrapolated walltimes per sol are approximately $\sim$3.8~min at 216~km, $\sim$1.2~h at 108~km, and $\sim$11.2~h at 54~km. Increasing the number of vertical levels from 21 to 51 approximately doubles the computational cost across the tested configurations.

\begin{table}[t]
\centering
\caption{MarsWRF computational cost from short (1-hour) integrations, extrapolated to one martian sol. Walltime scales steeply with horizontal resolution.}
\label{tab:marswrf_cost}
\begin{tabular}{@{}lll@{}}
\toprule
Horizontal grid spacing & Walltime per martian hour & Walltime per sol \\
\midrule
216 km & $\sim 9.3$ s & $\sim 3.8$ min \\
108 km & $\sim 3.0$ min & $\sim 1.2$ h \\
54 km & $\sim27.6$ min & $\sim 11.2$ h \\
\bottomrule
\end{tabular}
\end{table}

These scaling results have two implications for an MAFM. First, MarsWRF can in principle generate data at resolutions comparable to the FM's $2^\circ \times 2^\circ$ ($\sim 120$~km) target grid which one could use for pretraining. However, producing multi-Mars-year global simulations at this resolution is non-trivial. At 108~km resolution, a one-Mars-year integration requires approximately 669 sols of simulation, corresponding to nearly $\sim803$ wall-hours on 64 cores under the tested configuration. Generating a high resolution, year scale simulation archive is therefore a deliberate computational investment rather than a byproduct of standard modeling workflows. Second, MarsWRF provides a controlled validation environment. Independently of how one pretrains an MAFM, it can be evaluated against fine-resolution MarsWRF simulations for targeted case studies, such as frontal passage, regional dust storms, detached aerosol cloud (DAC) events, or other dynamically coherent atmospheric phenomena. In this role, MarsWRF serves a function analogous to operational NWP baselines in Earth atmosphere FM studies.

\subsubsection{Mars Global Climate Model}
\label{sec:mgcm}

The NASA Ames Mars Global Climate Model (MGCM) traces its lineage to the early Mars GCM work of Leovy and Mintz in 1969 and the Pollack-Haberle Ames model developed during the 1980s and 1990s. Version 3.2 couples an external finite-volume dynamical core from NOAA/GFDL \citep{harris2021gfdl} with the physics package from the legacy Ames MGCM \citep{haberle2019ames}. The model uses a cubed-sphere grid composed of six tiles, two centered at the poles and four around the equator. The cubed-sphere construction mitigates the polar singularity associated with latitude-longitude grids, which is particularly important for Mars given the central role of the polar caps in the \CO{} mass cycle introduced in Section~\ref{sec:massbudget}. The tile structure also supports efficient parallel computation.

Ames MGCM v3.2 supports horizontal grid configurations denoted C24, C48, C96, and C192, corresponding approximately to grid spacings of $3.75^\circ$ ($\sim 221$~km), $1.875^\circ$ ($\sim 110$~km), $0.938^\circ$ ($\sim 55$~km), and $0.469^\circ$ ($\sim 27.5$~km), respectively. The model also includes a stretching-grid capability, in which a base cubed-sphere grid is locally refined toward a region of interest. This enables focused higher resolution simulations without requiring the cost of a globally fine grid.

In the vertical, MGCM uses a terrain-following hybrid sigma-pressure coordinate. The default high resolution configuration has 56 sigma levels, with vertical spacing of $\sim 5$~m near the surface and $\sim 4$--$5$~km near the model top at $\sim 90$~km altitude ($\sim 0.1$~Pa). This vertical structure contrasts sharply with OpenMARS. In particular, MGCM's near-surface default vertical resolution is two orders of magnitude finer than that of OpenMARS, giving it a structural advantage for boundary-layer phenomena such as the LLJ (Section~\ref{sec:llj}).

The MGCM physics suite is comparatively comprehensive for Mars climate applications. It includes radiative transfer, radiatively interactive dust with lifting, transport, and sedimentation; \CO{} condensation and sublimation tied to the polar caps; water vapor and water-ice microphysics with radiatively active water ice; and gravity-wave parameterizations, in addition to the other components expected of a full GCM. Prognostic and diagnostic outputs include surface and atmospheric state variables (surface temperature, surface pressure, atmospheric temperature, winds), dust (opacity, mass mixing ratio, surface wind stress tied to lifting), surface \CO{} ice, water vapor mass mixing ratio, and water ice (radiatively active opacity, mass mixing ratio, surface frost). This output space is a strict superset of the OpenMARS variable set. In particular, MGCM provides $\HO$ frost and explicit water-ice microphysics which OpenMARS does not include, allowing MGCM output to supply training signal for downstream tasks that OpenMARS cannot structurally support.

MGCM is typically run as a free-running climate model, but it also underpins EMARS \citep{greybush2019ensemble} and is used with assimilation of MARCI, TES, MCS, and lander data to generate short-term, 2–5 sol operational forecasts for rover and entry, descent, and landing (EDL) support. Beyond computational cost, one important limitation is the reliance on a restricted set of prescribed dust scenarios, including MY24, MY30, MY31, MY34, and a background case, rather than an open-ended inventory of dust states.

\paragraph {Computational cost} 
The computational cost of MGCM depends strongly on horizontal resolution, physics package, timestep, and processor count. For 2 martian years of integration, dust-only benchmarks include $\sim 2.78$ h on 72 processors at C24 ($\sim 221$ km), $\sim 12.84$ h on 72 processors at C48 ($\sim 110$ km), and $\sim 7.64$ h on 72 processors for C24 with roughly threefold stretching. A dust-only C96 case, corresponding to $\sim 55$ km resolution, was estimated near 69 h on 96 processors by scaling from a shorter segment (25 h × 1336/483 sol). Adding microphysics substantially increases the cost, largely due to the smaller timesteps required for these extra calculations. At C48 the cost increases to $\sim 61$ h or $\sim 89.3$ h (73.67 h × 1336/1102 sol) depending on the segment normalization used in the benchmark.

In the context of an MAFM, MGCM can provide a complementary role to OpenMARS. Its cubed-sphere architecture, optional stretching grid, and comparatively rich physics suite make it useful for generating physically self-consistent training and validation fields in regimes where OpenMARS lacks either resolution or prognostic variables. In particular, MGCM can provide training signal for water-ice microphysics, surface frost, polar CO2 exchange, and near-surface boundary-layer structure. At the same time, its dependence on prescribed dust scenarios and its increasing cost at higher resolution mean that MGCM is best treated as a targeted augmentation and validation source.

\begin{figure}[!h]
    \centering
    \begin{minipage}{0.32\linewidth}
        \centering
        \includegraphics[width=\linewidth, keepaspectratio]{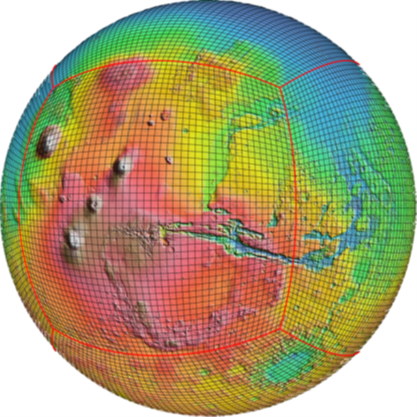}
    \end{minipage}
    \hfill
    \begin{minipage}{0.6\linewidth}
        \centering
        \includegraphics[width=\linewidth]{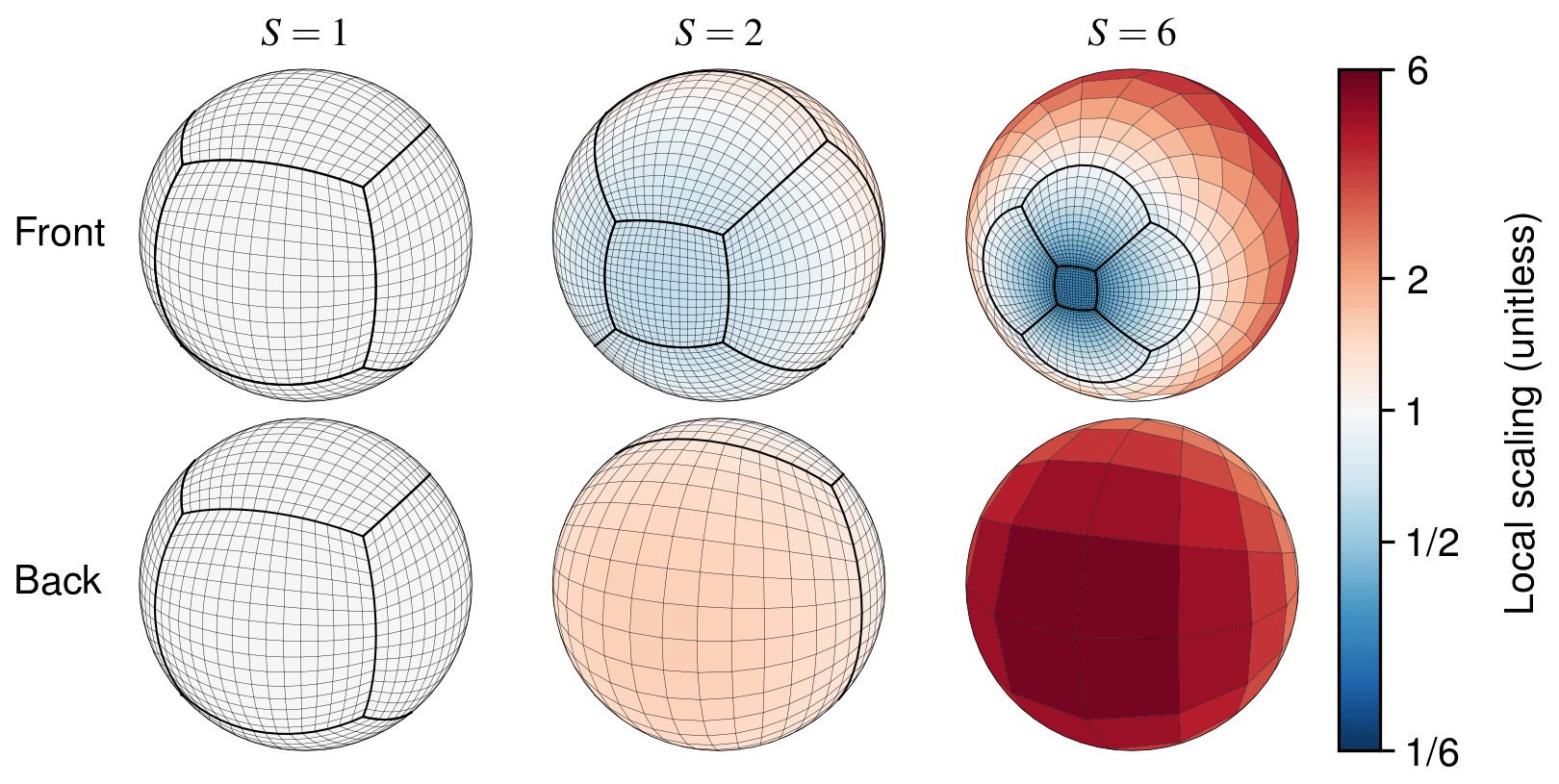}
    \end{minipage}
    \caption{(left) An illustration of Ames MGCM in c48 (1.875°) cubed-sphere grid over martian topography. Tile boundaries are drawn in red. (from Ames MGCM User Guide); (right) Three stretched grids that illustrate the effect of stretch factor (S) on stretching a C16 cubed-sphere. (From Bindle et al., 2021)}
\end{figure}

\begin{figure}[!h]
    \centering
    \begin{minipage}{0.32\linewidth}
        \centering
        \includegraphics[width=\linewidth, keepaspectratio]{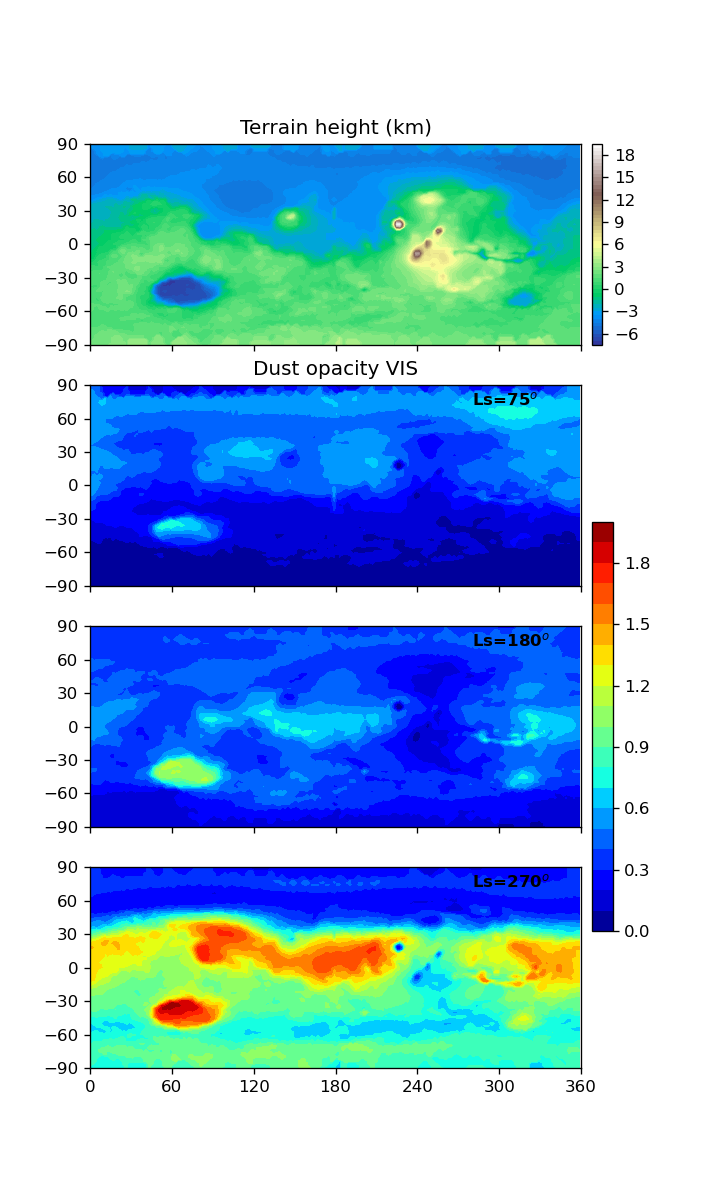}
    \end{minipage}
    \hfill
    \begin{minipage}{0.6\linewidth}
        \centering
        \includegraphics[width=\linewidth]{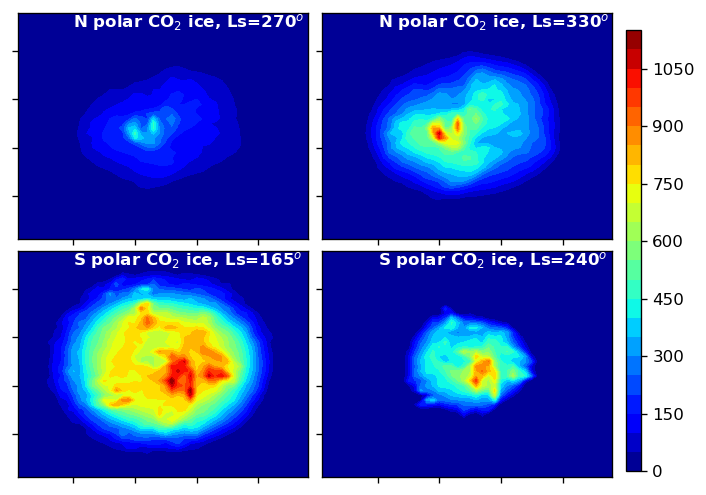}
    \end{minipage}
    \caption{(left) The simulation ran for 2 martian years using the same dust scenario for MY30. The first year run is used as spin-up. The second year simulation results are interpolated to a $2\degree$ $\times$ $2\degree$ latitude-longitude grid for analysis. The first panel is the Mars terrain used in the model. The bottom 3 panels show the simulated 5-day averaged column dust opacity centered at Ls= $75\degree$, $180\degree$, and $270\degree$, which correspond to the 139th, 334th, and 501th sols of the MY, respectively; (right) Simulated 5-day average surface CO$_2$ ice centered at  Ls=$270\degree$ and $330\degree$ for the north pole (top panels) and Ls=$165\degree$ and $240\degree$ for the south pole (bottom panels). These results are taken directly from the model output for the north and south polar tiles with interpolation to the latitude-longitude grid.}
\end{figure}

\subsection{Linking Downstream Tasks to Datasets}
\label{sec:task-data-link}

The phenomena introduced Section~\ref{sec:applications} and the datasets described above are coupled through a task-dataset mapping that determines, for each downstream task, which data sources supply the primary signal, which provide validation, and which are required rather than merely helpful. Table~\ref{tab:task-data} compiles this mapping. The structure of the table follows the ordering in Section~\ref{sec:applications}, so the first seven rows correspond to the phenomenon-centered tasks in Sections \ref{sec:dust}--\ref{sec:acb}, and the following four rows to the frost cycle, mass-budget forecasting, downscaling, and observation fusion.

Several patterns in the table are worth noting. Tasks tied to the water cycle (rows for Aphelion Cloud Belt (ACB), Arsia Mons Elongated Cloud (AMEC), orographic clouds) require OpenMARS v5 as the primary reanalysis, because v4 and earlier do not carry water vapor. Tasks tied to the \CO{} cycle and dust are well-served by v4's eight-Mars-year span and are less sensitive to the v4-versus-v5 choice. The Low-Level Jet (LLJ) is the only task for which spacecraft retrievals do not provide direct observational constraints; however, they contribute indirectly by improving the representation of the broader atmospheric state, which in turn influences the evolution and characteristics of the LLJ. The FM must therefore rely on reanalysis for explicit LLJ supervision, with performance evaluated against MGCM-resolved LLJs. And the final two rows, downscaling and fusion, are not phenomenon-specific: they are cross-cutting capabilities that the FM exercises in service of every other task.

\begin{sidewaystable}
\centering
\caption{Linkage between downstream tasks (Section~\ref{sec:applications}), scales, required reanalysis version, and supporting atmospheric retrievals. $\Delta t$ denotes the temporal resolution required to resolve the phenomenon.}
\label{tab:task-data}
\scriptsize
\begin{tabular}{@{}p{0.3cm}p{2.4cm}p{2.0cm}p{2.2cm}p{2.3cm}p{0.7cm}p{2.4cm}p{2.8cm}p{2.6cm}@{}}
\toprule
\# & Downstream task & Spatial scale & Lifespan / temporal scale & Season / window & Req. $\Delta t$ & Primary reanalysis & Supporting spacecraft data & Key variables \\
\midrule
1 & Dust storms (regional -- global) & $10^2$--$10^4$ km & 1--3 sol (local); $>3$ sol (reg.); $>30$ sol (global); global every $\sim 3$ MY & S. spring/summer, $\Ls{=}180^\circ$--$360^\circ$ & hours--sols & OpenMARS v4; v5 for hourly & MCS, TES, NOMAD, EMIRS (full-disk diurnal) & $\taud$, $T$, $T_\text{surf}$, $p_s$, $\vu$ \\
\addlinespace
2 & Frontal systems & $10^3$--$10^4$ km (synoptic) & $\sim 1$ week & N. hem. fall/winter, mid-to-high lat. & hours & v4; v5 for hourly & MCS ($T$, $\taud$), TES, EMIRS & $p_s$, $T$, $T_\text{surf}$, $\vu$, $\taud$, $\CO^\text{ice}$ \\
\addlinespace
3 & Double Annular Cyclones (DAC) & 1200--1600 km diameter, 100--300 km cloud-free core & seasonal recurrence; diurnal strengthening & N. hem. summer, $\Ls \approx 115^\circ$--$135^\circ$, $\sim 60^\circ$N & 1--2 h & v5 (hourly, with $\HO$) & MCS, NOMAD ($\HO$ ice), EMIRS & $\HO$ ice column, $\vu$, $T$, $p_s$ \\
\addlinespace
4 & Low-level jets (LLJ) & widespread; crater and plain-localized & nocturnal max at $\sim 1$ km, 03:00--07:00 LT & fall/winter peak; daily & 2 h & v4/v5 (reanalysis-only) & none (no wind retrievals) & $u, v$ on lower $\sigma$-levels, $T$, $p_s$ \\
\addlinespace
5 & Orographic volcano clouds & up to $\sim 10^3$ km (major volcanoes only) & daily recurrence & aphelion peak, $\Ls{=}50^\circ$--$150^\circ$ & 1--2 h & v5 (with $\HO$) & MCS, NOMAD, EMIRS ($\HO$ ice) & $\vu$, $T$, vmr\_h2ovap, $\tau_i$ topography \\
\addlinespace
6 & Arsia Mons Elongated Cloud (AMEC) & up to 1800 km tail & forms pre-dawn, dissipates by midday, $\sim 5$ h total & perihelion season & 1 h (v5 required) & v5 & NOMAD, MCS, EMIRS & $\vu$, $T$, vmr\_h2ovap, $\taud$ $\tau_i$ \\
\addlinespace
7 & Aphelion Cloud Belt (ACB) & $\sim 10^3$--$10^4$ km zonal, $10^\circ$S--$30^\circ$N & daily AM/PM, seasonal peak $\Ls{\approx}100^\circ$ & N. spring/summer, $\Ls{=}50^\circ$--$150^\circ$ & 1--2 h & v5 & MCS ($\HO$ ice column), NOMAD, EMIRS & vapcol, vmr\_h2ovap, $T$, $\vu$, $\taud$, $\tau_i$ \\
\midrule
8 & \CO{} and $\HO$ frost cycles & $\sim 10^3$ km (polar caps); transient on summits & seasonal (polar); sub-diurnal (summits) & polar winter; equatorial morning & hours--sols & v4 (long span) & MCS (\CO{} ice), TES ($T_\text{surf}$) & $\CO^\text{ice}$, $T_\text{surf}$, $p_s$,  vmr\_h2ovap, vapcol, $\tau_i$ \\
\addlinespace
9 & Global $p_s$ and \CO{} mass budget & global & seasonal, multi-MY & all $\Ls$, all MY & sols & v4 (MY 28--35) & MCS, TES & $p_s$, $\CO^\text{ice}$, $T_\text{surf}$ \\
\addlinespace
10 & Downscaling ($5^\circ \to 2^\circ / 1^\circ$) & reanalysis $5^\circ$ $\to$ spacecrafts 5--10 km & any & any & spacecraft cadence & v4/v5 as LR input & MCS, NOMAD, ACS, TES (HR targets); EMIRS (coarse HR) & $T$, $\taud$, vmr\_h2ovap, $T_\text{surf}$ \\
\addlinespace
11 & Reanalysis--observation fusion & global, multi-resolution & event-dependent & any & hours & v4/v5 baseline & MCS, NOMAD, ACS, EMIRS as targets & all state variables \\
\bottomrule
\end{tabular}
\end{sidewaystable}

\subsection{Scope Tradeoffs Imposed by the Data}
\label{sec:data-tradeoffs}

The structure of the available data imposes several constraints on the scope of an MAFM.

\paragraph{Heterogeneity of atmospheric retrievals.}
Table~\ref{tab:satellite} summarizes the five spacecraft instruments. Two key considerations arise from this comparison. First, the horizontal resolution varies substantially across instruments, ranging from $\sim 5$~km (NOMAD) to 100--300~km (EMIRS), with all observations finer than the reanalysis ($\sim 300$~km effective resolution). Any fusion framework must therefore explicitly account for these scale differences. Second, none of the instruments provide direct wind retrievals. As a result, the vertical wind structure relevant to features such as the low-level jet (LLJ), the Hadley circulation, and frontal dynamics is constrained solely by the reanalysis. Validation of the FM wind field must therefore rely either on internal consistency within the reanalysis or on indirect inference through retrieved thermal structure and cloud morphology. Landers and rovers provide surface observations of wind, pressure, and other meteorological variables in very limited quantities, but have been used in a small number of studies for comparison with numerical atmospheric models \cite{Corchete2025CO2, ruiz2025curiosity}

\paragraph{Water vapor mixing ratio is a v5-only quantity.}
The v2--v4 OpenMARS releases do not include atmospheric water vapor. Any downstream task centered on the hydrological cycle, including ACB climatology, AMEC forecasting, orographic cloud prediction, and frontal water-cycle diagnostics, is therefore restricted on the reanalysis side to the v5 span of MY~34--35, or roughly two Mars years. This is a severe data constraint: the model must learn the aphelion-season hydrological behavior from two realizations of the seasonal cycle rather than eight. Atmospheric retrievals from MCS, NOMAD, and EMIRS extend the usable record but they do not substitute for a long-span reanalysis.

\paragraph{Water ice is not provided by OpenMARS.}
The absence of water ice data in the OpenMARS limits its ability to represent the martian water cycle from water vapor, temperature, and pressure alone. Supplementing the reanalysis data with orbiter-retrieved water ice optical depth and vertical profiles is necessary for a more complete depiction of the water cycle.   

\paragraph{Vertical velocity is not provided by OpenMARS.}
OpenMARS provides only the horizontal wind components $u$ and $v$. The vertical velocity $w$, which is central to orographic  and frontal forcing, and DAC dynamics, is absent. An FM therefore cannot be trained or evaluated directly on $w$ using OpenMARS. Any vertical-velocity reasoning the model acquires must emerge from the dynamical consistency of the remaining state variables.

\paragraph{Electrification is not represented in OpenMARS.}
Dust electrification influences the martian atmosphere by strengthening boundary-layer convection, enhancing dust lifting  \citep{kok2006dustliftelec} and affecting dust transport \citep{Berthelier2000dustelec}. However, it is not present in any OpenMARS release and is not retrieved by any of the spacecraft measurements in Table~\ref{tab:satellite}. The associated processes, including dust-discharge lightning, are therefore out of scope. We note this not as a limitation to be worked around, but as a boundary of the scientific claims the MAFM can support.

\paragraph{The reanalysis retrieval resolution asymmetry is large.}
The $\sim 60$--fold gap between the $\sim 300$~km effective reanalysis resolution and the $\sim 5$~km best orbital resolution is a significant data asymmetry.

\paragraph{Mars-year coverage is fragmented across instruments.}
No single data source covers MY~24 through the present at uniform resolution. TES provides MY~24--27 at its native resolution; MCS begins in MY~28 and continues to the present; ACS and NOMAD begin in MY~34 on TGO; and EMIRS begins in MY~35 on Hope. Each transition between instruments introduces a different set of retrieval biases and observational geometries. Any MAFM must therefore generalize across these transitions, and the training protocol must be designed accordingly.

\paragraph{Orbital geometry and revisit times vary by mission.}
The sun-synchronous orbits of MGS and MRO fix the local solar time at which each geographic point is sampled for in-track scans, although cross-track and off-track scanning partially mitigates theses gaps. TGO is not sun-synchronous, and its $\sim 108$-sol revisit time at a given location means that the diurnal cycle at a point is reconstructed only over many orbits. Hope's highly elliptical orbit provides full-disk coverage per orbit but completes less than half an orbit per sol. Any fusion scheme must therefore handle observations that are not only sparse, but sparse in structurally different ways across instruments.

\paragraph{The role of physical models for MAFMs.} 

MGCM or MarsWRF can fill gaps that neither OpenMARS nor the spacecraft record can address on its own. OpenMARS is constrained by the TES and MCS observations assimilated into the reanalysis and by the representational limits of its parent GCM. It also lacks H$_{2}$O frost and explicit water-ice microphysics, has a coarse near-surface vertical resolution relative to the martian boundary layer, and is temporally bounded the available orbital observation record. 

MGCM and MarsWRF supply self-consistent global states with a broader set of prognostic variables, at resolutions that capture dynamics unresolved by OpenMARS and across Mars years and forcing scenarios outside the reanalysis envelope. For an AI model intended to learn the physics of the martian atmosphere, including the seasonal CO2 mass cycle, dust-radiation coupling, deep dry convection, and polar processes, these simulations provide dynamically consistent examples of phenomena that are only partially represented, or absent, in the reanalysis record.

This shifts the central question from data quantity alone to data fitness: what combination of reanalysis, free-running GCM output and high-resolution model simulations captures the martian atmosphere richly enough for an appropriate AI architecture to learn its dynamics? OpenMARS anchors the model to observations, MGCM extends coverage and adds prognostic variables and MarsWRF provides resolution and process detail at and below the mesoscale. The computational  cost estimates above define the practical envelope for this strategy: MGCM at C48 ($\sim 110$ km) and MarsWRF at 108 km are feasible for multi-Mars year archives, whereas finer resolution simulations are best reserved for targets case studies and downstream validation.

\section{The Applications Perspective}
\label{sec:applications}

We organize this section around the atmospheric phenomena that a Mars Atmospheric Foundation Model (MAFM) could be able to represent, forecast, or detect. Although it might seem natural to order these by increasing spatial, we prioritize phenomena by their centrality to the coupled \CO{}--dust--\HO{} system that governs martian weather. Dust storms come first because dust is the dominant radiative and dynamical actor on Mars \cite{fernandez1997duststorm,kulowski2017duststorm,wang2015duststorm,sanchezlavega2024dynamical}. Frontal systems and Double Annular Cyclones (DAC) follow as the principal synoptic-scale weather systems. Low-level jets and various water-ice cloud systems, including orographic volcano clouds, the Arsia Mons Elongated Cloud, and the Aphelion Cloud Belt, come next, each with its own diurnal and seasonal signature. We then turn to phenomena that have no direct analog in Earth foundation model literature: the \CO{} frost cycle, global surface-pressure and mass-budget forecasting. Finally we list spatial downscaling and reanalysis-observation fusion as a downstream task in their own right. We close the section with a discussion of what the foundation model paradigm specifically offers for Mars atmospheric science.

A note on scales is necessary because scale shapes every subsection that follows. The OpenMARS reanalysis is provided at $5^{\circ} \times 5^{\circ}$; at the martian radius of $\sim 3390$ km, one degree of latitude is $\sim 59$ km, so OpenMARS resolves features no finer than $\sim 300$ km. Given the scales of atmospheric retrievals, a foundation model with a resolution of $2^{\circ} \times 2^{\circ}$ ($\sim 120$ km) or even $1^{\circ} \times 1^{\circ}$ ($\sim 60$ km) could be achievable. This resolution expansion opens up a class of mesoscale phenomena, such as regional dust storms and DAC cores, that are unresolved or barely resolved at the native reanalysis grid. Microscale phenomena, including dust devils with diameters $<1$ km and planetary boundary layer turbulence, are not in scope.

\begin{figure}[t]
  \centering
  \includegraphics[width=0.9\textwidth]{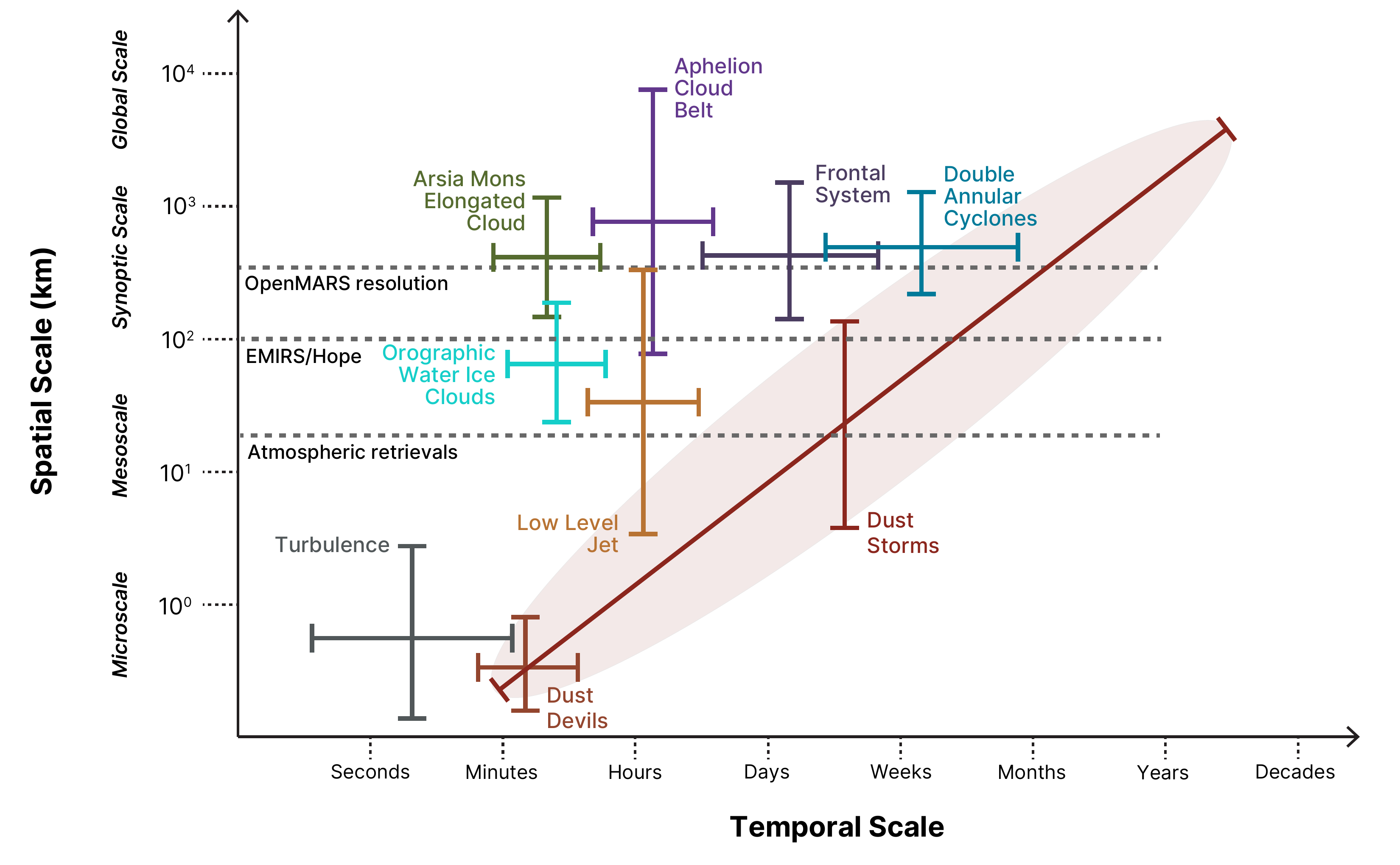}

  \caption{Spatial and temporal scales of martian atmospheric phenomena targeted by the MAFM. The dashed horizontal lines mark the OpenMARS native resolution ($5^\circ \times 5^\circ$) and a potential FM resolution of $2^\circ \times 2^\circ$. Dust devils and boundary-layer turbulence lie below the FM's resolvable scale and are out of scope. Adapted from \citet{sanchezlavega2024dynamical}. It should be noted that the orographic volcano clouds, Arsia Mon Elongated Clouds (AMEC), Aphelion Cloud Belt (ACB), and Low Level Jet (LLJ) all have a specific appearance time of the day, even though the phenomena are season long. Double Annular Cyclones (DAC) to some extent also falls in this category.}
  \label{fig:scales}
\end{figure}

Figure~\ref{fig:scales} summarizes the spatial and temporal scales of the phenomena discussed in this section, together with the envelopes covered by the native OpenMARS grid and the target FM resolution.

\subsection{Dust Storms}
\label{sec:dust}

Dust is the single most important radiatively and dynamically active tracer in the martian atmosphere. Its daytime absorption of solar radiation heats the convective boundary layer; at regional to global scales, its column optical depth modulates surface insolation and can reorganize the thermal structure of the entire atmosphere \citep{fernandez1997duststorm,kulowski2017duststorm,wang2015duststorm,wang2023duststormcycle,sanchezlavega2024dynamical}. Dust lifting, transport, and deposition are therefore not secondary effects of the weather but active components of the atmospheric system.

Dust storms span five orders of magnitude in horizontal extent, from microscale devils of $\sim 100$~m diameter lasting seconds, to local storms ($\sim 10$~km) scale lasting up to three sols, to regional storms ($10^2$--$10^3$~km) lasting more than three sols, to planet-encircling global dust storms lasting thirty sols or longer. The global events occur on average every $\sim 3$ Mars years (about 5.5 Earth years). The storm season is concentrated in southern spring and summer, $\Ls = 180^\circ$--$360^\circ$.

Local storms and dust devils are microscale phenomena; their aggregate effect, however, is represented byt the dust field $\taud$ in reanalysis. Storms from $\sim 100$ km upward are resolvable at the mesoscale, and can be associated with the following downstream modeling tasks:

\begin{itemize}
\item \emph{Detection and tracking} of regional dust storms from dust column optical depth and surface-pressure anomalies, validated against MCS, TES, and EMIRS retrievals.
\item \emph{Life-cycle prediction}, including forecasting the genesis, intensification, movement, and decay of an identified storm. This is a multi-day to multi-week task in the regional case and a multi-month task in the global case.
\item \emph{Climatological characterization} of storm frequency, size, and preferred genesis locations as a function of \Ls{} and Mars year, supporting long-term planning for landed missions.
\item \emph{Transport diagnostics}, including estimation of the dust emission, transport, and deposition budget implied by simulated storms, and comparison with retrieved column loading.
\end{itemize}

The scientific questions that drive these tasks are: what sets the size a storm can reach, what determines whether it stays local or transitions to regional or global scales, and to what degree the storm's radiative feedback on the atmosphere can be simulated consistently from reanalysis-scale dust fields. 

The $3$-Mars-year recurrence of global dust storms means that long-span reanalyses, such as  MY~24--35 described in Section~\ref{sec:openmars}) contain only a handful of such events, and generalization to new ones, including extrapolation of radiative effects, would be a severe out-of-distribution test for any model.

\subsection{Frontal Systems}
\label{sec:frontal}

Martian frontal systems are sub-synoptic- to synoptic-scale weather systems driven by the meridional temperature gradient between the polar vortex and the mid-latitudes. They are most prominent in the northern hemisphere in fall and winter, when the contrast is sharpest, and they typically propagate eastward around the polar cap. Occasionally, a front couples to the Hadley circulation and is advected southward through low-topography channels toward the equator \citep{wang2009north,hinson2010further}. Fronts are both tracers and agents: they carry water ice clouds and dust along their paths, so the dust field itself provides a detectable signature. At the same time, the passage of a front can rapidly lower local temperature, triggering significant water ice cloud condensation \citep{montmessin2004cloud}, freezing \CO{} out of the atmosphere as surface frost, and intensifying surface dust lifting through enhanced frontal wind stress \citep{wang2002frontdustco2cloud}.

From an FM perspective, frontal systems are attractive as a downstream target for three reasons. First, they are a multi-day phenomena at synoptic scales, so they are well-matched to the hourly cadence and target resolution of the model. Second, they couple the \CO, dust, and water cycles in a single event, so forecasting a front exercises the model's coupled dynamics rather than any single tracer channel. Third, they are well-documented in prior literature \citep{hunt1979frontal,banfield2004waves,wang2005dusteddies,wang2009north,Hinson2006transeddies,hollingsworth2010extratropical,hinson2010further}, providing a catalog of cases against which the model can be evaluated.

The downstream tasks are:

\begin{itemize}
\item \emph{Front detection and tracking} in OpenMARS and spacecraft retrievals, establishing a climatology of frontal passage frequency, latitude, and \Ls{} dependence.
\item \emph{Case-study forecasting}: given a front at time $t$, predict its track, intensification, and associated temperature and pressure anomalies over the following week.
\item \emph{Coupled-cycle diagnostics}: estimate the \CO, water, and dust fluxes associated with frontal passages, and test whether the FM's predictions conserve mass in the appropriate sense.
\item \emph{Extended-lead prediction}: determine far out can the FM track the evolution of a frontal system before forecast skill degrades to climatology.
\end{itemize}

The open scientific questions here parallel those on Earth. How do the three tracer cycles interact within a frontal system? How much does frontal passage modulate local weather, and for how long? Can a model resolve the dynamical processes, not merely the thermodynamic signatures, of the associated \CO, water, and dust transports?

\subsection{Double Annular Cyclones}
\label{sec:dac}

The Double Annular Cyclones (DAC) are a seasonally recurrent weather feature of the northern high latitudes, characterized by a double-cyclone structure in which water-ice clouds form a ring around a cloud-free core of radius $\sim 100$--$300$ km \citep{sanchezlavega2018seasonally}. The total diameter of a DAC is $1200$--$1600$ km, placing these systems at the synoptic scale. They appear in northern summer at $\Ls \approx 115^\circ$--$135^\circ$ and latitudes near $60^\circ$N. Like the other cloud-based phenomena on Mars, DACs have a diurnal signature: the vortices strengthen in the morning and weaken toward evening.

While DACs are relatively infrequent, they are also regular enough that each Mars year provides fresh validation cases. Their cloud-core geometry means they are well-characterized in orbital imagery and retrieval products, including MCS and NOMAD water-ice retrievals, and EMIRS full-disk coverage, providing a clean signal that is not heavily entangled with the dust or \CO{} cycles.

Downstream tasks include detection of DAC events in reanalysis data, characterization of their morphology and diurnal cycle from orbital retrievals, and short-range forecasting of the vortex intensification and decay. The DAC also serves as a valuable architectural probe: a model that fails to produce the annular structure under the right seasonal and latitudinal forcing likely has deficiencies in its representation of rotational flow at the relevant scales.

\subsection{Low-Level Jets}
\label{sec:llj}

The martian low-level jet (LLJ) is a nocturnal wind maximum at $\sim 1$~km above the surface, typically occurring from late night through early morning. It is a widespread phenomenon but is most prominent over mid- to high-latitude plains and inside craters \citep{sanchezlavega2024dynamical}. Fall and winter are the most active time of year for LLJs, which can occur daily. Although LLJs are common, they carry practical significance: the sheared environments they create are unfavorable to lander descents and favorable to sailplane operations, making reliable LLJ forecasting a mission-relevant downstream task.

The LLJ is unique among the phenomena in this section because it cannot currently be retrieved directly from present spacecraft observations as no current Mars orbiter or rover produces a wind profile product. Because the jet is a low altitude vertical wind-field structure, it is expected to be detectable by future spacecraft-based wind lidar \citep{abshire2023windlidar}. Crucially for this study, LLJs are commonly resolved by NASA's MGCM \citep{rou2022nasa-gcm-llj}. This this suggests that data generated from a model run or present in OpenMARS should serve as the primary LLJ signal for training or evaluation. Possible downstream tasks are limited by the absence of direct wind observations in atmospheric retrievals:

\begin{itemize}
\item \emph{Climatological characterization} of LLJ frequency, strength, and vertical structure as a function of region, topography, and season.
\item \emph{Short-range forecasting} of LLJ onset, peak magnitude, and shear at specific locations.
\item \emph{Topographic dependence studies}: quantifying how crater geometry and plain elevation control jet occurrence and strength, which in turn tests the FM's ability to represent boundary-layer flow over varying terrain.
\end{itemize}

Because the dynamical characteristics of LLJs depend on resolved boundary-layer dynamics, their accurate modeling requires terrain-following vertical coordinates and near-surface shear.

\subsection{Orographic Water-Ice Clouds over Volcanoes}
\label{sec:orographic}

When an air parcel is forced upward over a sufficiently tall volcano, adiabatic cooling can bring it to the water-ice condensation point, producing a lee-side or summit cloud \citep{sanchezlavega2024dynamical}. On Mars, the aphelion season ($\Ls \approx 50^\circ$--$150^\circ$) provides both the elevated water vapor and the low temperatures that favor such cloud formation. As a result, clouds consistently form above the volcano summits by late morning, and grow into a westward-stretching plumes during the afternoon in the aphelion season \citep{michaels2006volcano}. For the largest volcanoes, the plumes can extend over several hundred kilometers. During the perihelion season, the thermodynamic conditions are less favorable, but clouds can still form over the largest volcanoes where topographic lifting is strong enough to compensate.

Most volcano-induced clouds are mesoscale features with extents up to $\sim 1000$~km. At a resolution of $2^\circ \times 2^\circ$ ($\sim 120$~km) resolution, only the clouds over the largest volcanoes, notably Olympus Mons, Tharsis Montes, and Elysium Mons, are resolvable as coherent features; smaller orographic clouds will blur into the grid. The downstream tasks related to these major volcanic edifices are:
\begin{itemize}
    \item Characterization of where and when orographic clouds form as a function of volcano, season, and location time.
    \item Detection of individual cloud events in reanalysis products and spacecraft observations.
    \item Forecast cloud extent as a function of the incoming flow field. 
\end{itemize}
The related scientific questions are: what combinations of flow, temperature, and water vapor lead to cloud formation over a given volcano; how much water mass is involved in these events; and whether the orographic cloud population is modulated by dust activity, either through radiative effects on the near-surface temperature or through dust acting as a source of condensation nuclei.

\subsection{The Arsia Mons Elongated Cloud}
\label{sec:amec}

The Arsia Mons Elongated Cloud (AMEC) deserves its own subsection despite being a specific instance of the previous category. The cloud, which forms over the $\sim 20$-km tall Arsia Mons volcano, is a near-daily recurring feature during the perihelion season, when clouds are otherwise scarce on the planet \citep{hernandezbernal2021elongated}. It has a distinctive life cycle: formation before dawn, rapid westward extension of the plume to lengths up to 1800~km over the next 2--3 hours, and dissipation by midday. Unlike a typical volcano cloud, the AMEC's tail reaches far beyond the volcano itself, making it a mesoscale-to-synoptic scale phenomenon \citep{sanchezlavega2024dynamical}.

As a modeling target, the AMEC requires spatial fidelity near a specific topographic feature, together with accurate sub-scale dynamics to capture its strong diurnal variations.  Success here is tightly coupled to temporal resolution: the cloud's full life cycle occurs within a $\sim5$-hour window, so a model running at $2$-hour timesteps, as OpenMARS versions 1--4 do, will capture only two or three frames of the evolution. OpenMARS version 5, at hourly cadence, gives the FM its best chance to resolve AMEC dynamics temporally.  In numerical modelling studies of the AMEC, MGCMs have been used only to provide boundary conditions to mesoscale simulations \citep{hernandezbernal2021elongated}. Downstream tasks related to the AMEC are:
\begin{itemize}
    \item Daily event detection, estimation of cloud extent and optical depth as a function of local time and \Ls.
    \item Prediction of the cloud's plume length and orientation from the upstream flow field and water vapor column.
\end{itemize}
As with the general orographic case, we are interested in whether dust activity modulates the AMEC population, and in how much water vapor is processed through the cloud over the perihelion season.

\subsection{The Aphelion Cloud Belt}
\label{sec:acb}

The Aphelion Cloud Belt (ACB) is a quasi-continuous band of water-ice clouds that encircles the martian tropics from roughly $10^\circ$S to $30^\circ$N during the aphelion season, $\Ls = 50^\circ$--$150^\circ$, in northern spring and summer \citep{clancy1996acb,campbell2020estimating,sanchezlavega2024dynamical}. The cloud belt forms in the upward branch of the Hadley cell, due to lower ambient temperatures enabling water vapor to condense at lower altitudes compared to the warmer conditions during the perihelion season. Cloud formation peaks near $\Ls \approx 100^\circ$ and is concentrated in the morning and afternoon, driven by a combination of overnight cooling and daytime convective lifting, with daytime airborne dust supplying condensation nuclei. The belt extends thousands of kilometers zonally and is the most prominent water-ice cloud feature on Mars.

The ACB is central to the martian hydrological cycle: it mediates the inter-hemispheric transport of water vapor from the south to the north during the aphelion season, and its formation and dissipation modulate the availability of atmospheric water to other processes. Accurately representing the ACB is thus a prerequisite for any model that claims to address the water cycle at all. The related downstream tasks are:
\begin{itemize}
    \item Climatological characterization of the belt's latitudinal extent and diurnal cycle across Mars years
    \item Event-scale detection of when and where cloud cover densifies or thins
    \item  Forecasting of the belt's response to the omnipresent dust, which are hypothesized to interact with the water cycle through direct and indirect radiative effects, and as cloud condensation nuclei \citep{guha2021acbdust,navarro2014dustcloud}.
\end{itemize}
Quantifying the water vapor and ice mass involved in the ACB is a specific scientific target, and one that exposes a hard data constraint: OpenMARS versions 2--4 carry no water-vapor field, so data for model training is restricted to version 5 (MY~34--35) unless supplemented by orbital retrievals. We return to this constraint in Section~\ref{sec:data-tradeoffs}.

\subsection{\texorpdfstring{CO$_2$}{CO2} and \texorpdfstring{H$_2$O}{H2O} Frost Cycles}
\label{sec:frost}

Frosting is a seasonal and diurnal phenomenon on Mars, but at the planet's scale it becomes a first-order driver of the atmospheric mass budget. \CO{} frost caps a substantial fraction of the winter polar region each year and accounts for approximately a quarter of the total atmospheric mass of the planet as it cycles on and off the surface \citep{Kelly2006co2cycle}.  Transient $\HO$ frost appears on the summits of equatorial volcanoes in the morning \citep{valantinas2024waterfrost}, and \CO{} frost covers polar regions on the scale of $\sim 1000$ km during winter.

Two features distinguish the frost cycle from the phenomena above. First, it is simultaneously diurnal and seasonal. Second, the frost cycle couples surface state to atmospheric state in a way that pure atmospheric dynamics does not: the surface \CO{} ice mass is both a consequence of past atmospheric state and a boundary condition for future state. This provides a natural diagnostic for assessing whether a model's learned representation respects the coupled mass budget. The downstream tasks are: 
\begin{itemize}
    \item Detection of the seasonal growth and retreat of the polar caps in OpenMARS and spacecraft observations
    \item Diagnosis of transient morning-frost events over equatorial summits; and forecasting of pressure anomalies associated with rapid frost deposition events during frontal passages.
\end{itemize}
OpenMARS provides $p_s$, $T_\text{surf}$, and surface \CO{} ice mass directly, which could be used to provide the model with a tight supervisory signal. $\HO$ frost, in contrast, is not included in any standard OpenMARS output, and the frost-phase water cycle therefore remains out of scope for near-term scientific evaluation.

The related scientific questions are: what role does frosting play in the planetary \CO{} cycle, and what fraction of total \CO{} mass is involved in the seasonal cycle versus in transient frontal events? These are accessible with the reanalysis alone, and also potentially provide a high-leverage evaluation of a model's long-term rollout stability.

\subsection{Global Surface-Pressure and \texorpdfstring{CO$_2$}{CO2} Mass-Budget Forecasting}
\label{sec:massbudget}

The task described here has no direct analog in Earth weather foundation models, because Earth's atmospheric mass does not change appreciably on weather timescales. On Mars it does. The seasonal sublimation and deposition of \CO{} between the polar caps and the atmosphere drives a global surface-pressure cycle with a peak-to-peak amplitude of roughly 25--30\% of the mean \citep{hess1979co2pressure,hourdin1993psfc}. Any model that rolls out for more than a handful of sols must track this cycle explicitly, or it will drift.

Global-$p_s$ and \CO{} mass-budget forecasting could therefore function as a first-class downstream task, in which the model consumes the reanalysis state, including pressure, surface temperature, and surface \CO{} ice mass, and propagates it forward. The evaluation metric is its ability to reproduce the seasonal $p_s$ signature across multiple Mars years, accepting long-horizon skill degradation only to the extent that the underlying reanalysis itself degrades. OpenMARS version 4, spanning MY~28--35, provides the longest continuous span for this purpose.

This task is also a diagnostic. A model whose long rollouts fail to conserve global \CO{} mass, in the sense that it produces a $p_s$ cycle with the wrong amplitude or phase, is revealing a structural weakness that will corrupt every other downstream task it is asked to perform. For this reason, we view mass-budget forecasting as a prerequisite test rather than a separate application.

\subsection{Spatial Downscaling of the Reanalysis}
\label{sec:downscaling}

The gap between the $5^\circ \times 5^\circ$ OpenMARS grid and the $\sim 5$--$10$ km horizontal resolution of MCS, NOMAD, and ACS retrievals is roughly a factor of 30 in linear scale. If one chooses to leverage OpenMARS in the context of an MAFM, spatial downscaling from OpenMARS to instrument scale becomes a natural question. This would require fusion with orbital retrievals, which we discuss in Section~\ref{sec:fusion-downstream} as a task and in Section~\ref{sec:fusion-strategies} as a methodology.

The downscaling task itself is straightforward to state: given a reanalysis state at $5^\circ$, produce a physically consistent state at $2^\circ$ or finer, evaluated against orbital retrievals where they are available and against held-out reanalysis patches where they are not. The difficulties are substantial. The reanalysis and the orbital retrievals are not simply the same field at different resolutions; they carry different biases, different vertical representations, and different systematic errors. Training a downscaler that merely regresses orbital retrievals onto the reanalysis grid will bake both sources' biases into the model. We therefore believe the downscaling task is best posed as physically constrained fine-tuning, in which the FM's pretrained representation of martian dynamics is used as a strong prior and the orbital data supplies sparse supervisory signal, rather than as a direct super-resolution regression.

Direct downscaling of OpenMARS data to a finer resolution derived from orbital data is not straightforward; orbital retrievals are not gridded products but trajectory or point cloud samples with a variety of spatial resolution-related characteristics (e.g. field of view width, along-track sample length, sampling distance). Furthermore, the reanalysis and the orbital retrievals carry different biases, different vertical representations, and different systematic errors. It may therefore be more appropriate to pose the downscaling task as a physically constrained fine-tuning, in which orbital data provide a sparse supervisory signal on top of a reanalysis-based pretraining representation of martian dynamics.

\subsection{Reanalysis--Observation Fusion}
\label{sec:fusion-downstream}

Beyond downscaling, reanalysis--observation fusion is the broader task of combining multiple data streams into a single coherent atmospheric state. On Mars, the motivation is sharper than on Earth: no single spacecraft covers the full atmosphere, no spacecraft covers any given location continuously, and the reanalysis is already the product of a fusion process, with OpenMARS assimilating MCS, TES, ACS, or NOMAD depending on the version. The FM approach could add value in four ways that the underlying NWP-plus-assimilation system cannot easily match:

\begin{enumerate}
\item \emph{Ingestion of heterogeneous streams} at their native resolutions, rather than after interpolation onto the reanalysis grid.
\item \emph{Filling observational gaps} in time and space using a learned prior, which is particularly relevant for the $\sim 100$-sol revisit time of TGO at any given location.
\item \emph{Integration of instruments not present in the OpenMARS assimilation}, such as EMIRS from the UAE Hope mission, which was not included in the v5 release.
\item \emph{Uncertainty quantification} through ensemble prediction or diffusion-based sampling, providing an explicit distribution rather than a single best estimate.
\end{enumerate}

We consider the fusion task important enough to warrant a dedicated evaluation axis, and we return to candidate methodologies in Section~\ref{sec:fusion-strategies}.

\subsection{Benefits of Foundation Modeling for Mars}
\label{sec:fm-benefits}

Several benefits of the FM paradigm discussed in \citet{mukkavilli2023ai} for Earth carry over to Mars, but with shifts in emphasis that are worth making explicit. Some of what follows is specific to Mars; some is the common case stated in Mars-specific terms.

\begin{enumerate}

\item \emph{Data scarcity further motivates the use of self-supervised learning.} The central argument for foundation models on Earth is that supervised labels are scarce compared to the volume of unlabeled reanalysis data. On Mars, the disparity is larger: the reanalysis record is eight Mars years, the spacecraft retrievals are partial and instrument-dependent, and many of the downstream tasks, including storm tracking, DAC detection, AMEC climatology, have no standardized labeled dataset at all. A model that learns useful representations from the reanalysis without labels is the only tractable path for most of the downstream applications in this section.

\item \emph{Multi-instrument fusion is mandatory, not optional.} Earth FMs can in principle be trained on a single reanalysis such as ERA5; the reanalysis alone provides enough variety and coverage to be scientifically useful. On Mars, every mission has a finite lifetime, a specific orbital geometry, and a specific retrieval suite, and no instrument covers the full atmosphere. An FM that cannot ingest TES, MCS, NOMAD, ACS, and EMIRS with their varying resolutions, coverages, and physical retrievals is an FM that would underuse the available data.

\item \emph{Cross-Mars-year generalization is a hard test.} Supervised per-mission models have no mechanism for transferring knowledge across instruments or across years with different observational geometries. A pretrained FM, in contrast, can in principle absorb the shared physics across all Mars years and adapt to instrument-specific retrievals through fine-tuning. This is the FM's answer to the short mission lifetime problem.

\item \emph{Rare events make representation learning a necessity.} Global dust storms recur every $\sim 3$ Mars years. The reanalysis record thus contains only two or three well-characterized events, and any supervised detector is severely data-starved. A foundation model trained on all available atmospheric states, whether or not they contain storms, has a chance of generalizing to new global events through the learned representation of the underlying dynamics.

\item \emph{Multi-scale, multi-phenomenon coverage reduces engineering sprawl.} The alternative to a single foundation model is a separate supervised pipeline per phenomenon. Given the breadth of the phenomena in this section, including dust storms, fronts, DACs, LLJs, orographic clouds, AMEC, ACB, frost cycles, and given the small size of the Mars atmospheric science community, the maintenance cost of many specialized models is high. A single FM with fine-tuned heads is operationally simpler.

\item \emph{Coupling the \CO, dust, and water cycles in one model.} Every major martian weather event couples these three cycles. A pretraining objective that jointly models pressure, temperature, wind, dust, and, in v5, water vapor learns the cross-tracer relationships that a single-task supervised model is structurally unable to represent.

\item \emph{A natural bridge to planetary foundation models.} The group's prior work on heliophysics (Surya, \citet{surya2025arxiv}), Earth (Prithvi-WxC, \citet{prithviwxc2024}), and the Moon (Lunar FM dataset paper, in preparation) establishes a design lineage for planetary-scale foundation models. A Mars atmospheric FM fills a natural gap in that lineage and shares architectural components, including native-resolution tokenization, spherical-domain handling, multi-channel ingestion, with its siblings.

\item \emph{Human exploration relevance.} Accurate forecasting of dust storms, frontal passages, LLJs, and surface-pressure transients directly supports landed mission operations and long-horizon human-exploration planning, which is an applied driver that pure climate-science FMs on Earth lack.

\end{enumerate}

With the applications in view, we now turn to surveying recent developments in AI that help make them possible. 

\section{The AI Perspective}
\label{sec:ai_developments}

\subsection{Recent Developments}
\label{sec:ai_recent_developments}

\subsubsection{An introduction to AI models for atmospheric physics}
\label{sec:ai_atmospheric_physics}

The last few years have seen the emergence of operational weather forecasting systems based on purely data-driven approaches. The development started in industry \cite{keisler2022forecasting,pathak2022fourcastnet,bi2023pangu,lam2023graphcast}, but there are now major research efforts by organizations such as ECMWF \cite{lang2024aifs,moldovan2025update,lang2026aifs}, ECCC \cite{pereira2026learning} or NOAA \cite{abdi2026hrrrcast}. ECMWF's AIFS crossed the threshold from a research effort to an operational model in February 2025 \cite{ecmwf2025aifs}.

The fundamental paradigm has its roots in papers such as \cite{keisler2022forecasting} and \cite{pathak2022fourcastnet}. Here, the authors train deep neural networks $f_\theta$ that take two time stamps of ERA5 data as input and predict the next time stamp: $\hat{x}_{t+1} = f_\theta(x_t, x_{t-1})$. By feeding predictions back into the model one can predict further and further into the future
\begin{equation}\label{eq:ai_forecast_rollout}
    \hat{x}_{t+N} = f_\theta(\hat{x}_{t+N-1}, \hat{x}_{t+N-2}) = f_\theta \circ \hdots \circ f_\theta(x_t, x_{t-1}).
\end{equation}
Subsequently \cite{bi2023pangu,lam2023graphcast,bonev2023sfno,chen2023fengwu,chen2023fuxi} refined and scaled this approach such that prediction accuracy can rival or surpass operational weather forecasts (ECMWF IFS).

In \eqref{eq:ai_forecast_rollout} the state vector $x_t$ typically comprises five or six variables (q, t, u, v, z; sometimes also w) at maybe 13 vertical pressure levels as well as a number of surface variables. The latter are typically 2t, 10u, 10v and msl or sp. In addition, there are typically covariates such as orography or a land/sea mask, yet also latitude and longitude, hour of the day or time of the year. The latter allow the model to implicitly learn regional characteristics as well as (radiative) forcings. Models $f_\theta$ are then trained using an MSE objective with carefully chosen weights. Forecasting performance is further improved by a process referred to as ``rollout tuning''. Here, one predicts multiple steps ahead using \eqref{eq:ai_forecast_rollout} and then optimizes through the entire model. Technical challenges are frequently related to GPU memory management. This is due both to the high pixel count -- ERA5 data at 0.25 degrees resolution means that one is dealing with 721 by 1400 pixels. Yet also to accommodate to applying the model repeatedly during rollout tuning. Models trained at full resolution generally train with batch size one (one GPU per data sample) across 16 to 32 GPUs. Typically training schedules involve $\sim$300,000 gradient descent steps.

Following the success of deterministic models for medium range predictions, researchers started developing approaches for ensemble forecasting. Initially these were typically based on diffusion models \cite{price2023gencast}. Subsequent work explored alternatives. In particular by directly optimizing a proper scoring rule \cite{lang2026aifs,alet2025skillful,kossaifi2026demystifying}.

The above development happened somewhat in parallel with the foundation model revolution in the mainstream of AI. Foundation models are generally models trained at scale in a task agnostic, self-supervised manner that can then be tuned to address specific tasks. When doing so, they typically exhibit superior performance as well as data efficiency than a task specific model. As the above forecast emulators are typically large models trained on significant compute resources\footnote{What exactly is \emph{significant} is relative of course. The amount of compute resources used for models in atmospheric physics is dwarfed by that for state of the art language models.} they were sometimes referred to as foundation models as well. 

The foundation model paradigm was explored for atmospheric physics in \cite{lessig2023atmorep,nguyen2023climax,prithviwxc2024,bodnar2025foundation,wang2025orbit}. Regarding use cases, these models typically consider forecasting and downscaling. Prithvi WxC \cite{Schmude2024} has been fine-tuned for gravity wave parametrization \cite{gupta2025finetuning}. Aurora \cite{bodnar2025foundation} is fine-tuned for a series of challenging forecasting tasks.

\nocite{sun2025can}

\subsubsection{AI-Based Data Assimilation}
\label{sec:ai_data_assimilation}

AI models for atmospheric physics on Earth are typically trained on gridded datasets, most commonly reanalysis products such as ERA5 or MERRA2. In regional settings where long, high-quality reanalysis records may be unavailable, it is also common to use alternative gridded datasets, including analysis or forecast outputs. This approach isolates the forecasting problem from the additional complexities associated with observational data, which range from data acquisition, management, and preprocessing to modeling considerations such as architecture design and training strategy. While this work primarily focuses on the latter, it is important to note that incorporating observational data substantially increases the data engineering burden, requiring the acquisition, preprocessing, normalization, and integration of heterogeneous data sources rather than reliance on a single, regular gridded dataset.

At the same time, the rapid progress achieved in global medium-range forecasting has made it increasingly clear that integrating observations into AI-based workflows represents a key next step. There is a wide variety of approaches here. Some address data assimilation independently of the forecasting problem, others aim for a single end-to-end system. Some hew very closely to existing methods such as 4D-Var, others do not.

Table~\ref{tab:ai_data_assimiation_overview} provides an overview of recent work in this area. These approaches can be broadly categorized along three axes: the assimilation methodology, the scope of the approach, and the presence of an anchor dataset. We examine each of these dimensions in turn.

\paragraph{Anchor dataset.}
Many approaches rely on the presence of a single, dense dataset. The prototypical example is learning how to regress from sparse data -- whether observational or sparsified ERA5 -- to dense ERA5 data. See e.g.~\cite{andrychowicz2023deep,allen2025end,gupta2026healda}. The benefit of doing so is that one can leverage all the information in high quality datasets such as ERA5. Indeed, for purely data driven approaches, there is a question whether one can infer enough information about the system's underlying dynamics from sparse, noisy observational data. Using an anchor dataset is a way to ensure this.

Yet at the same time, this means that such datasets need to exist. Moreover, the performance of the approach is limited by the performance of said anchor dataset. Thus, when a method depends on the presence of a dense dataset, without which the method cannot be trained, we refer to this as the presence of an anchor dataset.

\paragraph{Scope.}
Scope differentiates between approaches which solely focus on data assimilation and those which aim for an end-to-end approach to forecasting.

\paragraph{Methodology.}
The conceptually simplest approaches are trained via a regression objective. This can happen by regressing from observation onto a gridded anchor dataset as outlined above; yet this is also the standard approach for so-called observation to observation forecasting models. The primary examples of these are EarthNet \cite{vandal2025global} and GraphDOP \cite{alexe2024graphdop}. Both take sparse observations as input and then regress onto sparse observations, side-stepping the presence of an anchor dataset. Obviously, there is no explicit trade-off between uncertainties in model physics, initial conditions and observations. One essentially assumes that the model can learn the relevant trade-offs from data. To our knowledge, all current approaches to direct to observation prediction are trained with deterministic (RMSE) objectives.

The other extreme is given by methodologies that start from the mathematics of an existing assimilation paradigm such as 4D-Var and infuse it with artificial intelligence. Backprop-4DVar \cite{solvik20254d} starts from the 4D-Var objective function
\begin{equation}\label{eq:4d_var_objective}
    J(\mathbf{x}_0) = \frac{1}{2} \left(\mathbf{x}_0 - \mathbf{x}^b\right)^T \mathbf{B}_0^{-1} \left(\mathbf{x}_0 - \mathbf{x}^b\right) + \frac{1}{2} \sum_{i=0}^N \left\lbrack\mathbf{y}_i^o-H(\mathbf{x}_i)\right\rbrack^T \mathbf{R}_i^{-1} \left\lbrack\mathbf{y}_i^o-H(\mathbf{x}_i)\right\rbrack.
\end{equation}
Here, $\mathbf{x}^b$ is the background state or first guess, $\mathbf{x}_0$ the analysis we are trying to obtain and $\mathbf{x}_i$ for forecasts based on $\mathbf{x}_0$. $\mathbf{y}_i^o$ encodes observations at step $i$, $H$ is the observation operator, $\mathbf{B}_0$ the background error covariance and $\mathbf{R}_i$ the observation covariance. Instead of solving \eqref{eq:4d_var_objective} via a double loop optimization, \cite{solvik20254d} uses models written in JAX or pytorch to predict $\mathbf{x}_i$ from $\mathbf{x}_0$ directly so that the gradient $\nabla J(\mathbf{x}_0)$ can be computed directly via automatic differentiation.

In our nomenclature, we refer to methods which entail some variational process that aims to balance the uncertainties of background, model physics and observations as \emph{variational}. This includes Backprop-4DVar. Another key example here is VAE-Var \cite{xiao2025vae}. Here, the authors train a generative model -- a variational autoencoder -- that can generate error fields from a latent normal distribution. If one adds such an error field to a background, one obtains an analysis. By comparing the resulting analysis to observations, one can optimize the latent state to fit said observations using automatic differentiation. Note that the approach depends on an anchor dataset in order to learn the characteristic error fields.

A borderline example between variational and regression approaches is given by AI-Var \cite{keller2024ai}. Here, train networks that use first guess and observations as inputs to predict directly the analysis. However, the approach uses the 3D-Var objective as loss function.

As should be obvious from table \ref{tab:ai_data_assimiation_overview}, an increasing number of papers leverages diffusion models in some form or another. This is obviously an attractive choice. Mainstream diffusion models can generate images and videos conditioned on other inputs such as other images or text. When thinking about gridded weather datasets as images, it is straightforward to envision a model that learns the likelihood of say ERA5 data given some observations: $p(\text{ERA5} \vert \text{observations})$. We refer to methods that directly adopt the machinery of mainstream AI as inpainting models. One considers a set of measurements as a sparse image and fills in the rest. Possibly conditioned on other data such as a first guess or a background forecast. On the other hand, there is work such as \cite{rozet2023score,manshausen2025generative} which adopt the mathematics of diffusion for a clear, mathematical model of data assimilation. We refer to these as score-based.

DiffDA \cite{huang2024diffda} is a prime example of the inpainting approach. DiffDA trains a diffusion model to generate ERA5 ground truth data conditioned on forecasts. During inference, an inpainting-inspired algorithm is used to nudge the generated values towards the observations. Naturally this depends on the ERA5 anchor dataset used to train the original diffusion model. This was addressed in Ambient Physics \cite{majid2026ambient}, which uses the concepts of Ambient Diffusion \cite{daras2023ambient} so that the original generative model can be trained using only sparse data. Note that even in the case of Ambient Physics, this data was generated by sampling from a gridded dataset. Which is not a criticism of the method, but rather a confirmation of the data engineering complexities when actually working with observational data mentioned above.

Score-based data assimilation (SDA) models \cite{rozet2023score} are based on the score-matching formulation of diffusion. If we have a data sample $x$ one can progressively add noise to said sample to eventually obtain random, Gaussian noise. If we denote the original sample as $x(0)$ and the end of the process as $x(1)$, we have a trajectory $x(t)$, $t\in \lbrack 0, 1\rbrack$, which can be formulated as a stochastic differential equation
\begin{equation}\label{eq:score_matching_process}
    dx(t) = f(t) x(t) dt + g(t) dw(t).
\end{equation}
Here, $w(t)$ is a Wiener process and $f(t)$ and $g(t)$ can be chosen. The above equation can be reversed to yield
\begin{equation}\label{eq:score_matching_reverse_process}
    dx(t) = \lbrack f(t) x(t) - g(t)^2 \nabla_{x(t)} \log p(x(t)) \rbrack dt + g(t) dw(t).
\end{equation}
When training a diffusion model, the score function $\nabla_{x(t)} \log p(x(t))$ is approximated with a neural network.  As $f(t)$ and $g(t)$ are known, one can then use the above process to generate a sample from the data distribution starting from pure noise.

The fundamental observation of score-based data assimilation is then the following: Given some observations $y^o$ one would like to formulate the above in terms of $\log p(x(t) \vert y^o)$. Instead of retraining the network, one can use the network trained to approximate the un-conditioned score function $\nabla_{x(t)} \log p(x(t))$ due to Bayes' rule
\begin{equation}\label{eq:score_based_data_assimilation}
    \nabla_{x(t)} \log p(x(t) \vert y) = \nabla_{x(t)} \log p(x(t)) + \nabla_{x(t)} \log p(y^o \vert x(t)).
\end{equation}
A key complication here is that while one generally has access to $p(y^o \vert x(0))$, one needs to approximate $p(y^0 \vert x(t))$.

SDA models can be trained independently of model dynamics as was done in \cite{rozet2023score,manshausen2025generative}. Yet it is also possible to use this methodology with a score-matching model capable of generating forecasts \cite{andry2025appa}. The latter results in an end-to-end method capable of assimilating and forecasting while the former is purely an approach to assimilation. To our knowledge all current SDA models leverage an anchor dataset.

One recent trend in diffusion-based models is to train model in such a way that they generate inputs and outputs at the same time. This allows for a lot of flexibility at inference time. If one assumes dense input and no pre-existing output, one has a forecasting model. Dense input with sparse output is a data assimilation setting. And sparse input and sparse output would mimic an observation-to-observation forecast.

\begin{sidewaystable}
\centering
\caption{AI-Based Data Assimilation}
\label{tab:ai_data_assimiation_overview}
\begin{tabular}{@{}llllll@{}}
\toprule
Name & Scope & Methodology & Anchor dataset & Ungridded obs. & Dataset \\
\midrule
Aardvark \cite{allen2025end} & End-to-end & Regression & ERA5 & Yes & ERA5\\
ADAF \cite{xiang2025adaf} & DA & Regression & RTMA & No & HRRR and observations \\
Appa \cite{andry2025appa} & End-to-end & Diffusion & ERA5 & No & ERA5 \\
AI-Var \cite{keller2024ai} & DA & Regression\footnote{Regression on 3D-Var objective.} & No & Yes & Observational data \\
Ambient Physics \cite{majid2026ambient} & DA & Diffusion (inpainting) & No & No & PDEs \\
Backprop-4DVar \cite{solvik20254d} & DA & Variational & No\footnote{Depends on existence of autograd-capable forward model.} & Yes  & PDEs\\
gappy POD \cite{xing2022fusing} & DA & Variational & Yes & No & CFD data \\
DiffDA \cite{huang2024diffda} & DA & Diffusion (inpainting) & Yes & No & ERA5 \\
DiffusionPDE \cite{huang2024diffusionpde} & End-to-end & Diffusion (score-based) & Yes & No & PDEs \\
EarthNet \cite{vandal2025global} & End-to-end & Regression & No & Yes & Observational data \\
FLRONet \cite{dang2026deep} & End-to-end & Regression (neural operator) & Yes & Yes & CFD \\
Graph-DOP \cite{alexe2024graphdop} & End-to-end & Regression & No & Yes & Observational data\\
HealDA \cite{gupta2026healda} & DA & Regression & ERA5 & Yes & ERA5 and observations \\
Latent DA \cite{fan2026physically} & DA & Variational & ERA5 & No & ERA5 and observations \\
LoRa-EnVar \cite{xiaolora} & DA & Variational & ERA5 & No & ERA5 \\
LO-SDA \cite{sun2025sda} & DA & Diffusion (score-based) & ERA5 & No & ERA5 \\
MetNet3 \cite{andrychowicz2023deep} & End-to-end & Regression & HRRR & No & HRRR and observations \\
PhyDA \cite{wang2025phyda} & DA & Diffusion (score-based) & ERA5 & No & ERA5 \\
RecFNO \cite{zhao2024recfno}& & Regression (neural operator) & Yes & Yes & PDEs and sea surface temp. \\
Score-Based DA \cite{rozet2023score} & DA & Diffusion (score-based) & Yes & No & PDEs \\
SDA \cite{manshausen2025generative} & DA & Diffusion (score-based) & Yes & No & HRRR \\
VAE-Var \cite{xiao2025vae} & DA & Variational & Yes & No & ERA5 \\
\bottomrule
\end{tabular}
\end{sidewaystable}

\subsubsection{Data Efficiency}
\label{sec:ai_data_efficiency}

As should have been clear from section \ref{sec:data}, training data for Mars is not abundant. This is particularly the case if one chooses not to found the model in the OpenMARS data. With this in mind we discuss approaches to train data-driven models in a data-efficient manner. To start, we define three distinct data failure modes: 
\begin{itemize}
    \item \textbf{Limited} data, where the dataset is short or small relative to model complexity (e.g., only 5 years worth of atmospheric retrievals or MGCM simulations for a large transformer)
    \item \textbf{Sparse} data, where data exist but are thinly and/or irregularly distributed across space or time (e.g., too coarse spatial or temporal resolution for the problem at hand, or no sampling on certain regions of Mars and dense sampling on others), and
    \item \textbf{Partial} data, where only a subset of the variables can be observed (e.g. no available wind observations).
\end{itemize}
Each failure mode calls for a different approach. 
We briefly address all three, but organize the discussion by technique rather than by data type.
The techniques differ primarily in \textit{how} they compensate for missing information: through explicit physics, transferable priors, architectural bias, or data-efficient training.
Our goal therefore is to discuss the latest advances in building data-driven and data-efficient models using (a) observational data that may be limited and/or partial; (b) reanalysis data that are typically limited in duration and sparse in spatial resolution; or (c) a combination of the two.
We note that parameter estimation for MGCM's also face limited, sparse, and partial observations.
We do not address those methods here.

Most existing approaches for building data-driven models from limited, sparse, or partial data rely on some form of prior structure.
Physics-informed methods use known equations to provide direct supervision or regularization, enabling learning from limited measurements but requiring accurate model specification. 
Equation-discovery methods relax this requirement by attempting to infer the governing equations from data, but instead demand (fewer) observations that are sufficiently dense and well resolved to reliably estimate derivatives.
Fully data-driven approaches reduce their dependence on explicitly defined physics by pretraining on large collections of prior simulations or related datasets, and fine-tuning on the limited, sparse, or partial data available.
This reframes data-efficiency around transfer and adaptation: whether pre-training has captured physics that transfers to the target system, and whether sparse observations are sufficient to condition the model at inference time.

More recent work also improves data efficiency by embedding generic PDE biases directly into model architectures, such as locality, continuity, conservation, variational structure, multiscale interactions, or operator-like information flow. 
These biases are especially useful when full governing equations are unavailable, but they usually encode only selected aspects of the physics. 
Training, sampling, and augmentation strategies provide an additional practical layer by making partial observations more useful, selecting informative measurements, or generating physically meaningful augmented views.

\paragraph{Learning with Known Equations.}
Approaches using known equations are highly data-efficient as the governing equations directly constrain the learned solution, reducing the amount of observational data required.
Assuming the governing equations are accurate, strong-form physics-informed methods can learn from sparse measurements by enforcing PDE residuals, boundary conditions, and/or initial conditions throughout the space-time domain~\cite{raissi2019pinns,wang2021pidon,Li2024pino}. Data primarily anchor the solution, while the equations provide the dominant constraint. 

These methods can work for limited and partial datasets.
Additionally, several extensions improve robustness to sparse and noisy observations. 
Bayesian formulations incorporate uncertainty in the learned solution and/or model parameters~\cite{yang2021bpinns}, while weak-form and variational methods enforce the governing equations in an integral or averaged sense rather than pointwise~\cite{wang2025wfpinns,kharazmi2019variational,kharazmi2021hp,rojas2024robust,eshaghi2025vino,de2024wpinns}.
There is precedent for applying such approaches to ERA5 data, where simplified primitive equations or low‑order physical constraints are enforced (diffusion, continuity, etc)~\cite{eusebi2024cyclone,park2025pint}; full circulation models are not incorporated for tractability.
However, these approaches fundamentally rely on (nearly) accurate specification of the governing physics, which is not available for Mars or readily usable as a learning constraint. 
In practice, only limited proxies or low‑order constraints may be feasible.

\paragraph{Pre-training \& Transfer Learning.}
Pre-training offers a data-efficient path when target data are limited but related data are available. 
Data-driven models (neural operators, transformer-based models, generative models e.g., \cite{li2020fourier,lu2021learning,wang2024cvit,cho2026mbno,pathak2022fourcastnet,Bi2023,lam2023graphcast}) can learn reusable solution structure from related PDE systems, then adapt to the target problem with relatively few samples (i.e. limited data)~\cite{soares2025towards,subramanian2023towards,rahman2024pretraining,herde2024poseidon,mccabe2024multiple,mccabe2025walrus}. 
In this sense, the problem shifts from learning Mars weather directly from limited observations to determining whether related data encode transferable atmospheric dynamics.
Pre-training has seen success in weather, climate, and geospatial systems~\cite{nguyen2023climax,prithviwxc2024,lentz2025oceantransfer}. Initial results towards leveraging pre-training on generic PDEs for the Martian atmosphere appeared in \cite{schmude2026pde}. See section \ref{sec:ai_poc_pde_pretraining_results} for a summary.

A distinct challenge is that the input observations at inference time may themselves be sparse or partial. Sparse-conditioning methods use sensor-aware encoders, masked training, reconstruction operators, or generative inpainting to condition learned priors on incomplete measurements~\cite{zhao2024recfno,dang2026deep,koupai2025enma}; some of these methods can also be considered data assimilation techniques.
Thus, pre-training reduces the amount of target-system data needed for learning, while sparse-conditioning reduces the amount of observed state information needed for prediction. For Mars weather, both are relevant: data are limited overall, and forecasts must often be conditioned on sparse input observations.

\paragraph{Architectural Inductive Bias.}
Architectural inductive bias can improve data efficiency in limited-data regimes by restricting models toward physically plausible solution operators/spaces. 
Rather than prescribing the full governing equations, these methods encode selected PDE structure directly into the architecture.

Some approaches impose known physical constraints. ClawNO builds conservation laws into neural operators through flux-based representations~\cite{liu2023clawno}, while Phase-Field DeepONet incorporates variational and thermodynamic structure, including irreversibility, for phase-field dynamics~\cite{li2023phasefielddeeponet}. 
Other methods encode more general field-level structure. Space-time continuous neural PDE models represent sparse, partially observed dynamics as continuous space-time fields with local spatiotemporal structure and latent PDE-like evolution~\cite{knigge2024spacetime}, while Multipole Graph Neural Operators impose hierarchical local-to-global message passing to capture multiscale interactions~\cite{li2020multipole}.

A further class modifies operator-learning architectures through structured attention or latent interaction spaces. 
Galerkin/Fourier attention treats attention more like operator approximation or numerical projection than generic sequence attention~\cite{cao2021galerkin}, Latent Neural Operators use physics-cross-attention to map between physical coordinates and latent operator spaces~\cite{wang2024lno}, and CoDA-NO~\cite{rahman2024pretraining}, Transolver~\cite{hu2026transolver}, and LinearNO~\cite{zhong2025linearattention} use codomain, physics-space, or efficient attention mechanisms respectively to capture interactions across PDE variables and spatial locations. 
These architectural choices improve data efficiency by reducing the burden on data to learn generic physical structure from scratch and suggest that building general PDE biases into the architecture is worthwhile.

\paragraph{Training, Sampling, and Augmentation Strategies.}
Data efficiency can also be improved through training procedures, sampling strategies, and augmentations that make each observation more informative. 
Ambient Physics, for example, learns directly from partial measurements rather than requiring complete system snapshots. During training, it randomly hides some already-observed data points and tasks the model with predicting them, encouraging plausible field reconstructions beyond the measured locations~\cite{majid2026ambient} (this is also a form of data assimilation).
This is directly relevant for Mars weather, where available measurements provide only partial observations of the underlying atmospheric state.

Other strategies are more naturally suited to limited-data regimes, where additional simulations, measurements, or labeled examples are expensive. 
One-shot operator learning addresses an extreme version of this setting by learning a local solution operator from a single, dense input-output PDE solution pair, using sliding local domains to generate many local training examples and then coupling the learned local operator through fixed-point or neural-informed iteration for new inputs~\cite{jiao2021one}; in this sense they artificially increase the size of their dataset. 
This is relevant for Mars if there is at least one sufficiently informative (i.e. spatially dense but temporally limited) simulation that can be used to learn and exploit local PDE structure.
Active operator learning uses predictive uncertainty to select informative training samples or observation locations, improving operator learning with fewer data points~\cite{winovich2026active}. 
For modeling the martian atmosphere, active learning could be relevant to choosing which MGCM simulation cases to generate next and at which spatial resolution.
Lie-symmetry-based self-supervised learning provides a physically meaningful data-augmentation strategy by using known PDE symmetries to generate equivalent views of the same system, improving learned representations without requiring additional labeled data~\cite{mialon2023liesymmetry}.
An example of symmetry for the martian atmosphere could be time translation at comparable seasons and dust conditions; other symmetries may be hard to justify due to symmetry-breaking fixed topography and dust variability.

\subsection{Candidate Data-Fusion Strategies}
\label{sec:fusion-strategies}

From our discussion in section \ref{sec:data} it should be clear that if one wants to leverage the OpenMARS dataset for pretraining, one needs to deal with the significant asymmetry in resolution between the reanalysis data and atmospheric retrievals. We sketch three candidate strategies for doing so. Each has tradeoffs that depend on the prioritized downstream task, the target resolution, and the available compute.

\paragraph{Reanalysis-prior, spacecraft-supervised super-resolution.}
Pretrain the FM on the full OpenMARS corpus at native $5^\circ$ resolution. At fine-tuning time, expose the model to paired low-resolution reanalysis and high-resolution orbital samples, with a super-resolution head that produces output at $2^\circ$ or $1^\circ$ and is supervised against interpolated or native-resolution atmospheric retrievals wherever they are spatially and temporally coincident. This is the most direct path, is well-understood methodologically \citep{harder2022physics,yang2023fourier}, and has the virtue of keeping the FM backbone untouched. The principal cost is bias transfer. The super-resolution head learns both the physics of the higher resolution and the biases of whichever orbital retrieval is used for supervision. A single-instrument supervisor, such as MCS alone, encodes instrument specific biases; a multi-instrument supervisor mixes biases that may not be dynamically consistent with one another. Ablations against held-out instruments are essential.

\paragraph{Multi-modal tokenization with cross-resolution attention.}
Rather than treating the atmospheric retrievals as a supervision target, this strategy treats them as additional input modalities tokenized at their native resolutions. Each instrument's retrievals enter the model as a separate token stream with position encodings that reflect orbital geometry, local time, and retrieval resolution. The FM's attention mechanism then fuses the reanalysis tokens and the orbital tokens explicitly, producing a state that is informed by both sources. This is closer in spirit to techniques used in direct to observation prediction (DOP) models such as GraphDOP \cite{alexe2024graphdop}, but adapted to the Mars multi-instrument case. See also \cite{vandal2025global}. The advantage is that the model can learn instrument-specific corrections and exploit the orbital data even where those observations are sparse. The cost is architectural complexity: the attention mechanism must scale to the combined token count, and positional encoding for non-sun-synchronous orbits is not a solved problem.

\paragraph{Diffusion-based resolution refinement.}
Treat fine resolution as a denoising target. At pretraining time, the FM learns a low-resolution state; at fine-tuning time, a diffusion-based refinement stage generates high-resolution states conditioned on the low-resolution FM output and on available atmospheric retrievals, following the spirit of PDE-Refiner \citep{lippe2023pde} adapted for cross-resolution rather than cross-time dynamics. The diffusion process gives an explicit distribution over fine-resolution states, providing built-in uncertainty quantification, and is a natural fit for sparsely observed systems. The cost is sampling expense at inference time for a state that spans five to seven orders of magnitude in its key quantities (pressure from $\sim 600$~Pa at the surface to $\sim 10^{-2}$~Pa at 105~km, dust opacity from $\sim 0.1$ in quiescent conditions to $>3$ in a global storm).

Each strategy is compatible with, rather than exclusive of, the others: The reanalysis-prior could leverage diffusion as the super-resolution head. The output of multi-modal tokenizers can serve as input to a supervised super-resolution stage. The choice answer depends on downstream task priorities. Moreover, note that the resolution imbalance arises primarily from the presence of OpenMARS data. If one chooses to train without that data, the fusion problem becomes considerably simpler.

\subsection{Initial Results towards an MAFM}
\label{sec:ai_initial_results}

\subsubsection{Predictive Models for the OpenMARS Data}
\label{sec:ai_initial_result_predictive_models}

\paragraph{Spectral-LS Transformer.}
\subparagraph{Data.}
The Mars SpectFormer is trained on OpenMARS reanalysis data~\cite{Holmes2020} spanning Mars Years~28--35 (MY28--MY35), corresponding to approximately 5{,}300 martian sols of atmospheric state observations at $5^{\circ}$ horizontal
resolution ($36 \times 72$ latitude--longitude grid).  Six surface-level prognostic variables are used: zonal wind ($u$), meridional wind ($v$), atmospheric temperature ($T$), surface pressure ($p_s$), surface skin temperature  $T_{\mathrm{surf}}$), and dust column opacity
($\tau_{\mathrm{dust}}$).  All variables are normalized to zero mean and unit standard deviation using statistics computed from the training partition.
 
The dataset is partitioned chronologically within the concatenated MY28--35 record: 80\% training, 10\% validation, and 10\% test.  Each timestep corresponds to approximately 2\,h of Martian atmospheric evolution.  The model receives 2~input frames (${\approx}4$\,h of context) and predicts one step
ahead (${\approx}2$\,h); during Phase~2 rollout training, 3 successive predictions are chained, corresponding to a ${\approx}6$\,h training horizon.
At inference, autoregressive rollouts of up to 12~steps (${\approx}24$\,h) are evaluated.  A persistence baseline-repeating the last input frame at every forecast lead time--is computed over the test partition and serves as the
no-skill reference for all metrics.
\begin{figure}[!h]
  \centering
  \includegraphics[width=0.95\textwidth]{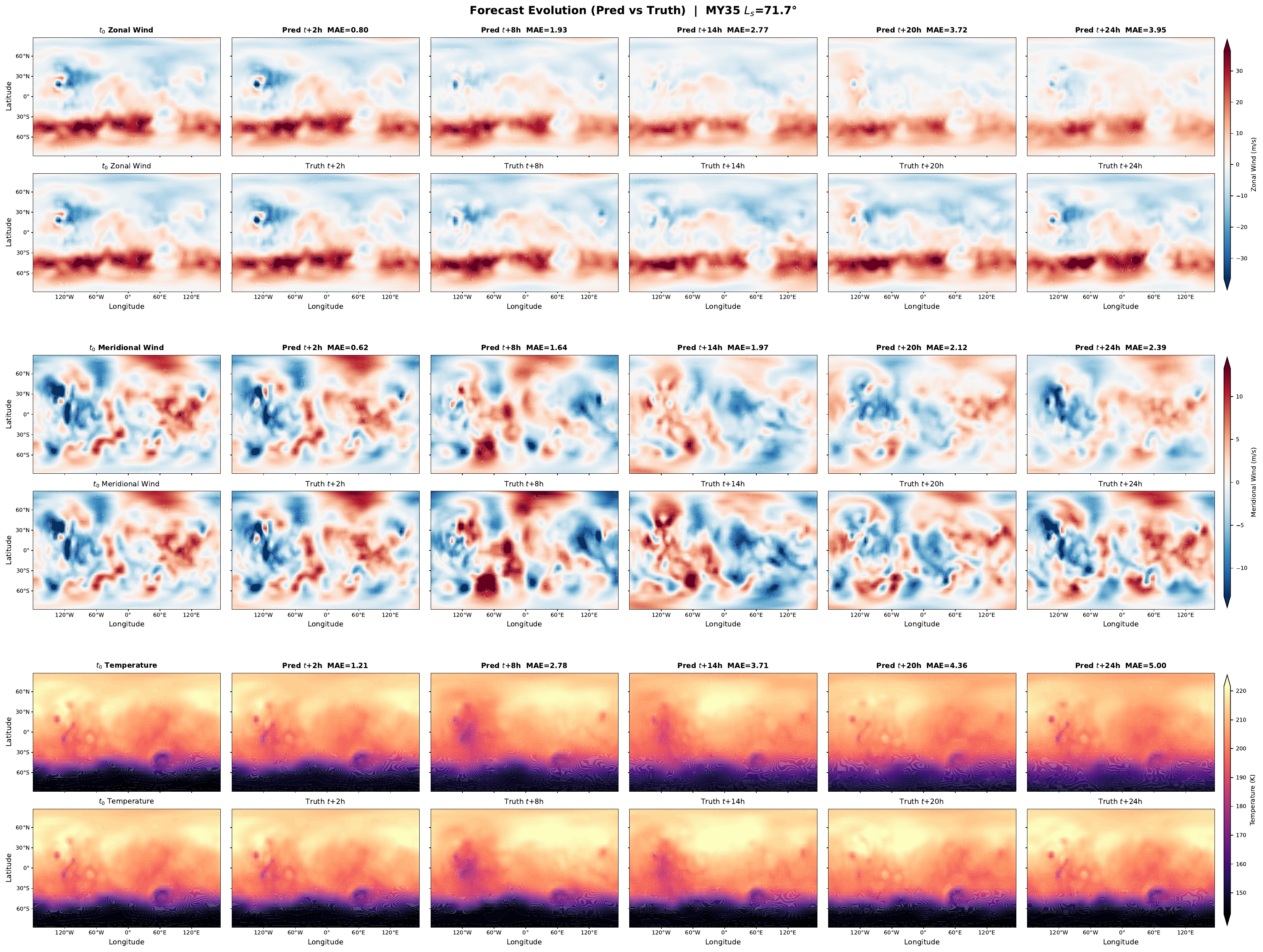}
  \caption{Forecast evolution of the Mars SpectFormer model with MAE error for Mars Year~35, $L_s = 71.7^{\circ}$ (northern spring). Each variable shows two rows: model predictions
(top) and OpenMARS reanalysis ground truth (bottom), from the initial condition ($t_0$) through a 24-hour autoregressive rollout at 2-hour resolution. Columns show selected lead times ($t{+}2$\,h, $t{+}8$\,h, $t{+}14$\,h, $t{+}20$\,h, $t{+}24$\,h). Per-panel MAE is annotated on prediction frames. Zonal wind ($u$)
and meridional wind ($v$) are in m\,s$^{-1}$; temperature in K. The model preserves large-scale spatial structure through 24\,h but accumulates smoothing at finer scales, particularly in the mid-latitude jet and polar temperature gradients. All fields are denormalized to physical units using training-set statistics.}
  \label{fig:mae_72}
\end{figure}
\subparagraph{Architecture.}
The Mars SpectFormer is a hybrid spectral--attention transformer that combines global frequency-domain processing with local spatial attention for Mars atmospheric forecasting.  The model treats the $36\times72$ atmospheric field as a 2D image, padded to $72\times72$ for compatibility with windowed attention, and partitions it into non-overlapping $2\times2$ patches (producing
$36\times36 = 1{,}296$ tokens).  The backbone consists of 8 transformer blocks:
the first 3 are \emph{spectral gating} blocks, whose mixing layer applies alearnable element-wise filter in the 2D Fourier domain (via \texttt{rfft2}/\texttt{irfft2}),followed by 5 \emph{long-short term attention} (LS) blocks~\cite{zhu2021long} with windowed local attention (window size~2), low-rank global projections (rank~2), and 8 heads.  Each block uses pre-norm LayerNorm, GELU activation, an MLP expansion ratio of 4, and stochastic depth (linearly increasing drop-path rate up to 0.05).

Patch embeddings are produced by a 2D convolution with kernel and stride equal to the patch size, mapping $C_{\mathrm{in}} \times T$ input channels to the embedding dimension $d = 256$.  Positional information is encoded via a deterministic multi-frequency Fourier encoding-sinusoidal and cosinusoidal
functions of latitude and longitude at $d/4$ logarithmically spaced frequencies-registered as a non-learnable buffer.  The decoder is a $1\times1$ convolution followed by PixelShuffle (factor 2), mapping tokens back to $C_{\mathrm{out}}$ output channels at the original spatial resolution.
After decoding, the padded region is cropped to recover the native $36\times72$ grid.  The model ingests 6-channel fields (\texttt{u}, \texttt{v},
\texttt{temp}, \texttt{ps}, \texttt{tsurf}, \texttt{dustcol}) over 2 input timesteps and predicts all 6 channels one step ahead, yielding approximately 17\,M trainable parameters.

\subparagraph{Training.}
The model is optimized with AdamW (lr $= 3\times10^{-4}$) using a sequential schedule: a 500-step linear warmup from $10^{-6}$ to the peak learning rate, followed by cosine annealing to a minimum of $10^{-6}$.  Training uses mixed precision (\texttt{bfloat16}) and Fully Sharded Data Parallel (FSDP) for
multi-GPU scaling, with gradient norms clipped to 1.0.

The training objective for the baseline configuration is a combined MSE\,+\,MAE loss:
\begin{equation}
  \mathcal{L}_{\mathrm{combined}}
  = \frac{w_{\mathrm{MSE}}}{B H W C}
    \sum_{b,i,j,c} \bigl(\hat{y}_{b,i,j,c} - y_{b,i,j,c}\bigr)^{2}
  + \frac{w_{\mathrm{MAE}}}{B H W C}
    \sum_{b,i,j,c} \bigl|\hat{y}_{b,i,j,c} - y_{b,i,j,c}\bigr|,
  \label{eq:loss_spectformer}
\end{equation}
where $w_{\mathrm{MSE}} = 1.0$ and $w_{\mathrm{MAE}} = 0.5$; $B$ is the batch size; $H = 36$, $W = 72$ are the latitude and longitude grid dimensions;
$C = 6$ is the number of output channels; and $\hat{y}$, $y$ denote prediction and ground truth, respectively.  No latitude weighting is applied.

Training follows a two-phase \emph{curriculum}:
\begin{itemize}
  \item \textbf{Phase~1} (epochs 1--40): single-step prediction.  The model receives 2 input frames and predicts one step ahead; the loss is computed on this single forecast.
  \item \textbf{Phase~2} (epochs 41--80): autoregressive rollout.  The model performs 3 successive forward passes, feeding each prediction back into the input window (sliding-window rollout).  The loss is accumulated over all rollout steps with equal weighting, so the model learns to correct its own error accumulation.  Gradients flow through the full rollout chain.
\end{itemize}
Dropout (rate 0.1) is applied after patch embedding and within the MLP blocks.
Training proceeds for 80 epochs with a batch size of 32.

\paragraph{Model Performance}
 \begin{figure}[!h]
  \centering
  \includegraphics[width=0.7\textwidth]{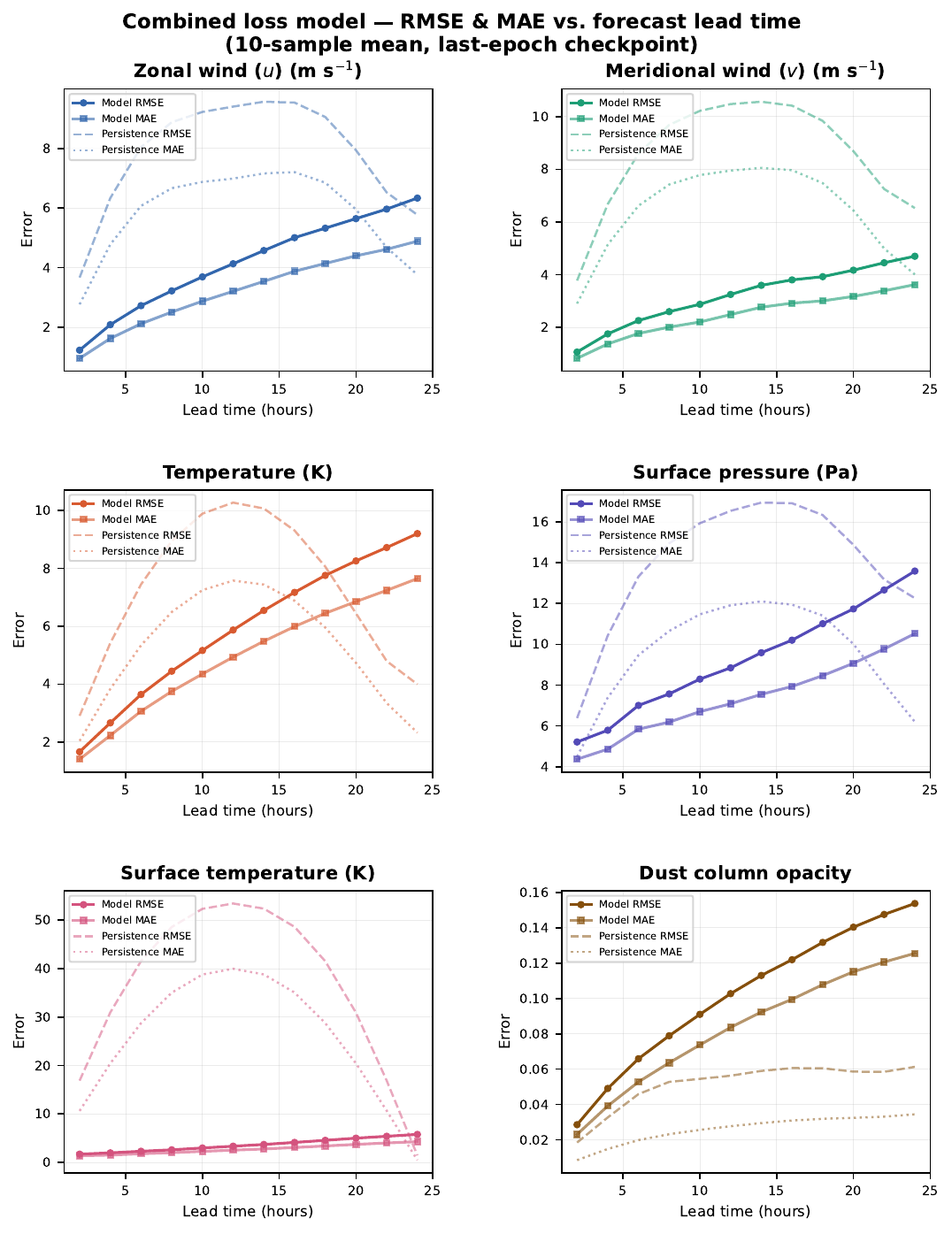}
  \caption{Autoregressive rollout performance of the combined-loss Mars SpectFormer model over a 24\,h forecast horizon (12 steps at ${\sim}2$\,h cadence), averaged over 10 test samples at pressure level 13 (${\sim}4$mbar). Each panel shows one prognostic variable: zonal wind ($u$, m\,s$^{-1}$), meridional wind ($v$, m\,s$^{-1}$), temperature (K), surface pressure (Pa), surface temperature (K), and dust column opacity (dimensionless). Solid lines with markers denote model RMSE (circles) and MAE (squares), while dashed and dotted lines show the persistence baseline at each lead time. The persistence error peaks near half a sol (${\sim}12$\,h) and then declines as the ground truth returns toward the same phase of the diurnal cycle, whereas the model error grows monotonically with lead time. The model stays below persistence RMSE for the full 24\,h window for $u$, $v$, and $p_s$, and through approximately 20\,h for temperature. Surface temperature persistence is dominated by the diurnal thermal wave (peaking at ${\sim}53$\,K), making the model error of ${\sim}5.8$\,K at 24\,h comparatively small. Dust column opacity is the only variable where the model exceeds persistence at all lead times.}
  \label{fig:mae_72}
\end{figure}

We evaluate the combined-loss (MSE\,+\,MAE) Mars SpectFormer on held-out test samples over a 12-step (24\,h) autoregressive rollout, comparing against a per-step persistence baseline that repeats the last input frame at every lead time.
 
\paragraph{Overall skill.}
The model substantially outperforms persistence for the dynamical variables $u$ and $v$ across the full 24\,h horizon. At 2\,h lead time the zonal wind RMSE is ${\sim}1.2$\,m\,s$^{-1}$ versus $3.7$\,m\,s$^{-1}$ for persistence, a threefold improvement. For meridional wind, the model never crosses the persistence baseline even at 24\,h (4.7\,m\,s$^{-1}$ versus 6.5\,m\,s$^{-1}$). Surface pressure follows a similar pattern, remaining below persistence through ${\sim}22$\,h before converging near the baseline at 24\,h (13.6\,Pa versus 12.3\,Pa).
 
\paragraph{Diurnal structure in the persistence baseline.}
The per-step persistence baseline reveals strong diurnal modulation. For all variables, persistence error rises steeply through the first ${\sim}12$\,h (half a sol) and then decreases as the ground truth returns toward the same phase of the diurnal cycle. This is most extreme for $T_{\mathrm{surf}}$, where persistence peaks at ${\sim}53$\,K RMSE near 12\,h before collapsing to ${\sim}1.4$\,K at 24\,h. The model error grows monotonically to ${\sim}5.8$\,K, vastly outperforming persistence through the first ${\sim}22$\,h but exceeding it at the 24\,h point where persistence benefits from diurnal wrap-around. Temperature and surface pressure show similar asymmetric behavior in their persistence curves (2.9 $\to$ 10.3 $\to$ 4.0\,K and 6.4 $\to$ 17.0 $\to$ 12.3\,Pa, respectively).
 
\paragraph{Temperature and dust: persistent challenges.}
Temperature is the most difficult variable for the model. It crosses the persistence RMSE at ${\sim}20$\,h, reaching 9.2\,K at 24\,h compared to 4.0\,K for persistence. This suggests the model accumulates systematic thermal bias during autoregressive rollout, likely in the polar regions where sharp meridional temperature gradients are difficult to maintain. Dust column opacity is the only variable where the model is worse than persistence at \emph{every} lead time (0.029 versus 0.019 at 2\,h, growing to 0.154 versus 0.061 at 24\,h). This is consistent with the intermittent, spatially localized nature of Martian dust events, which are poorly captured by a deterministic model trained on smooth reconstruction objectives.

\paragraph{Mars-Adapted GraphCast}

The models of this study are trained on OpenMARS reanalysis data~\cite{Holmes2020} spanning Mars
Years 24--28 (MY24--MY28), corresponding to approximately 3{,}500 Martian sols of atmospheric
state observations at $5^{\circ}$ horizontal resolution ($36 \times 72$ latitude--longitude
grid). The six surface-level input variables are zonal wind ($u$), meridional wind
($v$), atmospheric temperature ($T$), surface pressure ($p_s$), surface skin temperature
($T_{\mathrm{surf}}$), and dust column opacity ($\tau_{\mathrm{dust}}$). All variables are normalized to zero mean and unit standard deviation
using statistics computed from the training corpus. The models are evaluated on MY29
($L_s \approx 278^{\circ}$--$297^{\circ}$, Southern Summer), a held-out test year unseen
during training. Each forecast step advances the atmosphere by 3 OpenMARS timesteps (${{\approx}}6$~h), and
autoregressive rollouts of up to 12 steps (72~h) are evaluated.

\begin{figure}[!h]
  \centering
\includegraphics[width=0.95\textwidth]{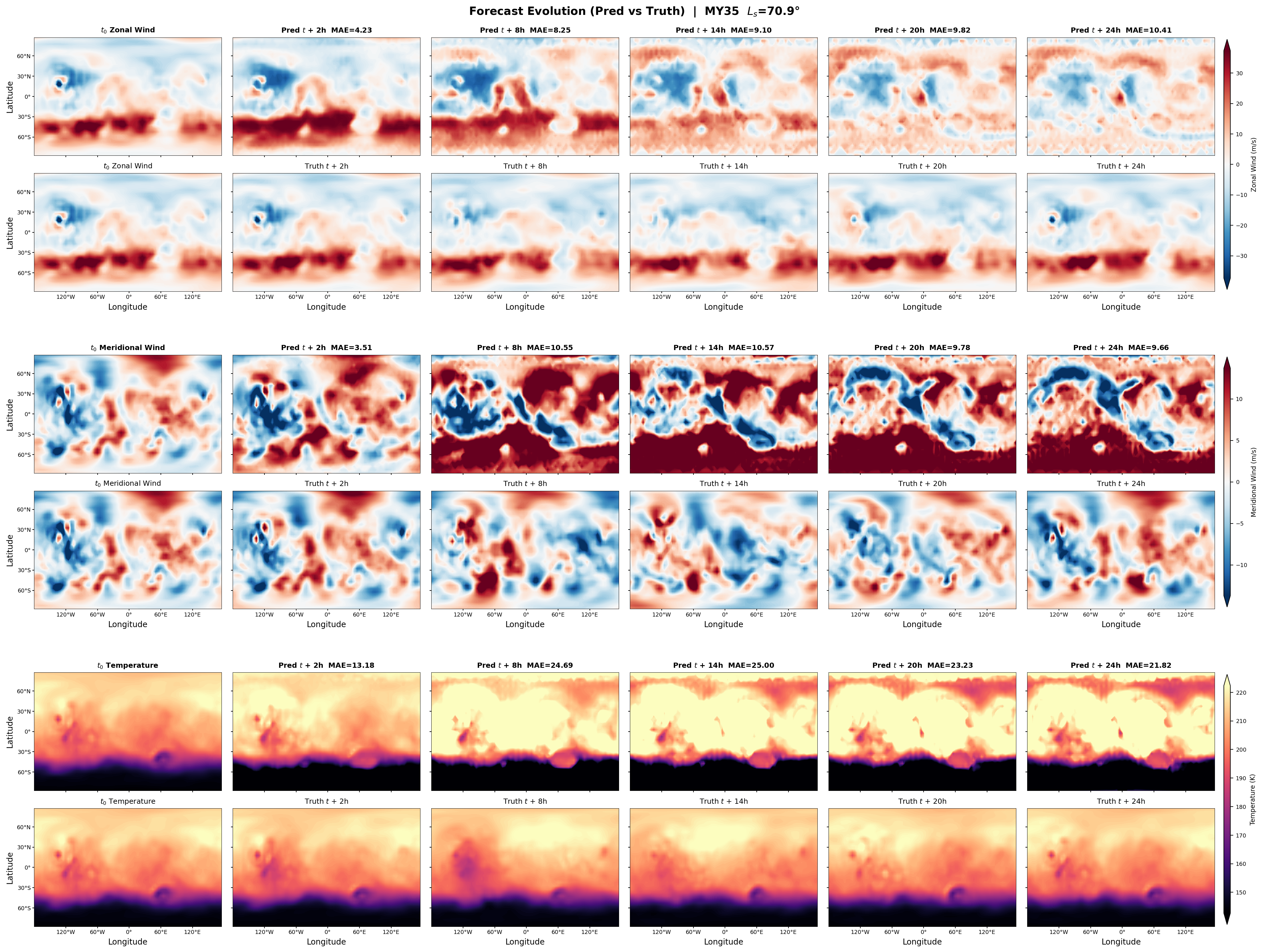}
  \caption{Forecast evolution of the Mars GraphCast model with MAE error for Mars Year~35, $L_s = 71.7^{\circ}$ (northern spring). Each variable shows two rows: model predictions
(top) and OpenMARS reanalysis ground truth (bottom), from the initial condition ($t_0$) through a 24-hour autoregressive rollout at 2-hour resolution. Columns show selected lead times ($t{+}2$\,h, $t{+}8$\,h, $t{+}14$\,h, $t{+}20$\,h, $t{+}24$\,h). Per-panel MAE is annotated on prediction frames. Zonal wind ($u$)
and meridional wind ($v$) are in m\,s$^{-1}$; temperature in K. The model exhibits significant errors in temperature forecasting from the earliest time, with temperature MAE exceeding 16\,K at $t{+}2$\,h, suggesting a systematic bias rather than gradual error growth. Wind components show rapid spatial pattern divergence, with the model failing to reproduce mesoscale structures beyond short horizons. All fields are denormalized to physical units using training-set statistics. All fields are denormalized to physical units using training-set statistics.}
  \label{fig:graphcast_24hr}
\end{figure}

\begin{figure}[!h]
  \centering
\includegraphics[width=0.7\textwidth]{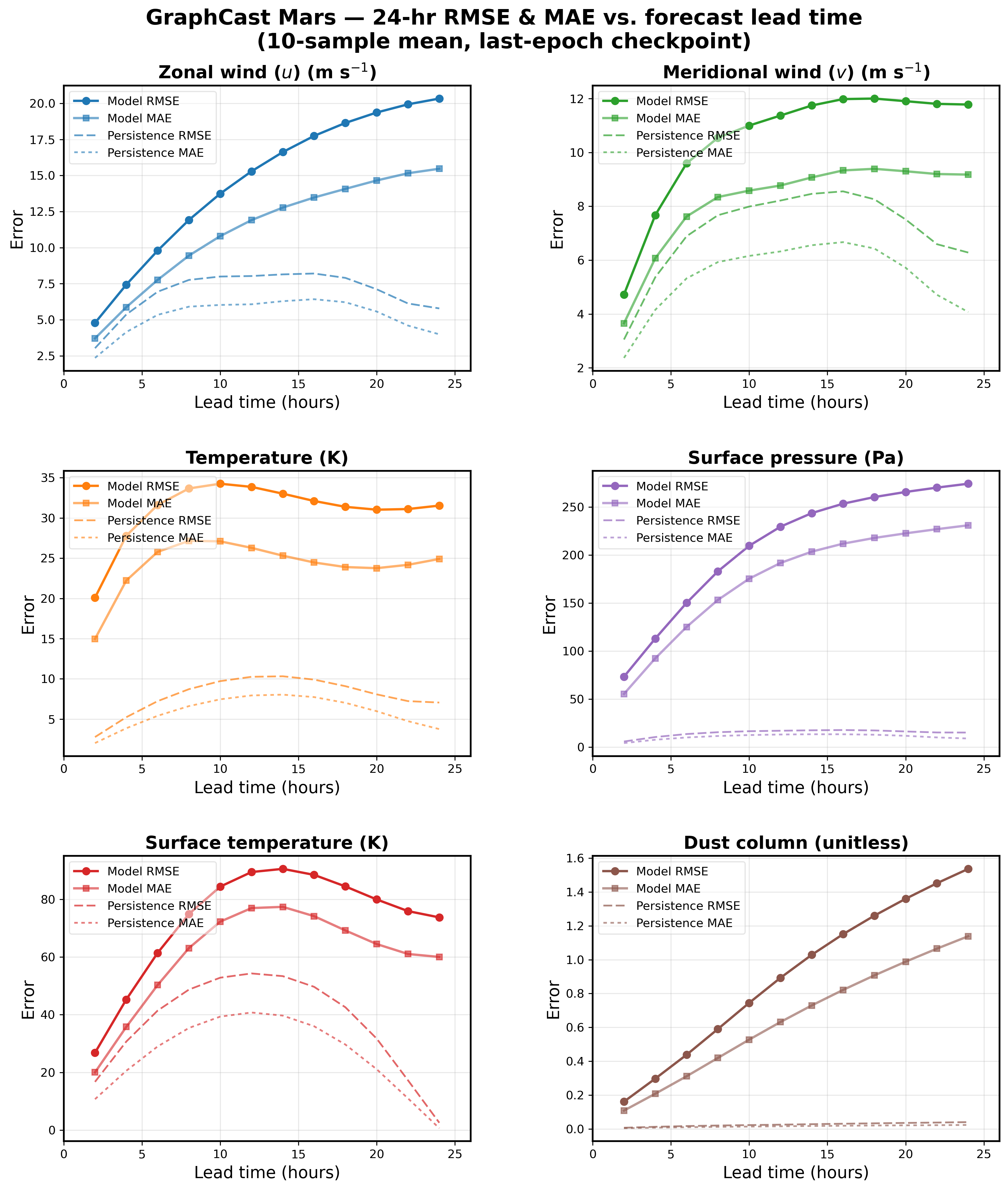}
  \caption{Autoregressive rollout performance of the combined-loss Mars GraphCast model over a 24\,h forecast horizon (12 steps at ${\sim}2$\,h cadence), averaged over 10 test samples at pressure level 13 (${\sim}4$mbar). Each panel shows one prognostic variable: zonal wind ($u$, m\,s$^{-1}$), meridional wind ($v$, m\,s$^{-1}$), temperature (K), surface pressure (Pa), surface temperature (K), and dust column opacity (dimensionless). Solid lines with markers denote model RMSE (circles) and MAE (squares); dashed lines show the corresponding persistence baseline. The persistence error for dynamical variables ($u$, $v$, temperature, surface temperature) peaks near half a sol (${\sim}12$\,h) and subsequently declines as the ground truth returns toward the same phase of the diurnal cycle. In contrast, model errors grow monotonically or peak at large amplitude and do not recover, indicating that the model fails to track the diurnal cycle. The model exceeds persistence RMSE for $u$ and temperature from the earliest lead time, and for $v$ as well. Surface pressure errors grow continuously to ${\sim}270$\,Pa at 24\,h, far above the near-flat persistence baseline (${\sim}15$\,Pa), suggesting a systematic drift in the pressure field. Dust column opacity errors likewise grow monotonically and substantially exceed persistence at all lead times. These results indicate that the current checkpoint has not learned to reproduce the dominant diurnal and synoptic variability of the Mars mid-atmosphere, motivating further training or architectural refinement.}
  \label{fig:graphcast_72hr}
\end{figure}

\subparagraph{Architecture.}
GraphCast~\cite{Lam2023} is a graph neural network (GNN) weather model that represents the
atmosphere on an icosahedral multi-mesh. The Mars-adapted variant encodes the $36\times72$
latitude--longitude grid onto an icosahedral mesh with 4 refinement levels (2{,}562 mesh
nodes), processes information through 8 rounds of message passing with a latent dimension of
256, and decodes back to the grid. Beyond the five prognostic variables, the model ingests
six forcing variables encoding martian seasonality: solar longitude ($L_s$), and the sine
and cosine of sol-of-day and sol-of-year progress as well as a static surface-altitude
field representing martian topography. This design allows GraphCast to distinguish diurnal
and seasonal atmospheric regimes without explicit physics parameterization.

\begin{figure}[!h]
  \centering
\includegraphics[width=1.0\textwidth]{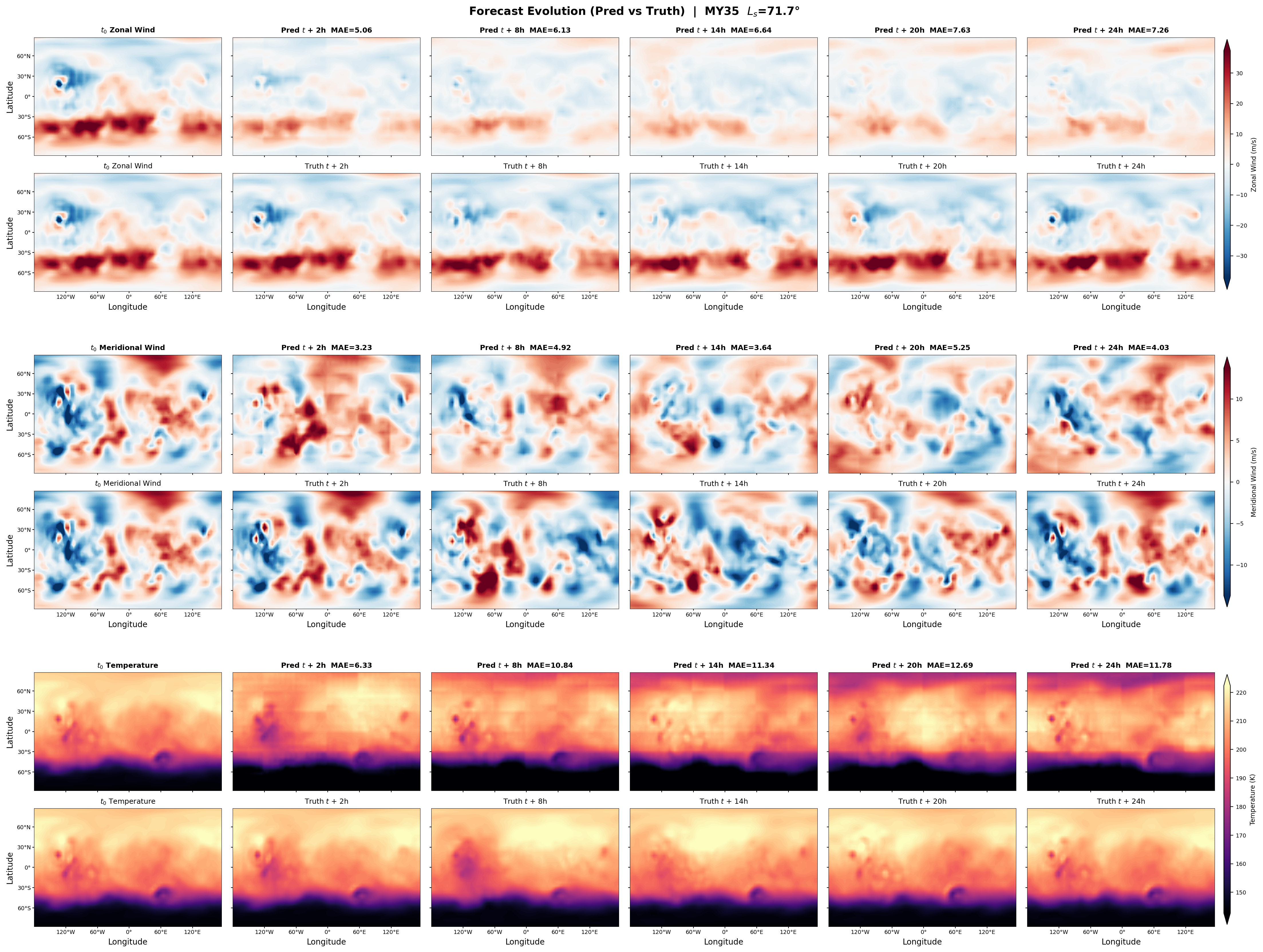}
  \caption{Forecast evolution of the Mars Prithvi-WxC model with MAE error for Mars Year~35, $L_s = 71.7^{\circ}$ (northern spring). Each variable shows two rows: model predictions
(top) and OpenMARS reanalysis ground truth (bottom), from the initial condition ($t_0$) through a 24-hour autoregressive rollout at 2-hour resolution. Columns show selected lead times ($t{+}2$\,h, $t{+}8$\,h, $t{+}14$\,h, $t{+}20$\,h, $t{+}24$\,h). Per-panel MAE is annotated on prediction frames. Zonal wind ($u$) and meridional wind ($v$) are in m\,s$^{-1}$; temperature in K. The model maintains reasonable skill for zonal wind (MAE $\sim$5--7\,m\,s$^{-1}$) and shows moderate errors for meridional wind (MAE $\sim$2--4\,m\,s$^{-1}$), with both wind components preserving large-scale circulation patterns through 24\,h. Temperature forecasts exhibit larger errors (MAE $\sim$5--11\,K), with the model slightly underestimating polar temperature gradients at longer lead times but capturing the dominant latitudinal structure. Overall, the Prithvi-WxC model demonstrates improved mid-atmosphere forecast skill compared to GraphCast, particularly for temperature and wind fields. All fields are denormalized to physical units using training-set statistics.}
  \label{fig:prithvi_24hr}
\end{figure}

\subparagraph{Training.}
The model is optimized with AdamW ($\mathrm{lr} = 1\times10^{-4}$, weight decay $= 0.01$)
for 100 epochs with a batch size of 8. To account for the unequal area representation of
latitude--longitude grids, the MSE loss is latitude-weighted:
\begin{equation}
    \mathcal{L}_{\mathrm{GC}}
    = \frac{1}{B \cdot N_{\mathrm{lat}} \cdot N_{\mathrm{lon}} \cdot C}
      \sum_{b,i,j,c} \cos(\phi_i)
      \left(\hat{y}_{b,i,j,c} - y_{b,i,j,c}\right)^2,
    \label{eq:loss_gc}
\end{equation}
where $B$ is the batch size; $N_{\mathrm{lat}} = 36$ and $N_{\mathrm{lon}} = 72$ are the
number of latitude and longitude grid points, respectively; $C$ is the number of predicted
variables; $b \in \{1,\ldots,B\}$ indexes the sample in the batch; $i \in \{1,\ldots,N_{\mathrm{lat}}\}$
and $j \in \{1,\ldots,N_{\mathrm{lon}}\}$ index the spatial grid point; $c \in \{1,\ldots,C\}$
indexes the output variable; $\phi_i$ is the latitude (in radians) of the $i$-th row, so
that $\cos(\phi_i)$ gives the fractional surface-area weight of that latitude band;
$\hat{y}_{b,i,j,c}$ is the model's predicted value; and $y_{b,i,j,c}$ is the corresponding
ground-truth value. Gradients are clipped to unit norm.
A 90/10 chronological train--validation split over MY24--28 yields approximately 3{,}150
training and 350 validation samples.

\paragraph{Mars-Adapted Prithvi-WxC}

\begin{figure}[!h]
  \centering
\includegraphics[width=0.7\textwidth]{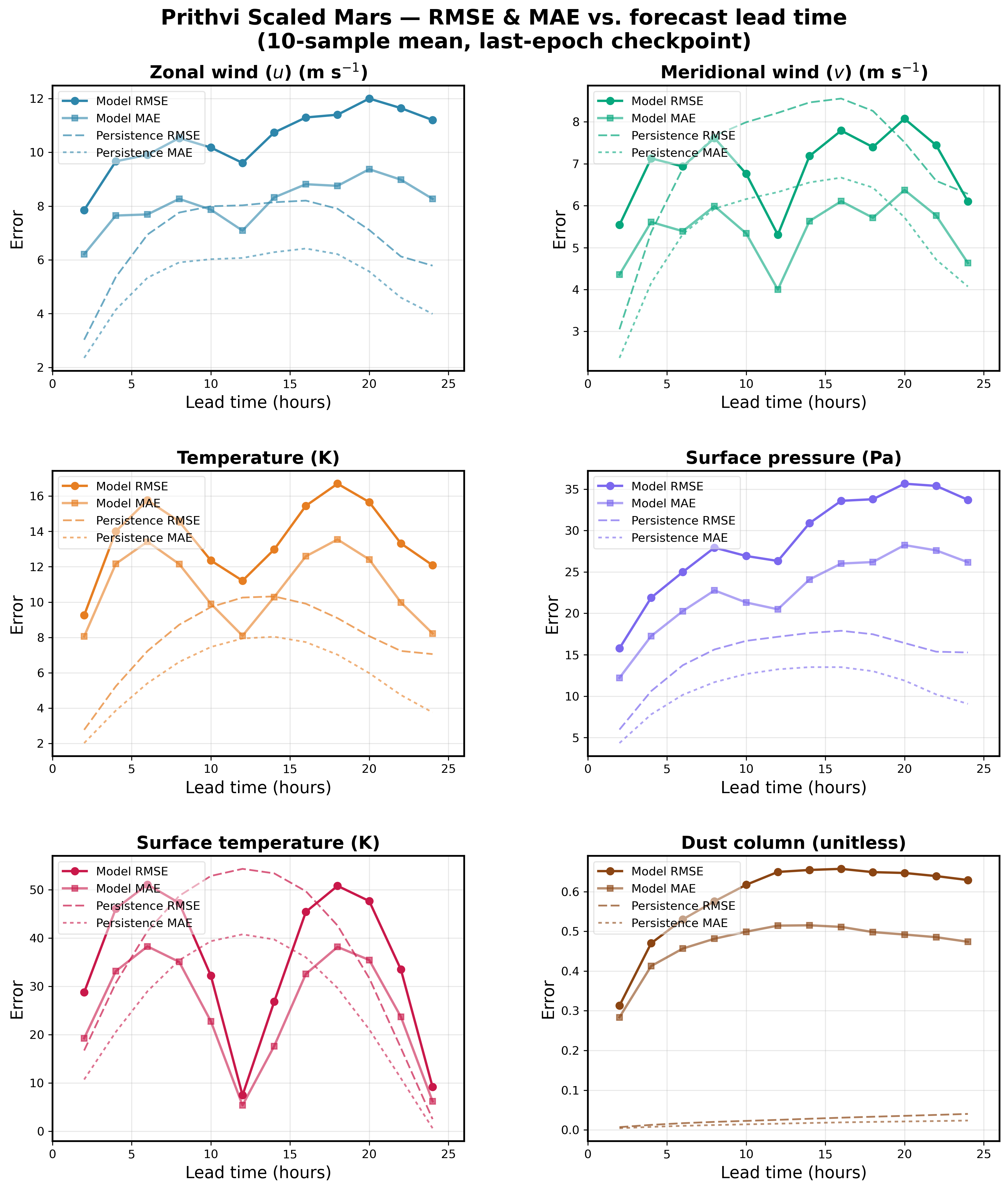}
  \caption{Autoregressive rollout performance of the combined-loss Mars Prithvi-WxC model over a 24\,h forecast horizon (12 steps at ${\sim}2$\,h cadence), averaged over 10 test samples at pressure level 13 (${\sim}4$mbar). Each panel shows one prognostic variable: zonal wind ($u$, m\,s$^{-1}$), meridional wind ($v$, m\,s$^{-1}$), temperature (K), surface pressure (Pa), surface temperature (K), and dust column opacity (dimensionless). Solid lines with markers denote model RMSE (circles) and MAE (squares); dashed lines show the corresponding persistence baseline. The persistence error for surface temperature and temperature exhibits a pronounced diurnal oscillation peaking near $t{+}12$\,h and declining as the ground truth revisits the same phase of the diurnal cycle; the model errors follow a similar oscillatory structure, indicating that the model partially captures the diurnal thermal forcing. For zonal and meridional wind, model errors grow quasi-monotonically but meridional wind remains comparable to persistence ($v$ RMSE $\sim$5--8\,m\,s$^{-1}$), suggesting skill close to but not clearly below the persistence baseline at this pressure level. Surface pressure shows a steadily growing RMSE reaching ${\sim}35$\,Pa at 24\,h, well above the persistence baseline, indicating accumulating systematic drift in the pressure field. Dust column opacity errors grow rapidly in the first 10\,h (${\sim}0.6$ RMSE) before plateauing, substantially exceeding the near-zero persistence baseline throughout the forecast window. Overall, the Prithvi-WxC model demonstrates partial skill for wind components while exhibiting systematic biases in surface pressure and dust column that warrant further investigation.}
  \label{fig:prithvi_24hr}
\end{figure}

\subparagraph{Architecture.}
A Mars-adapted version of Prithvi-WxC~\cite{Schmude2024} is a transformer (ViT) \cite{dosovitskiy2020vit} based foundation model for weather
and climate, adapted here to the Mars surface. The model treats the $36\times72$ atmospheric
field as a 2D image, partitioned into non-overlapping $2\times2$ patches (producing 648
tokens), and processes them via 12 layers (8 encoder + 4 decoder layers) of multi-head self-attention (8 heads) with local
windowed attention in windows of size $3\times6$. The embedding dimension is 384 with an MLP
expansion ratio of 4, yielding approximately 17~M trainable parameters. The model ingests
six-channel atmospheric fields ($u$, $v$, $T$, $p_s$, $T_{\mathrm{surf}}$, $\tau_{\mathrm{dust}}$) and outputs six prognostic channels ($u$, $v$, $T$, $p_s$, $T_{\mathrm{surf}}$, $\tau_{\mathrm{dust}}$); dust column opacity serves as a diagnostic input only and is not predicted.

\subparagraph{Training.}
Prithvi-WxC is optimized with AdamW ($\mathrm{lr} = 5\times10^{-5}$, weight decay $= 0.05$)
using a cosine annealing schedule with a 5-epoch linear warmup. The training objective is the
standard MSE:
\begin{equation}
    \mathcal{L}_{\mathrm{Prithvi}}
    = \frac{1}{B \cdot H \cdot W \cdot C_{\mathrm{out}}}
      \sum_{b,i,j,c} \left(\hat{y}_{b,i,j,c} - y_{b,i,j,c}\right)^2.
    \label{eq:loss_prithvi}
\end{equation}
Here $B$ is the batch size; $H = 36$ and $W = 72$ are the spatial height and width of the
input image (latitude and longitude grid points); $C_{\mathrm{out}} = 6$ is the number of
predicted output channels ($u$, $v$, $T$, $p_s$, $T_{\mathrm{surf}}$, $\tau_{\mathrm{dust}}$); $b$, $i$, $j$, $c$ index the batch, latitude row,
longitude column, and output channel, respectively; $\hat{y}_{b,i,j,c}$ and $y_{b,i,j,c}$
are the predicted and ground-truth values. Unlike Eq.~\eqref{eq:loss_gc}, no latitude
weighting is applied, so all grid cells contribute equally to the loss. Stochastic depth (drop-path rate 0.2) and input dropout (0.1) serve as regularizers.
Gradient norms are clipped to 1.0. Training proceeds for 50 epochs with a batch size of 2.



\subsubsection{The Efficacy of PDE Pretraining}
\label{sec:ai_poc_pde_pretraining_results}

For atmospheric physics on Earth, it is by now well established that one can obtain skillful forecast models with purely data-driven approaches. See our summary in Section \ref{sec:ai_atmospheric_physics}.\footnote{One cannot overemphasize that this does not mean that AI is replacing HPC or other conventional methods. It simply means that AI is a valid additional approach that can be run as an alternative to or combined with existing methods.} The arguably biggest driver in this development has been the availability of large-scale data to train such models on. In particular global reanalysis datasets such as ERA5. While our discussion in Section \ref{sec:ai_initial_result_predictive_models} suggests that one can leverage similar approaches for Mars when working with the OpenMARS dataset at that dataset's native resolution, it is important to consider what happens when sufficient data is not available. Which we did in section \ref{sec:ai_data_efficiency}.

\begin{figure}[!h]
\begin{center}
\includegraphics[width=1.0\textwidth]{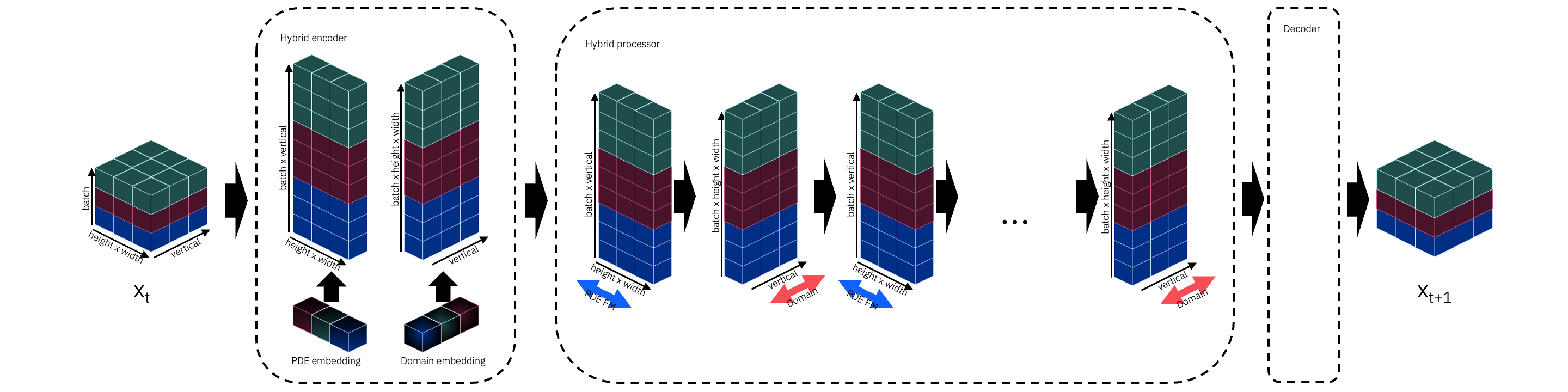}
\end{center}
\caption{Extending Poseidon (scOT) to three dimension.  (Source: \cite{schmude2026pde}.)}
\label{fig:pde_3d_extension}
\end{figure}

\begin{figure}[!h]
\begin{center}
\includegraphics[width=0.5\textwidth]{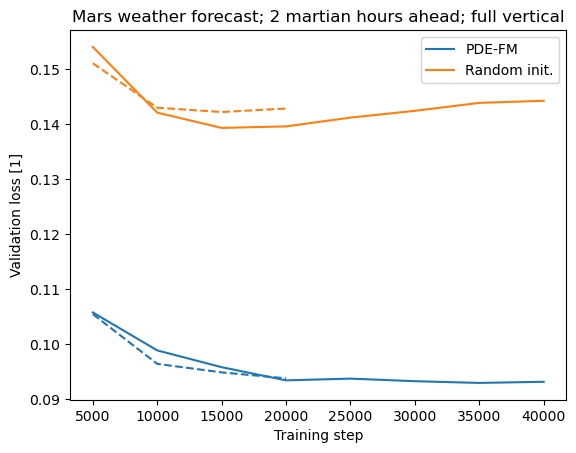}
\end{center}
\caption{When extending the training to 40,000 steps, the randomly initialized model not only does not close the performance gap, but rather shows clear signs of overfitting. The differences between the 20,000 and 40,000 step runs stem from the learning rate scheduler.  (Source: \cite{schmude2026pde}.)}
\label{fig:pde_loss_curves}
\end{figure}

\begin{figure}[!h]
\begin{center}
\includegraphics[width=0.5\textwidth]{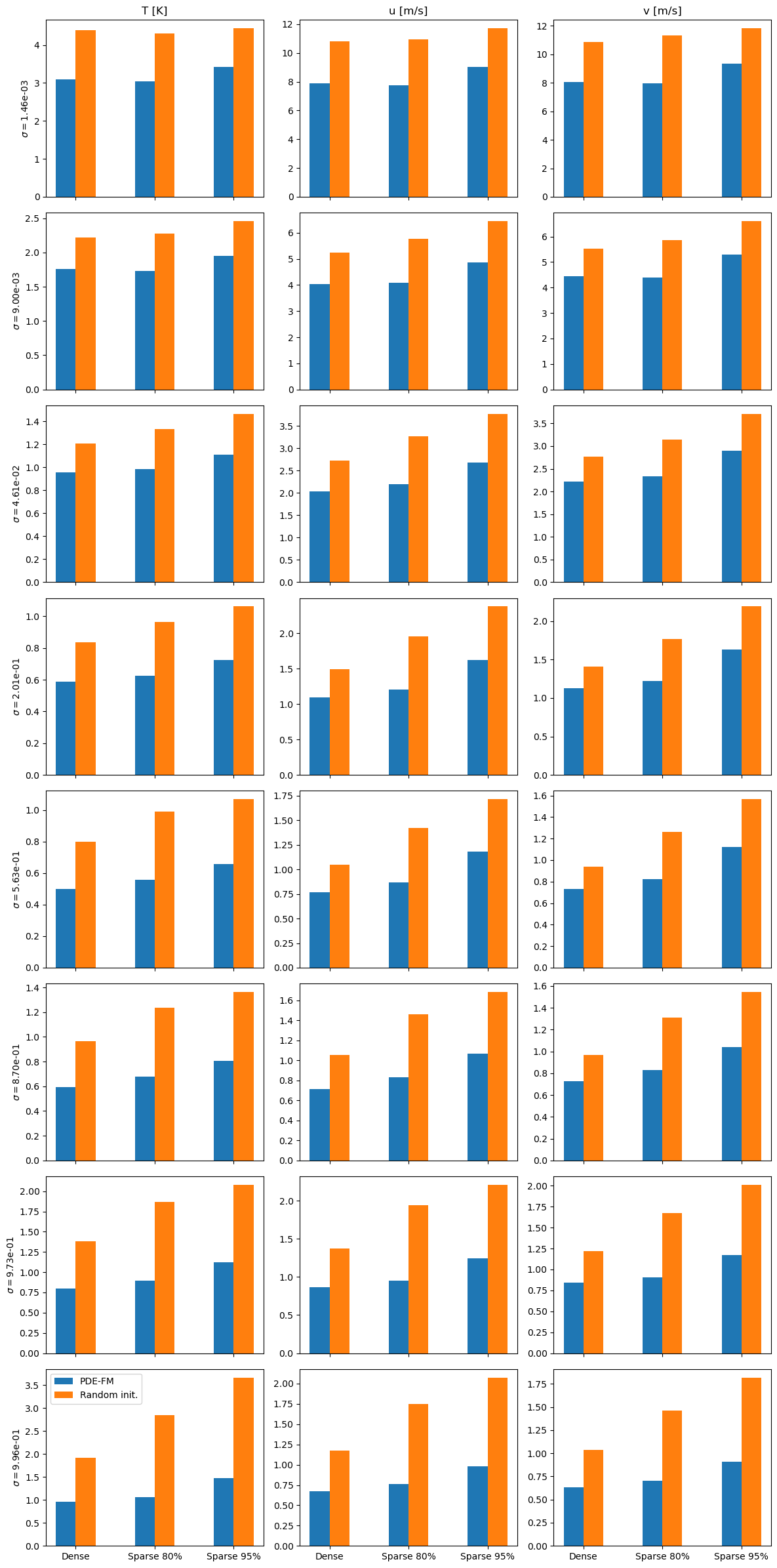} 
\end{center}
\caption{Model performance (MAE) after 20,000 training steps. The figure shows $T$, $u$ and $v$ at different $\sigma$-levels. Moreover, it also shows changes in performance when sparsifying the initial conditions. (Source: \cite{schmude2026pde}.)}
\label{fig:pde_pretraining_results}
\end{figure}

As discussed there, one possibility is to consider transfer learning from or pre-training with other systems. Indeed, the last few years have seen the rise of so-called PDE foundation models. That is, AI model pre-trained on heterogeneous corpora of numerical solutions to a variety of partial differential equations with changing initial conditions and parameters (\cite{herde2024poseidon,mccabe2024multiple,mccabe2025walrus}).

The authors of \cite{schmude2026pde} considered whether one can leverage such models for applications beyond the idealized settings of numerical solutions to PDEs. In particular, they considered whether one can accelerate the training of predictive models trained on the OpenMARS data by using a model pretrained on PDEs. Starting point was the \emph{Poseidon} PDE foundation model \cite{herde2024poseidon}. Poseidon was pretrained on two-dimensional PDE data. In particular solutions to the Navier-Stokes and Compressible Euler equations. A key innovation of \cite{schmude2026pde} was the extension of this two-dimensional model to three dimensions. The core idea, as outlined in figure \ref{fig:pde_3d_extension}, is to let the pretrained model act along the horizontal dimensions while introducing new components that act along the vertical. Only the parts of the model acting along the horizontal are pretrained; the remainder is randomly initialized.

With this 3D extension of a 2D model, the authors considered pressure level data from the OpenMARS dataset. In particular, the model was trained with data form 18 levels from sol $2674.416748$ to sol $5348.750000$ and validated using sol $5348.833496$ to $6031.000000$. This corresponds to Mars Years 28 through 31 for training and Mars Year 32 for validation. The training and validation loss for two model runs can be seen in figure \ref{fig:pde_loss_curves}. Crucially, not only does the pretrained model significantly outperform the randomly initialized one; the latter starts overfitting before even approaching the performance of the former.

Figure \ref{fig:pde_pretraining_results} shows results in physical units. Note that it not only shows results for prediction from dense initial conditions. Yet also from sparsely sampled initial condition data. As should be clear from the plot, the impact of PDE pretraining is further enhanced in the case where initial conditions become sparse. This should be relevant for MAFM pretraining from sparse observations.

\subsubsection{Explainability for MARS Foundation Model}
Mechanistic Interpretability (MI) \cite{olah2022} has recently emerged as a powerful framework for understanding the internal computation of Large Language Models (LLMs). At a high level, MI aims to move beyond black-box evaluation by identifying the internal structures, representations, and pathways that give rise to model behavior. Broadly, existing MI techniques can be grouped into three complementary strands, each of which offers promising opportunities for application in weather foundation models such as MARS.

First, causal component analysis focuses on identifying the internal units (e.g., attention heads, neurons, or intermediate representations) that are causally responsible for specific model behaviors. Techniques such as activation patching \cite{zheng2023towards} and edge attribution \cite{syed2023attribution} enable the isolation of components whose intervention significantly alters the model’s output. In the context of weather foundation models, these methods can be used to investigate whether predictions—such as extreme weather events or regional forecasts—are driven by localized causal pathways or distributed representations. Furthermore, such analysis can reveal whether different prediction tasks (e.g., temperature forecasting vs. precipitation modeling) rely on shared or distinct internal mechanisms, analogous to domain-specific expertise (e.g., geological vs. atmospheric reasoning). Importantly, identifying these causal components also enables targeted interventions, allowing for localized editing of model behavior in a controlled and interpretable manner.

Second, representation-level analysis aims to uncover the latent features encoded within the model’s internal activations. Recent work in mechanistic interpretability has explored the concept of monosemantic features through sparse autoencoders(SAE) \cite{huben2024sparse}, where individual neurons or directions in activation space correspond to interpretable concepts. Extending this idea to Mars foundation models, such techniques can help identify whether the model encodes meaningful physical variables (e.g., pressure gradients, humidity patterns, or atmospheric circulation structures) in its internal representations. Similar approaches have already shown promise in biological foundation models \cite{biology}, where latent features correspond to biologically meaningful signals. Applying these methods to weather models could provide insights into whether the model learns physically grounded abstractions or relies on spurious correlations.

Third, steering and control methods \cite{haon2025mechanistic} aim to modify model behavior by intervening on internal representations. These approaches provide efficient mechanisms to guide model outputs without requiring full retraining. In weather forecasting settings, such techniques could be used to correct undesirable behaviors in failure scenarios, such as systematic biases, instability in extreme event prediction, or sensitivity to noisy inputs. By steering internal activations, it may be possible to enforce desirable properties—such as physical consistency or robustness—while preserving overall model performance.

Taken together, these three strands of mechanistic interpretability—causal component analysis, representation discovery, and steering—offer a comprehensive toolkit for understanding, diagnosing, and improving weather foundation models. Applying these methods has the potential not only to enhance model transparency but also to enable more reliable and controllable forecasting systems in high-stakes environmental applications.

\section{Design Considerations for an MAFM}
\label{sec:discussion}

The existing datasets and machine learning methodology narrows which research questions can be addressed. The emerging themes are that climatological characterization of atmospheric dynamics (DAC, LLJs, AMEC, ACB) is hindered by the coarse resolution and limited temporal coverage of existing reanalyses. Further, existing reanalyses have incomplete understanding of the water and CO$_2$ cycle and radiative effects. 

An MAFM could likely improve spatial resolution from 5$^\circ$ up to $1^\circ$ and maintain hourly resolution. This could improve characterization of the atmospheric dynamics in medium-scale (orographic clouds, DACs, AMEC) and diurnal processes (ACB). Significantly harder to improve would be representation of black swan events (global dust cycles), very high-resolution phenomena (microscale and local storms). In comparison to Earth, Mars has increased scientific importance of the water and CO$_2$ cycle and radiation. Accurately emulating state transitions and low-magnitude high-variability variables is a problem that is not yet fully resolved in Earth machine learning models, and poses an interesting challenge to an MAFM.

To contribute to Mars reanalyses, both approaches are promising: Fusing reanalysis and observations or observation-to-observation prediction. The first might lead to higher spatial resolution and bias correction of Mars atmospheric models. The second would be helpful for generating multiple lines of evidence, for example, for a sensitivity analysis of AMEC to water: Agreement between both models would validate process understanding and disagreement could point out areas for model improvement.

We distinguish between short-term and long-term rollouts of atmospheric flow. As short-term rollouts are already well-validated on Earth they are the more approachable target. Longer-term rollouts will be challenged by accumulating drift and possible divergence in ML models. Both, reanalysis runs that incorporate observations throughout a martian year or free running ensemble simulations for studying climatological characteristics are challenging tasks, but it seems that incorporating observations could resolve the issue of stability on reanalysis timescales.

Regarding the MAFM architecture, vision transformers are well-validated on Earth and development will likely start there and iteratively incorporate more assumptions and inductive biases to overcome data limitations and increase interpretability. Likewise, deterministic models are well-validated but more experimental probabilistic models provide the promise of increased skill for data assimilation and accurately capturing statistics of chaotic high-resolution events.

\newpage
\bibliography{references}

@article{abdi2026hrrrcast,
  title={HRRRCast: A Data-Driven Emulator for Regional Weather Forecasting at Convection-Allowing Scales},
  author={Abdi, Daniel and Jankov, Isidora and Madden, Paul and Vargas, Vanderlei and Smith, Timothy A and Frolov, Sergey and Flora, Montgomery and Potvin, Corey},
  journal={Artificial Intelligence for the Earth Systems},
  volume={5},
  number={2},
  pages={250061},
  year={2026},
  publisher={American Meteorological Society}
}

@article{alexe2024graphdop,
  title={GraphDOP: Towards skilful data-driven medium-range weather forecasts learnt and initialised directly from observations},
  author={Alexe, Mihai and Boucher, Eulalie and Lean, Peter and Pinnington, Ewan and Laloyaux, Patrick and McNally, Anthony and Lang, Simon and Chantry, Matthew and Burrows, Chris and Chrust, Marcin and others},
  journal={arXiv preprint arXiv:2412.15687},
  year={2024}
}

@article{alet2025skillful,
  title={Skillful joint probabilistic weather forecasting from marginals},
  author={Alet, Ferran and Price, Ilan and El-Kadi, Andrew and Masters, Dominic and Markou, Stratis and Andersson, Tom R and Stott, Jacklynn and Lam, Remi and Willson, Matthew and Sanchez-Gonzalez, Alvaro and others},
  journal={arXiv preprint arXiv:2506.10772},
  year={2025}
}

@article{allen2025end,
  title={End-to-end data-driven weather prediction},
  author={Allen, Anna and Markou, Stratis and Tebbutt, Will and Requeima, James and Bruinsma, Wessel P and Andersson, Tom R and Herzog, Michael and Lane, Nicholas D and Chantry, Matthew and Hosking, J Scott and others},
  journal={Nature},
  volume={641},
  number={8065},
  pages={1172--1179},
  year={2025},
  publisher={Nature Publishing Group UK London}
}

@article{andry2025appa,
  title={Appa: Bending weather dynamics with latent diffusion models for global data assimilation},
  author={Andry, G{\'e}r{\^o}me and Lewin, Sacha and Rozet, Fran{\c{c}}ois and Rochman, Omer and Mangeleer, Victor and Pirlet, Matthias and Faulx, Elise and Gr{\'e}goire, Marilaure and Louppe, Gilles},
  journal={arXiv preprint arXiv:2504.18720},
  year={2025}
}

@article{andrychowicz2023deep,
  title={Deep learning for day forecasts from sparse observations},
  author={Andrychowicz, Marcin and Espeholt, Lasse and Li, Di and Merchant, Samier and Merose, Alexander and Zyda, Fred and Agrawal, Shreya and Kalchbrenner, Nal},
  journal={arXiv preprint arXiv:2306.06079},
  year={2023}
}

@article{abshire2023windlidar,
	author = {Abshire, James and Cremons, Daniel and Sun, Xiaoli and Guzewich, Scott and Smith, Michael and Hovis, Floyd},
	journal = {Bulletin of the AAS},
	number = {8},
	year = {2023},
	month = {oct 23},
	url = {https://baas.aas.org/pub/2023n8i317p04},
	title = {MARLI: Mars {Lidar} for {Measuring} {Global} {Wind} and {Aerosol} {Profiles} from {Orbit}},
	volume = {55},
}

@article{banfield2004waves,
  title={Traveling waves in the Martian atmosphere from MGS TES nadir data},
  author={Banfield, D and Conrath, BJ and Gierasch, PJ and Wilson, R John and Smith, MD},
  journal={Icarus},
  volume={170},
  number={2},
  pages={365--403},
  year={2004},
  publisher={Elsevier}
}

@article{Berthelier2000dustelec,
title = {ARES, atmospheric relaxation and electric field sensor, the electric field experiment on NETLANDER},
author = {J.J Berthelier and R Grard and H Laakso and M Parrot},
journal = {Planetary and Space Science},
volume = {48},
number = {12},
pages = {1193-1200},
year = {2000},
note = {Mars Exploration Program},
issn = {0032-0633},
doi = {https://doi.org/10.1016/S0032-0633(00)00103-3},
url = {https://www.sciencedirect.com/science/article/pii/S0032063300001033},
}

@article{bi2023pangu,
  author  = {Bi, K. and Xie, L. and Zhang, H. and Chen, X. and Gu, X. and Tian, Q.},
  title   = {Accurate medium-range global weather forecasting with {3D} neural networks},
  journal = {Nature},
  year    = {2023},
  doi     = {10.1038/s41586-023-06185-3}
}

@article{bodnar2025foundation,
  title={A foundation model for the Earth system},
  author={Bodnar, Cristian and Bruinsma, Wessel P and Lucic, Ana and Stanley, Megan and Allen, Anna and Brandstetter, Johannes and Garvan, Patrick and Riechert, Maik and Weyn, Jonathan A and Dong, Haiyu and others},
  journal={Nature},
  volume={641},
  number={8065},
  pages={1180--1187},
  year={2025},
  publisher={Nature Publishing Group UK London}
}

@inproceedings{bonev2023sfno,
  title={Spherical fourier neural operators: Learning stable dynamics on the sphere},
  author={Bonev, Boris and Kurth, Thorsten and Hundt, Christian and Pathak, Jaideep and Baust, Maximilian and Kashinath, Karthik and Anandkumar, Anima},
  booktitle={International conference on machine learning},
  pages={2806--2823},
  year={2023},
  organization={PMLR}
}

@article{campbell2020estimating,
  author  = {Campbell, C. L. and Kling, A. M. and Guzewich, S. D. and Smith, C. L. and Kloos, J. L. and Lemmon, M. T. and Moore, C. A. and Cooper, B. A. and Haberle, R. M. and Moores, J. E.},
  title   = {Estimating the altitudes of {M}artian water-ice clouds above the {M}ars {S}cience {L}aboratory rover landing site},
  journal = {Planetary and Space Science},
  volume  = {182},
  pages   = {104785},
  year    = {2020},
  doi     = {10.1016/j.pss.2019.104785}
}

@article{chen2023fuxi,
  title={FuXi: A cascade machine learning forecasting system for 15-day global weather forecast},
  author={Chen, Lei and Zhong, Xiaohui and Zhang, Feng and Cheng, Yuan and Xu, Yinghui and Qi, Yuan and Li, Hao},
  journal={npj climate and atmospheric science},
  volume={6},
  number={1},
  pages={190},
  year={2023},
  publisher={Nature Publishing Group UK London}
}

@article{Christensen1992tesMission,
author = {Christensen, Philip R. and Anderson, Donald L. and Chase, Stillman C. and Clark, Roger N. and Kieffer, Hugh H. and Malin, Michael C. and Pearl, John C. and Carpenter, James and Bandiera, Nuno and Brown, F. Gerald and Silverman, Steven},
title = {Thermal emission spectrometer experiment: Mars Observer mission},
journal = {Journal of Geophysical Research: Planets},
volume = {97},
number = {E5},
pages = {7719-7734},
doi = {https://doi.org/10.1029/92JE00453},
url = {https://agupubs.onlinelibrary.wiley.com/doi/abs/10.1029/92JE00453},
year = {1992}
}

@article{Christensen2001tesRes,
author = {Christensen, P. R. and Bandfield, J. L. and Hamilton, V. E. and Ruff, S. W. and Kieffer, H. H. and Titus, T. N. and Malin, M. C. and Morris, R. V. and Lane, M. D. and Clark, R. L. and Jakosky, B. M. and Mellon, M. T. and Pearl, J. C. and Conrath, B. J. and Smith, M. D. and Clancy, R. T. and Kuzmin, R. O. and Roush, T. and Mehall, G. L. and Gorelick, N. and Bender, K. and Murray, K. and Dason, S. and Greene, E. and Silverman, S. and Greenfield, M.},
title = {Mars Global Surveyor Thermal Emission Spectrometer experiment: Investigation description and surface science results},
journal = {Journal of Geophysical Research: Planets},
volume = {106},
number = {E10},
pages = {23823-23871},
doi = {https://doi.org/10.1029/2000JE001370},
url = {https://agupubs.onlinelibrary.wiley.com/doi/abs/10.1029/2000JE001370},
year = {2001}
}

@article{clancy1996acb,
title = {Water Vapor Saturation at Low Altitudes around Mars Aphelion: A Key to Mars Climate?},
author = {R.T. Clancy and A.W. Grossman and M.J. Wolff and P.B. James and D.J. Rudy and Y.N. Billawala and B.J. Sandor and S.W. Lee and D.O. Muhleman},
journal = {Icarus},
volume = {122},
number = {1},
pages = {36-62},
year = {1996},
issn = {0019-1035},
doi = {https://doi.org/10.1006/icar.1996.0108},
}

@article{dang2026deep,
  title={Deep operator learning for high-fidelity fluid flow field reconstruction from sparse sensor measurements},
  author={Dang, Hiep Vo and Nguyen, Phong CH},
  journal={Journal of Computing and Information Science in Engineering},
  volume={26},
  number={1},
  pages={011007},
  year={2026},
  publisher={American Society of Mechanical Engineers}
}

@article{daras2023ambient,
  title={Ambient diffusion: Learning clean distributions from corrupted data},
  author={Daras, Giannis and Shah, Kulin and Dagan, Yuval and Gollakota, Aravind and Dimakis, Alex and Klivans, Adam},
  journal={Advances in Neural Information Processing Systems},
  volume={36},
  pages={288--313},
  year={2023}
}

@article{dosovitskiy2020vit,
  author  = {Dosovitskiy, A. and Beyer, L. and Kolesnikov, A. and Weissenborn, D. and Zhai, X. and Unterthiner, T. and Dehghani, M. and Minderer, M. and Heigold, G. and Gelly, S. and Uszkoreit, J. and Houlsby, N.},
  title   = {An Image Is Worth 16$\times$16 Words: Transformers for Image Recognition at Scale},
  journal = {arXiv preprint arXiv:2010.11929},
  year    = {2020}
}

@online{ecmwf2025aifs,
    author = "ECMWF",
    title = "ECMWF’s AI forecasts become operational",
    url  = "https://www.ecmwf.int/en/about/media-centre/news/2025/ecmwfs-ai-forecasts-become-operational",
    addendum = "(accessed: 2025-05-02)",
}

@article{edwards2021emirs,
  title={The emirates mars mission (EMM) emirates mars infrared spectrometer (EMIRS) instrument},
  author={Edwards, Christopher S and Christensen, Philip R and Mehall, Greg L and Anwar, Saadat and Tunaiji, Eman Al and Badri, Khalid and Bowles, Heather and Chase, Stillman and Farkas, Zoltan and Fisher, Tara and others},
  journal={Space Science Reviews},
  volume={217},
  number={7},
  pages={77},
  year={2021},
  publisher={Springer}
}

@article{fan2026physically,
  title={Physically consistent global atmospheric data assimilation with machine learning in latent space},
  author={Fan, Hang and Bai, Lei and Fei, Ben and Xiao, Yi and Chen, Kun and Liu, Yubao and Qu, Yongquan and Ling, Fenghua and Gentine, Pierre},
  journal={Science Advances},
  volume={12},
  number={1},
  pages={eaea4248},
  year={2026},
  publisher={American Association for the Advancement of Science}
}

@article{fernandez1997duststorm,
title = {Martian Dust Storms: A Review.},
author = {Fernández, W.},
journal = {Earth, Moon, and Planets},
volume = {77},
pages = {19-46},
year = {1997},
doi = {https://doi.org/10.1023/A:1006134805153},
}

@article{greybush2019ensemble,
  title={The ensemble Mars atmosphere reanalysis system (EMARS) version 1.0},
  author={Greybush, Steven J and Kalnay, Eugenia and Wilson, R John and Hoffman, Ross N and Nehrkorn, Thomas and Leidner, Mark and Eluszkiewicz, Janusz and Gillespie, Hartzel E and Wespetal, Matthew and Zhao, Yongjing and others},
  journal={Geoscience data journal},
  volume={6},
  number={2},
  pages={137--150},
  year={2019},
  publisher={Wiley Online Library}
}

@article{guha2021acbdust,
title = {Observation of aphelion cloud belt over Martian tropics, its evolution, and associated dust distribution from MCS data},
author = {Bijay Kumar Guha and Jagabandhu Panda and Zhaopeng Wu},
journal = {Advances in Space Research},
volume = {67},
number = {4},
pages = {1392-1411},
year = {2021},
issn = {0273-1177},
doi = {https://doi.org/10.1016/j.asr.2020.11.010},
}

@article{gupta2025finetuning,
  title={Finetuning AI foundation models to develop subgrid-scale parameterizations: A case study on atmospheric gravity waves},
  author={Gupta, Aman and Sheshadri, Aditi and Roy, Sujit and Schmude, Johannes and Gaur, Vishal and Leong, Wei Ji and Maskey, Manil and Ramachandran, Rahul},
  journal={Journal of Advances in Modeling Earth Systems},
  volume={17},
  number={11},
  pages={e2025MS005075},
  year={2025},
  publisher={Wiley Online Library}
}

@article{gupta2026healda,
  title={HealDA: Highlighting the importance of initial errors in end-to-end AI weather forecasts},
  author={Gupta, Aayush and Subramaniam, Akshay and Pritchard, Michael S and Kashinath, Karthik and Frolov, Sergey and Lieberman, Kelsey and Miller, Christopher and Silverman, Nicholas and Brenowitz, Noah D},
  journal={arXiv preprint arXiv:2601.17636},
  year={2026}
}

@article{haberle2019ames,
  author  = {Haberle, R. M. and Kahre, M. A. and Hollingsworth, J. L. and Montmessin, F. and Wilson, R. J. and Urata, R. A. and Brecht, A. S. and Wolff, M. J. and Kling, A. M. and Schaeffer, J. R.},
  title   = {Documentation of the {NASA}/{A}mes legacy {M}ars global climate model: Simulations of the present seasonal water cycle},
  journal = {Icarus},
  volume  = {333},
  pages   = {130--164},
  year    = {2019},
  doi     = {10.1016/j.icarus.2019.03.026}
}

@article{harder2022physics,
  author  = {Harder, P. and Yang, Q. and Ramesh, V. and others},
  title   = {Generating physically-consistent high-resolution climate data with hard-constrained neural networks},
  journal = {arXiv preprint arXiv:2208.05424},
  year    = {2022}
}

@techreport{harris2021gfdl,
  author      = {Harris, L. and Chen, X. and Putman, W. and Zhou, L. and Chen, J.-H.},
  title       = {A Scientific Description of the {GFDL} Finite-Volume Cubed-Sphere Dynamical Core},
  institution = {NOAA Technical Memorandum OAR GFDL ; 2021-001},
  year        = {2021},
  doi         = {10.25923/6nhs-5897}
}

@article{herde2024poseidon,
  title={Poseidon: Efficient foundation models for pdes},
  author={Herde, Maximilian and Raoni{\'c}, Bogdan and Rohner, Tobias and K{\"a}ppeli, Roger and Molinaro, Roberto and De Bezenac, Emmanuel and Mishra, Siddhartha},
  journal={Advances in Neural Information Processing Systems},
  volume={37},
  pages={72525--72624},
  year={2024}
}

@article{hernandezbernal2021elongated,
  author  = {Hern{\'a}ndez-Bernal, J. and S{\'a}nchez-Lavega, A. and del R{\'\i}o-Gaztelurrutia, T. and Ravanis, E. and Cardes{\'\i}n-Moinelo, A. and Connour, K. and others},
  title   = {An extremely elongated cloud over {A}rsia {M}ons volcano on {M}ars: {I}. {L}ife cycle},
  journal = {Journal of Geophysical Research: Planets},
  volume  = {126},
  year    = {2021},
  note    = {e2020JE006517},
  doi     = {10.1029/2020JE006517}
}

@article{hess1979co2pressure,
title = {The seasonal variation of atmospheric pressure on Mars as affected by the south polar cap},
author = {Hess, Seymour L. and Henry, Robert M. and Tillman, James E.},
journal = {Journal of Geophysical Research: Solid Earth},
volume = {84},
number = {B6},
pages = {2923-2927},
doi = {https://doi.org/10.1029/JB084iB06p02923},
url = {https://agupubs.onlinelibrary.wiley.com/doi/abs/10.1029/JB084iB06p02923},
year = {1979}
}

@article{Hinson2006transeddies,
title = {Radio occultation measurements of transient eddies in the northern hemisphere of Mars},
author = {Hinson, D. P.},
journal = {Journal of Geophysical Research},
volume = {111},
number = {E05002},
year = {2006},
doi = {10.1029/2005JE002612},
}

@article{hinson2010further,
  author  = {Hinson, D. P. and Wang, H.},
  title   = {Further observations of regional dust storms and baroclinic eddies in the northern hemisphere of {M}ars},
  journal = {Icarus},
  volume  = {206},
  pages   = {290--305},
  year    = {2010},
  doi     = {10.1016/j.icarus.2009.08.019}
}

@article{hollingsworth2010extratropical,
  author  = {Hollingsworth, J. L. and Kahre, M. A.},
  title   = {Extratropical cyclones, frontal waves, and {M}ars dust: {M}odeling and considerations},
  journal = {Geophysical Research Letters},
  volume  = {37},
  number  = {L22202},
  year    = {2010},
  doi     = {10.1029/2010GL044262}
}

@article{Holmes2018openmarsOzone,
title = {A reanalysis of ozone on Mars from assimilation of SPICAM observations},
author = {James A. Holmes and Stephen R. Lewis and Manish R. Patel and Franck Lefèvre},
journal = {Icarus},
volume = {302},
pages = {308-318},
year = {2018},
issn = {0019-1035},
doi = {https://doi.org/10.1016/j.icarus.2017.11.026},
url = {https://www.sciencedirect.com/science/article/pii/S0019103517302889},
}

@article{holmes2020openmars,
title = {OpenMARS: A global record of martian weather from 1999 to 2015},
author = {James A. Holmes and Stephen R. Lewis and Manish R. Patel},
journal = {Planetary and Space Science},
volume = {188},
pages = {104962},
year = {2020},
issn = {0032-0633},
doi = {https://doi.org/10.1016/j.pss.2020.104962},
url = {https://www.sciencedirect.com/science/article/pii/S0032063319303617},
}

@article {hourdin1993psfc,
      author = "Frédéric  Hourdin and Phu  Le Van and François  Forget and Olivier  Talagrand",
      title = "Meteorological Variability and the Annual Surface Pressure Cycle on Mars",
      journal = "Journal of Atmospheric Sciences",
      year = "1993",
      publisher = "American Meteorological Society",
      address = "Boston MA, USA",
      volume = "50",
      number = "21",
      pages=      "3625 - 3640",
      doi = "10.1175/1520-0469(1993)050<3625:MVATAS>2.0.CO;2",
      url = "https://journals.ametsoc.org/view/journals/atsc/50/21/1520-0469_1993_050_3625_mvatas_2_0_co_2.xml"
}

@article{huang2024diffda,
  title={Diffda: a diffusion model for weather-scale data assimilation},
  author={Huang, Langwen and Gianinazzi, Lukas and Yu, Yuejiang and Dueben, Peter D and Hoefler, Torsten},
  journal={arXiv preprint arXiv:2401.05932},
  year={2024}
}

@article{huang2024diffusionpde,
  title={DiffusionPDE: Generative PDE-solving under partial observation},
  author={Huang, Jiahe and Yang, Guandao and Wang, Zichen and Park, Jeong Joon},
  journal={Advances in Neural Information Processing Systems},
  volume={37},
  pages={130291--130323},
  year={2024}
}

@article{hunt1979frontal,
title = {Martian extratropical cyclones},
author = {Hunt GE, James PB},
journal = {Nature},
volume = {278},
pages = {531-532},
year = {1979},
doi = {https://doi.org/10.1038/278531a0},
}

@article{keisler2022forecasting,
  title={Forecasting global weather with graph neural networks},
  author={Keisler, Ryan},
  journal={arXiv preprint arXiv:2202.07575},
  year={2022}
}

@article{keller2024ai,
  title={AI-based data assimilation: Learning the functional of analysis estimation},
  author={Keller, Jan D and Potthast, Roland},
  journal={arXiv preprint arXiv:2406.00390},
  year={2024}
}

@article{Kelly2006co2cycle,
author = {Kelly, N. J. and Boynton, W. V. and Kerry, K. and Hamara, D. and Janes, D. and Reedy, R. C. and Kim, K. J. and Haberle, R. M.},
title = {Seasonal polar carbon dioxide frost on Mars: CO2 mass and columnar thickness distribution},
journal = {Journal of Geophysical Research: Planets},
volume = {111},
number = {E3},
pages = {},
doi = {https://doi.org/10.1029/2006JE002678},
url = {https://agupubs.onlinelibrary.wiley.com/doi/abs/10.1029/2006JE002678},
year = {2006}
}

@article{kleinbohl2009mcsThermaltide,
    title={Mars Climate Sounder limb profile retrieval of atmospheric temperature, pressure, and dust and water ice opacity},
  author={Kleinb{\"o}hl, Armin and Schofield, John T and Kass, David M and Abdou, Wedad A and Backus, Charles R and Sen, Bhaswar and Shirley, James H and Lawson, W Gregory and Richardson, Mark I and Taylor, Fredric W and others},
  journal={Journal of Geophysical Research: Planets},
  volume={114},
  number={E10},
  year={2009},
  publisher={Wiley Online Library}
}

@article{kok2006dustliftelec,
author = {Kok, Jasper F. and Renno, Nilton O.},
title = {Enhancement of the emission of mineral dust aerosols by electric forces},
journal = {Geophysical Research Letters},
volume = {33},
number = {19},
doi = {https://doi.org/10.1029/2006GL026284},
url = {https://agupubs.onlinelibrary.wiley.com/doi/abs/10.1029/2006GL026284},
year = {2006}
}

@article{korablev2018acs,
  author  = {Korablev, O. and Montmessin, F. and Trokhimovskiy, A. and Fedorova, A. A. and Shakun, A. V. and Grigoriev, A. V. and others},
  title   = {The {A}tmospheric {C}hemistry {S}uite ({ACS}) of three spectrometers for the {E}xo{M}ars 2016 {T}race {G}as {O}rbiter},
  journal = {Space Science Reviews},
  volume  = {214},
  number  = {7},
  year    = {2018},
  doi     = {10.1007/s11214-017-0437-6}
}

@article{kossaifi2026demystifying,
  title={Demystifying Data-Driven Probabilistic Medium-Range Weather Forecasting},
  author={Kossaifi, Jean and Kovachki, Nikola and Mardani, Morteza and Leibovici, Daniel and Ravuri, Suman and Shokar, Ira and Calvello, Edoardo and Abbas, Mohammad Shoaib and Harrington, Peter and Subramaniam, Ashay and others},
  journal={arXiv preprint arXiv:2601.18111},
  year={2026}
}

@article{kulowski2017duststorm,
title = {The seasonal and spatial distribution of textured dust storms observed by Mars Global Surveyor Mars Orbiter Camera},
author = {Laura Kulowski and Huiqun Wang and Anthony D. Toigo},
journal = {Advances in Space Research},
volume = {59},
number = {2},
pages = {715-721},
year = {2017},
issn = {0273-1177},
doi = {https://doi.org/10.1016/j.asr.2016.10.028},
url = {https://www.sciencedirect.com/science/article/pii/S0273117716306226},
}

@article{lam2023graphcast,
  author  = {Lam, R. and Sanchez-Gonzalez, A. and Willson, M. and Wirnsberger, P. and Fortunato, M. and Alet, F. and Ravuri, S. and Ewalds, T. and Eaton-Rosen, Z. and Hu, W. and Merose, A. and Hoyer, S. and Holland, G. and Vinyals, O. and Stott, J. and Pritzel, A. and Mohamed, S. and Battaglia, P.},
  title   = {Learning skillful medium-range global weather forecasting},
  journal = {Science},
  volume  = {382},
  number  = {6677},
  pages   = {1416--1421},
  year    = {2023},
  doi     = {10.1126/science.adi2336}
}

@article{lang2024aifs,
  title={AIFS--ECMWF's data-driven forecasting system},
  author={Lang, Simon and Alexe, Mihai and Chantry, Matthew and Dramsch, Jesper and Pinault, Florian and Raoult, Baudouin and Clare, Mariana CA and Lessig, Christian and Maier-Gerber, Michael and Magnusson, Linus and others},
  journal={arXiv preprint arXiv:2406.01465},
  year={2024}
}

@article{lang2026aifs,
  title={Aifs-crps: ensemble forecasting using a model trained with a loss function based on the continuous ranked probability score},
  author={Lang, Simon and Alexe, Mihai and Clare, Mariana CA and Roberts, Christopher and Adewoyin, Rilwan and Ben Bouall{\`e}gue, Zied and Chantry, Matthew and Dramsch, Jesper and Dueben, Peter D and Hahner, Sara and others},
  journal={npj Artificial Intelligence},
  volume={2},
  number={1},
  pages={18},
  year={2026},
  publisher={Nature Publishing Group UK London}
}

@article{lee2011dart,
  author  = {Lee, Christopher and Lawson, W. Gregory and Richardson, Mark I. and Anderson, Jeffrey L. and Collins, Nancy and Hoar, Tim and Mischna, Michael A.},
  title   = {Demonstration of Ensemble Data Assimilation for {M}ars Using {DART}, {M}ars{WRF}, and Radiance Observations from {MGS} {TES}},
  journal = {Journal of Geophysical Research: Planets},
  volume  = {116},
  number  = {E11},
  pages   = {E11011},
  year    = {2011},
  doi     = {10.1029/2011JE003815}
}

@article{lessig2023atmorep,
  author  = {Lessig, C. and Luise, I. and Gong, B. and Langguth, M. and Stadler, S. and Schultz, M.},
  title   = {{A}tmo{R}ep: A stochastic model of atmosphere dynamics using large scale representation learning},
  journal = {arXiv preprint arXiv:2308.13280},
  year    = {2023}
}

@article{lippe2023pde,
  author  = {Lippe, P. and Veeling, B. S. and Perdikaris, P. and Turner, R. E. and Brandstetter, J.},
  title   = {{PDE}-{R}efiner: Achieving accurate long rollouts with neural {PDE} solvers},
  journal = {arXiv preprint arXiv:2308.05732},
  year    = {2023}
}

@article{lorenc1991meteorological,
  title={The Meteorological Office analysis correction data assimilation scheme},
  author={Lorenc, AC and Bell, RS and Macpherson, B},
  journal={Quarterly Journal of the Royal Meteorological Society},
  volume={117},
  number={497},
  pages={59--89},
  year={1991},
  publisher={Wiley Online Library}
}

@article{majid2026ambient,
  title={Ambient Physics: Training Neural PDE Solvers with Partial Observations},
  author={Majid, Harris Abdul and Daras, Giannis and Tudisco, Francesco and McDonagh, Steven},
  journal={arXiv preprint arXiv:2602.13873},
  year={2026}
}

@article{manshausen2025generative,
  title={Generative data assimilation of sparse weather station observations at kilometer scales},
  author={Manshausen, Peter and Cohen, Yair and Harrington, Peter and Pathak, Jaideep and Pritchard, Mike and Garg, Piyush and Mardani, Morteza and Kashinath, Karthik and Byrne, Simon and Brenowitz, Noah},
  journal={Journal of Advances in Modeling Earth Systems},
  volume={17},
  number={10},
  pages={e2024MS004505},
  year={2025},
  publisher={Wiley Online Library}
}

@article{mccabe2024multiple,
  title={Multiple physics pretraining for spatiotemporal surrogate models},
  author={McCabe, Michael and R{\'e}galdo-Saint Blancard, Bruno and Parker, Liam and Ohana, Ruben and Cranmer, Miles and Bietti, Alberto and Eickenberg, Michael and Golkar, Siavash and Krawezik, Geraud and Lanusse, Francois and others},
  journal={Advances in Neural Information Processing Systems},
  volume={37},
  pages={119301--119335},
  year={2024}
}

@article{mccabe2025walrus,
  title={Walrus: A cross-domain foundation model for continuum dynamics},
  author={McCabe, Michael and Mukhopadhyay, Payel and Marwah, Tanya and Blancard, Bruno Regaldo-Saint and Rozet, Francois and Diaconu, Cristiana and Meyer, Lucas and Wong, Kaze WK and Sotoudeh, Hadi and Bietti, Alberto and others},
  journal={arXiv preprint arXiv:2511.15684},
  year={2025}
}

@article{mccleese2007mcs,
  title={Mars Climate Sounder: An investigation of thermal and water vapor structure, dust and condensate distributions in the atmosphere, and energy balance of the polar regions},
  author={McCleese, DJ and Schofield, JT and Taylor, FW and Calcutt, SB and Foote, MC and Kass, DM and Leovy, CB and Paige, DA and Read, PL and Zurek, RW},
  journal={Journal of Geophysical Research: Planets},
  volume={112},
  number={E5},
  year={2007},
  publisher={Wiley Online Library}
}

@article{michaels2006volcano,
author = {Michaels, T. I. and Colaprete, A. and Rafkin, S. C. R.},
title = {Significant vertical water transport by mountain-induced circulations on Mars},
journal = {Geophysical Research Letters},
volume = {33},
number = {16},
pages = {},
doi = {https://doi.org/10.1029/2006GL026562},
year = {2006}
}

@article{mischna2012radiation,
  author  = {Mischna, Michael A. and Lee, Christopher and Richardson, Mark I.},
  title   = {Development of a Fast, Accurate Radiative Transfer Model for the {M}artian Atmosphere, Past and Present},
  journal = {Journal of Geophysical Research: Planets},
  volume  = {117},
  number  = {E10},
  pages   = {E10009},
  year    = {2012},
  doi     = {10.1029/2012JE004110}
}

@article{moldovan2025update,
  title={An update to ECMWF's machine-learned weather forecast model AIFS},
  author={Moldovan, Gabriel and Pinnington, Ewan and Nemesio, Ana Prieto and Lang, Simon and Bouall{\`e}gue, Zied Ben and Dramsch, Jesper and Alexe, Mihai and Cruz, Mario Santa and Hahner, Sara and Cook, Harrison and others},
  journal={arXiv preprint arXiv:2509.18994},
  year={2025}
}

@article{montmessin2004cloud,
author = {Montmessin, F. and Forget, F. and Rannou, P. and Cabane, M. and Haberle, R. M.},
title = {Origin and role of water ice clouds in the Martian water cycle as inferred from a general circulation model},
journal = {Journal of Geophysical Research: Planets},
volume = {109},
number = {E10},
doi = {https://doi.org/10.1029/2004JE002284},
url = {https://agupubs.onlinelibrary.wiley.com/doi/abs/10.1029/2004JE002284},
year = {2004}
}

@article{mukkavilli2023ai,
  author  = {Mukkavilli, S. K. and Civitarese, D. S. and Schmude, J. and Jakubik, J. and Jones, A. and Nguyen, N. and Phillips, C. and Roy, S. and Singh, S. and Watson, C. and Ganti, R. and Hamann, H. and Nair, U. and Ramachandran, R. and Weldemariam, K.},
  title   = {{AI} Foundation Models for Weather and Climate: Applications, Design, and Implementation},
  journal = {arXiv preprint arXiv:2309.10808},
  year    = {2023}
}

@article{navarro2014dustcloud,
title = {Global climate modeling of the Martian water cycle with improved microphysics and radiatively active water ice clouds},
author = {Navarro, T. and Madeleine, J.-B and Forget, F. and Spiga, A. and Millour, E. and Montmessin, F. and Määttänen, A.},
journal = {Journal of Geophysical Research: Planets},
volume = {119},
number = {7},
pages = {1479-1495},
doi = {https://doi.org/10.1002/2013JE004550},
url = {https://agupubs.onlinelibrary.wiley.com/doi/abs/10.1002/2013JE004550},
year = {2014}
}

@article{navarro2017challenge,
  title={The challenge of atmospheric data assimilation on Mars},
  author={Navarro, T and Forget, F and Millour, E and Greybush, SJ and Kalnay, E and Miyoshi, T},
  journal={Earth and Space Science},
  volume={4},
  number={12},
  pages={690--722},
  year={2017},
  publisher={Wiley Online Library}
}

@article{newman2017bagnold,
  author  = {Newman, Claire E. and G{\'o}mez-Elvira, Javier and Marin, M. and Navarro, S. and Torres, J. and Richardson, Mark I. and Battalio, J. M. and Guzewich, Scott D. and Sullivan, R. and de la Torre Ju{\'a}rez, M. and Vasavada, Ashwin R. and Bridges, Nathan T.},
  title   = {Winds Measured by the {R}over {E}nvironmental {M}onitoring {S}tation ({REMS}) During the {M}ars {S}cience {L}aboratory ({MSL}) Rover's {B}agnold {D}unes Campaign and Comparison with Numerical Modeling Using {M}ars{WRF}},
  journal = {Icarus},
  volume  = {291},
  pages   = {203--231},
  year    = {2017},
  doi     = {10.1016/j.icarus.2016.12.016}
}

@article{pereira2026learning,
  title={Learning to Advect: A Neural Semi-Lagrangian Architecture for Weather Forecasting},
  author={Pereira, Carlos A and Gaudreault, St{\'e}phane and Dallerit, Valentin and Subich, Christopher and Panday, Shoyon and Wei, Siqi and Zhang, Sasa and Rout, Siddharth and Haber, Eldad and Spiteri, Raymond J and others},
  journal={arXiv preprint arXiv:2601.21151},
  year={2026}
}

@article{price2023gencast,
  title={Gencast: Diffusion-based ensemble forecasting for medium-range weather},
  author={Price, Ilan and Sanchez-Gonzalez, Alvaro and Alet, Ferran and Andersson, Tom R and El-Kadi, Andrew and Masters, Dominic and Ewalds, Timo and Stott, Jacklynn and Mohamed, Shakir and Battaglia, Peter and others},
  journal={arXiv preprint arXiv:2312.15796},
  year={2023}
}

@article{prithviwxc2024,
  title={Prithvi wxc: Foundation model for weather and climate},
  author={Schmude, Johannes and Roy, Sujit and Trojak, Will and Jakubik, Johannes and Civitarese, Daniel Salles and Singh, Shraddha and Kuehnert, Julian and Ankur, Kumar and Gupta, Aman and Phillips, Christopher E and others},
  journal={arXiv preprint arXiv:2409.13598},
  year={2024}
}

@article{rafkin2001mrams,
  author  = {Rafkin, Scot C. R. and Haberle, Robert M. and Michaels, Timothy I.},
  title   = {The {M}ars {R}egional {A}tmospheric {M}odeling {S}ystem: Model Description and Selected Simulations},
  journal = {Icarus},
  volume  = {151},
  number  = {2},
  pages   = {228--256},
  year    = {2001},
  doi     = {10.1006/icar.2001.6605}
}

@article{richardson2007planetwrf,
  author  = {Richardson, M. I. and Toigo, A. D. and Newman, C. E.},
  title   = {{P}lanet{WRF}: A general purpose, local to global numerical model for planetary atmospheric and climate dynamics},
  journal = {Journal of Geophysical Research: Planets},
  volume  = {112},
  number  = {E9},
  year    = {2007}
}

@article{rozet2023score,
  title={Score-based data assimilation},
  author={Rozet, Fran{\c{c}}ois and Louppe, Gilles},
  journal={Advances in Neural Information Processing Systems},
  volume={36},
  pages={40521--40541},
  year={2023}
}

@article{sanchezlavega2018seasonally,
  author  = {S{\'a}nchez-Lavega, A. and Garro, A. and del R{\'\i}o-Gaztelurrutia, T. and Hueso, R. and Ordo{\~n}ez-Etxeberria, I. and Chen Chen, H. and others},
  title   = {A seasonally recurrent annular cyclone in {M}ars northern latitudes and observations of a companion vortex},
  journal = {Journal of Geophysical Research: Planets},
  volume  = {123},
  pages   = {3020--3034},
  year    = {2018},
  doi     = {10.1029/2018JE005740}
}

@article{sanchezlavega2024dynamical,
  author  = {S{\'a}nchez-Lavega, A. and del R{\'\i}o-Gaztelurrutia, T. and Spiga, A. and Hern{\'a}ndez-Bernal, J. and Larsen, E. and Tirsch, D. and Cardesin-Moinelo, A. and Machado, P.},
  title   = {Dynamical Phenomena in the {M}artian Atmosphere Through {M}ars {E}xpress Imaging},
  journal = {Space Science Reviews},
  volume  = {220},
  number  = {16},
  year    = {2024},
  doi     = {10.1007/s11214-024-01047-4}
}

@article{schmude2026pde,
  title={PDE foundation models are skillful AI weather emulators for the Martian atmosphere},
  author={Schmude, Johannes and Roy, Sujit and Wang, Liping and van Kessel, Theodore and Klein, Levente and Freitag, Marcus and Bentivegna, Eloisa and Manson-Sawko, Robert and Lutjens, Bjorn and Maskey, Manil and others},
  journal={arXiv preprint arXiv:2602.15004},
  year={2026}
}

@article{skamarock2008wrf,
  author  = {Skamarock, W. C. and Klemp, J. B.},
  title   = {A time-split nonhydrostatic atmospheric model for weather research and forecasting applications},
  journal = {Journal of Computational Physics},
  volume  = {227},
  number  = {7},
  pages   = {3465--3485},
  year    = {2008}
}

@article{smith2004tes,
  title={Interannual variability in TES atmospheric observations of Mars during 1999--2003},
  author={Smith, Michael D},
  journal={Icarus},
  volume={167},
  number={1},
  pages={148--165},
  year={2004},
  publisher={Elsevier}
}

@article{solvik20254d,
  title={4D-Var using hessian approximation and backpropagation applied to automatically differentiable numerical and machine learning models},
  author={Solvik, Kylen and Penny, Stephen G and Hoyer, Stephan},
  journal={Journal of Advances in Modeling Earth Systems},
  volume={17},
  number={4},
  pages={e2024MS004608},
  year={2025},
  publisher={Wiley Online Library}
}

@article{sun2025can,
  title={Can AI weather models predict out-of-distribution gray swan tropical cyclones?},
  author={Sun, Y Qiang and Hassanzadeh, Pedram and Zand, Mohsen and Chattopadhyay, Ashesh and Weare, Jonathan and Abbot, Dorian S},
  journal={Proceedings of the National Academy of Sciences},
  volume={122},
  number={21},
  pages={e2420914122},
  year={2025},
  publisher={National Academy of Sciences}
}

@article{sun2025sda,
  title={LO-SDA: Latent Optimization for Score-based Atmospheric Data Assimilation},
  author={Sun, Jing-An and Fan, Hang and Gong, Junchao and Fei, Ben and Chen, Kun and Ling, Fenghua and Zhang, Wenlong and Xu, Wanghan and Yan, Li and Gentine, Pierre and others},
  journal={arXiv preprint arXiv:2510.22562},
  year={2025}
}

@article{surya2025arxiv,
title={Surya: Foundation model for heliophysics},
  author={Roy, Sujit and Schmude, Johannes and Lal, Rohit and Gaur, Vishal and Freitag, Marcus and Kuehnert, Julian and van Kessel, Theodore and Hegde, Dinesha V and Mu{\~n}oz-Jaramillo, Andr{\'e}s and Jakubik, Johannes and others},
  journal={arXiv preprint arXiv:2508.14112},
  year={2025}
}

@article{temel2021les,
  author  = {Temel, Orkun and Senel, Cem M. and Porchetta, Stefano and Mu{\~n}oz-Esparza, Domingo and Mischna, Michael A. and Van Hoolst, Tim and van Beeck, Jeroen and Karatekin, {\"O}zg{\"u}r},
  title   = {Large Eddy Simulations of the {M}artian Convective Boundary Layer: Towards Developing a New Planetary Boundary Layer Scheme},
  journal = {Atmospheric Research},
  volume  = {250},
  pages   = {105381},
  year    = {2021},
  doi     = {10.1016/j.atmosres.2020.105381}
}

@article{toigo2012resolution,
  author  = {Toigo, Anthony D. and Lee, Christopher and Newman, Claire E. and Richardson, Mark I.},
  title   = {The Impact of Resolution on the Dynamics of the {M}artian Global Atmosphere: Varying Resolution Studies with the {M}ars{WRF} {GCM}},
  journal = {Icarus},
  volume  = {221},
  number  = {1},
  pages   = {276--288},
  year    = {2012},
  doi     = {10.1016/j.icarus.2012.07.020}
}

@article{valantinas2024waterfrost,
  title={Evidence for transient morning water frost deposits on the Tharsis volcanoes of Mars},
  author={Valantinas, A and Thomas, N and Pommerol, A and Karatekin, O and Ruiz Lozano, L and Senel, Cem Berk and Temel, Orkun and Hauber, Ernst and Tirsch, Daniela and Bickel, VT and others},
  journal={Nature Geoscience},
  volume={17},
  number={7},
  pages={608--616},
  year={2024},
  publisher={Nature Publishing Group UK London}
}

@article{vandaele2018nomad,
  author  = {Vandaele, A. C. and Lopez-Moreno, J.-J. and Patel, M. R. and Bellucci, G. and Daerden, F. and Ristic, B. and others},
  title   = {{NOMAD}, an integrated suite of three spectrometers for the {E}xo{M}ars {T}race {G}as {M}ission: technical description, science objectives and expected performance},
  journal = {Space Science Reviews},
  volume  = {214},
  number  = {80},
  year    = {2018},
  doi     = {10.1007/s11214-018-0517-2}
}

@inproceedings{vandal2025global,
  title={Global atmospheric data assimilation with multi-modal masked autoencoders},
  author={Vandal, Thomas J and Duffy, Kate and Nachmany, Yoni and McDuff, Daniel},
  booktitle={2025 IEEE International Conference on Data Mining Workshops (ICDMW)},
  pages={863--872},
  year={2025},
  organization={IEEE}
}

@article{wang2002frontdustco2cloud,
author = {Wang, Huiqun and Ingersoll, Andrew P.},
title = {Martian clouds observed by Mars Global Surveyor Mars Orbiter Camera},
journal = {Journal of Geophysical Research: Planets},
volume = {107},
number = {E10},
pages = {8-1-8-16},
doi = {https://doi.org/10.1029/2001JE001815},
url = {https://agupubs.onlinelibrary.wiley.com/doi/abs/10.1029/2001JE001815},
year = {2002}
}

@article{wang2005dusteddies,
  title={Relationship between frontal dust storms and transient eddy activity in the northern hemisphere of Mars as observed by Mars Global Surveyor},
  author={Wang, Huiqun and Zurek, Richard W and Richardson, Mark I},
  journal={Journal of Geophysical Research: Planets},
  volume={110},
  number={E7},
  year={2005},
  publisher={Wiley Online Library}
}

@article{wang2009north,
  author  = {Wang, H. and Fisher, J. A.},
  title   = {North polar frontal clouds and dust storms on {M}ars during spring and summer},
  journal = {Icarus},
  volume  = {204},
  number  = {1},
  pages   = {103--113},
  year    = {2009},
  doi     = {10.1016/j.icarus.2009.05.028}
}

@article{wang2015duststorm,
title = {The origin, evolution, and trajectory of large dust storms on Mars during Mars years 24–30 (1999–2011)},
author = {Huiqun Wang and Mark I. Richardson},
journal = {Icarus},
volume = {251},
pages = {112-127},
year = {2015},
note = {Dynamic Mars},
issn = {0019-1035},
doi = {https://doi.org/10.1016/j.icarus.2013.10.033},
url = {https://www.sciencedirect.com/science/article/pii/S0019103513004624},
}

@article{wang2023duststormcycle,
title = {Martian dust storm distribution and annual cycle from Mars daily global map observations},
author = {Huiqun Wang and Morgan Saidel and Mark I. Richardson and Anthony D. Toigo and J. Michael Battalio},
journal = {Icarus},
volume = {394},
pages = {115416},
year = {2023},
issn = {0019-1035},
doi = {https://doi.org/10.1016/j.icarus.2022.115416},
url = {https://www.sciencedirect.com/science/article/pii/S0019103522005085},
}

@inproceedings{wang2025orbit,
  title={ORBIT-2: Scaling Exascale Vision Foundation Models for Weather and Climate Downscaling},
  author={Wang, Xiao and Choi, Jong-Youl and Kurihaya, Takuya and Lyngaas, Isaac and Yoon, Hong-Jun and Xiao, Xi and Pugmire, David and Fan, Ming and Nafi, Nasik Muhammad and Tsaris, Aristeidis and others},
  booktitle={Proceedings of the International Conference for High Performance Computing, Networking, Storage and Analysis},
  pages={86--98},
  year={2025}
}

@article{wang2025phyda,
  title={Phyda: Physics-guided diffusion models for data assimilation in atmospheric systems},
  author={Wang, Hao and Han, Jindong and Fan, Wei and Zhang, Weijia and Liu, Hao},
  journal={arXiv preprint arXiv:2505.12882},
  year={2025}
}

@article{wu2021les,
  author  = {Wu, Zhaopeng and Richardson, Mark I. and Zhang, Xi and Cui, Jun and Heavens, Nicholas G. and Lee, Christopher and Li, Tao and Lian, Yuan and Newman, Claire E. and Soto, Alejandro and Toigo, Anthony D. and Witek, Marcin L. and Yang, Jun},
  title   = {Large Eddy Simulations of the Dusty {M}artian Convective Boundary Layer with {M}ars{WRF}},
  journal = {Journal of Geophysical Research: Planets},
  volume  = {126},
  number  = {9},
  pages   = {e2020JE006752},
  year    = {2021},
  doi     = {10.1029/2020JE006752}
}

@article{xiang2025adaf,
  title={ADAF: An artificial intelligence data assimilation framework for weather forecasting},
  author={Xiang, Yanfei and Jin, Weixin and Dong, Haiyu and Weyn, Jonathan and Bai, Mingliang and Fang, Zuliang and Zhao, Pengcheng and Sun, Hongyu and Thambiratnam, Kit and Zhang, Qi and others},
  journal={Journal of Advances in Modeling Earth Systems},
  volume={17},
  number={9},
  pages={e2024MS004839},
  year={2025},
  publisher={Wiley Online Library}
}

@inproceedings{xiao2025vae,
  title={VAE-Var: Variational autoencoder-enhanced variational methods for data assimilation in meteorology},
  author={Xiao, Yi and Jia, Qilong and Chen, Kun and Bai, Lei and Xue, Wei},
  booktitle={The Thirteenth International Conference on Learning Representations},
  year={2025}
}

@inproceedings{xiaolora,
  title={LoRA-EnVar: Parameter-Efficient Hybrid Ensemble Variational Assimilation for Weather Forecasting},
  author={Xiao, Yi and Fan, Hang and Chen, Kun and Cao, Ye and Fei, Ben and Xue, Wei and Bai, Lei},
  booktitle={The Thirty-ninth Annual Conference on Neural Information Processing Systems},
  year={2025}
}

@article{xing2022fusing,
  title={Fusing sensor data with CFD results using gappy POD},
  author={Xing, Xiuqing and Dao, My Ha and Zhang, Baili and Lou, Jing and Tan, Wei Siang and Cui, Yongdong and Khoo, Boo Cheong},
  journal={Ocean Engineering},
  volume={246},
  pages={110549},
  year={2022},
  publisher={Elsevier}
}

@article{yang2023fourier,
  author  = {Yang, Q. and Hernandez-Garcia, A. and Harder, P. and others},
  title   = {{F}ourier neural operators for arbitrary resolution climate data downscaling},
  journal = {arXiv preprint arXiv:2305.14452},
  year    = {2023}
}

@article{Holmes2020,
  author    = {Holmes, James A. and Lewis, Stephen R. and Patel, Manish R.},
  title     = {{OpenMars}: A global record of {Martian} weather from 1999 to 2015},
  journal   = {Planetary and Space Science},
  volume    = {188},
  pages     = {104962},
  year      = {2020},
  doi       = {10.1016/j.pss.2020.104962},
  publisher = {Elsevier}
}

@article{Lam2023,
  author    = {Lam, Remi and Sanchez-Gonzalez, Alvaro and Willson, Matthew and
               Wirnsberger, Peter and Fortunato, Meire and Alet, Ferran and
               Ravuri, Suman and Ewalds, Timo and Eaton-Rosen, Zach and
               Hu, Weihua and Merose, Alexander and Hoyer, Stephan and
               Holland, George and Vinyals, Oriol and Stott, Jacklynn and
               Pritzel, Alexander and Mohamed, Shakir and Battaglia, Peter},
  title     = {Learning skillful medium-range global weather forecasting},
  journal   = {Science},
  volume    = {382},
  number    = {6677},
  pages     = {1416--1421},
  year      = {2023},
  doi       = {10.1126/science.adi2336},
  publisher = {American Association for the Advancement of Science}
}

@article{Schmude2024,
  author    = {Schmude, Johannes and Roy, Sujit and Trojak, Will and
               Jakubik, Johannes and Civitarese, Daniela S. and
               Watson, Campbell D. and Mukkavilli, S. Karthik and
               Kalariya, Vishal and Stoney, Anne and Ganti, Raghu and
               Ramachandran, Pavithra and Khandelwal, Ankur and
               Jones, Andrew and Freitag, Johannes and Pierson, Mairead and
               Maskey, Manil and Oloso, Arif and Nguyen, Tung},
  title     = {{Prithvi WxC}: Foundation Model for Weather and Climate},
  journal   = {arXiv preprint arXiv:2409.13598},
  year      = {2024},
  url       = {https://arxiv.org/abs/2409.13598}
}

@article{Bi2023,
  author    = {Bi, Kaifeng and Xie, Lingxi and Zhang, Hengheng and
               Chen, Xin and Gu, Xiaotao and Tian, Qi},
  title     = {Accurate medium-range global weather forecasting with {Pangu-Weather}},
  journal   = {Nature},
  volume    = {619},
  pages     = {533--538},
  year      = {2023},
  doi       = {10.1038/s41586-023-06185-3},
  publisher = {Springer Nature}
}

@article{hess1979seasonal,
  author  = {Hess, S. L. and Henry, R. M. and Tillman, J. E.},
  title   = {The seasonal variation of atmospheric pressure on {M}ars as affected by the south polar cap},
  journal = {Journal of Geophysical Research: Solid Earth},
  volume  = {84},
  number  = {B6},
  pages   = {2923--2927},
  year    = {1979},
  doi     = {10.1029/JB084iB06p02923}
}

@article{hourdin1993meteorological,
  author  = {Hourdin, F. and Le Van, P. and Forget, F. and Talagrand, O.},
  title   = {Meteorological variability and the annual surface pressure cycle on {M}ars},
  journal = {Journal of the Atmospheric Sciences},
  volume  = {50},
  number  = {21},
  pages   = {3625--3640},
  year    = {1993},
  doi     = {10.1175/1520-0469(1993)050<3625:MVATAS>2.0.CO;2}
}

@article{kelly2006seasonal,
  author  = {Kelly, N. J. and Boynton, W. V. and Kerry, K. and Hamara, D. and Janes, D. and Reedy, R. C. and Kim, K. J. and Haberle, R. M.},
  title   = {Seasonal polar carbon dioxide frost on {M}ars: {CO}$_2$ mass and columnar thickness distribution},
  journal = {Journal of Geophysical Research: Planets},
  volume  = {111},
  number  = {E3},
  year    = {2006},
  doi     = {10.1029/2006JE002678}
}

@article{fernandez1997martian,
  author  = {Fern{\'a}ndez, W.},
  title   = {Martian dust storms: A review},
  journal = {Earth, Moon, and Planets},
  volume  = {77},
  pages   = {19--46},
  year    = {1997},
  doi     = {10.1023/A:1006134805153}
}

@article{kulowski2017seasonal,
  author  = {Kulowski, L. and Wang, H. and Toigo, A. D.},
  title   = {The seasonal and spatial distribution of textured dust storms observed by {M}ars {G}lobal {S}urveyor {M}ars {O}rbiter {C}amera},
  journal = {Advances in Space Research},
  volume  = {59},
  number  = {2},
  pages   = {715--721},
  year    = {2017},
  doi     = {10.1016/j.asr.2016.10.028}
}

@article{wang2015origin,
  author  = {Wang, H. and Richardson, M. I.},
  title   = {The origin, evolution, and trajectory of large dust storms on {M}ars during {M}ars years 24--30 (1999--2011)},
  journal = {Icarus},
  volume  = {251},
  pages   = {112--127},
  year    = {2015},
  doi     = {10.1016/j.icarus.2013.10.033}
}

@article{wang2023martian,
  author  = {Wang, H. and Saidel, M. and Richardson, M. I. and Toigo, A. D. and Battalio, J. M.},
  title   = {Martian dust storm distribution and annual cycle from {M}ars daily global map observations},
  journal = {Icarus},
  volume  = {394},
  pages   = {115416},
  year    = {2023},
  doi     = {10.1016/j.icarus.2022.115416}
}

@article{clancy1996water,
  author  = {Clancy, R. T. and Grossman, A. W. and Wolff, M. J. and James, P. B. and Rudy, D. J. and Billawala, Y. N. and Sandor, B. J. and Lee, S. W. and Muhleman, D. O.},
  title   = {Water vapor saturation at low altitudes around {M}ars aphelion: A key to {M}ars climate?},
  journal = {Icarus},
  volume  = {122},
  number  = {1},
  pages   = {36--62},
  year    = {1996},
  doi     = {10.1006/icar.1996.0108}
}

@article{christensen1992tes,
  author  = {Christensen, P. R. and Anderson, D. L. and Chase, S. C. and Clark, R. N. and Kieffer, H. H. and Malin, M. C. and Pearl, J. C. and Carpenter, J. and Bandiera, N. and Brown, F. G. and Silverman, S.},
  title   = {Thermal emission spectrometer experiment: {M}ars {O}bserver mission},
  journal = {Journal of Geophysical Research: Planets},
  volume  = {97},
  number  = {E5},
  pages   = {7719--7734},
  year    = {1992},
  doi     = {10.1029/92JE00453}
}

@article{christensen2001tes,
  author  = {Christensen, P. R. and Bandfield, J. L. and Hamilton, V. E. and Ruff, S. W. and Kieffer, H. H. and Titus, T. N. and Malin, M. C. and Morris, R. V. and Lane, M. D. and Clark, R. L. and others},
  title   = {{M}ars {G}lobal {S}urveyor {T}hermal {E}mission {S}pectrometer experiment: Investigation description and surface science results},
  journal = {Journal of Geophysical Research: Planets},
  volume  = {106},
  number  = {E10},
  pages   = {23823--23871},
  year    = {2001},
  doi     = {10.1029/2000JE001370}
}

@article{holmes2018reanalysis,
  author  = {Holmes, J. A. and Lewis, S. R. and Patel, M. R. and Lef{\`e}vre, F.},
  title   = {A reanalysis of ozone on {M}ars from assimilation of {SPICAM} observations},
  journal = {Icarus},
  volume  = {302},
  pages   = {308--318},
  year    = {2018},
  doi     = {10.1016/j.icarus.2017.11.026}
}

@article{greybush2019emars,
  author  = {Greybush, S. J. and Kalnay, E. and Wilson, R. J. and Hoffman, R. N. and Nehrkorn, T. and Leidner, M. and Eluszkiewicz, J. and Gillespie, H. E. and Wespetal, M. and Zhao, Y. and others},
  title   = {The {E}nsemble {M}ars {A}tmosphere {R}eanalysis {S}ystem ({EMARS}) version 1.0},
  journal = {Geoscience Data Journal},
  volume  = {6},
  number  = {2},
  pages   = {137--150},
  year    = {2019}
}

@article{pathak2022fourcastnet,
  author  = {Pathak, J. and Subramanian, S. and Harrington, P. and Raja, S. and Chattopadhyay, A. and Mardani, M. and Kurth, T. and Hall, D. and Li, Z. and Azizzadenesheli, K. and others},
  title   = {{F}our{C}ast{N}et: A global data-driven high-resolution weather model using adaptive {F}ourier neural operators},
  journal = {arXiv preprint arXiv:2202.11214},
  year    = {2022}
}

@article{chen2023fengwu,
  author  = {Chen, K. and Han, T. and Gong, J. and Bai, L. and Ling, F. and Luo, J.-J. and Chen, X. and Ma, L. and Zhang, T. and Su, R. and others},
  title   = {{F}eng{W}u: Pushing the skillful global medium-range weather forecast beyond 10 days lead},
  journal = {arXiv preprint arXiv:2304.02948},
  year    = {2023}
}

@article{haberle2001possibility,
  title={On the possibility of liquid water on present-day Mars},
  author={Haberle, Robert M and McKay, Christopher P and Schaeffer, James and Cabrol, Nathalie A and Grin, Edmon A and Zent, Aaron P and Quinn, Richard},
  journal={Journal of Geophysical Research: Planets},
  volume={106},
  number={E10},
  pages={23317--23326},
  year={2001},
  publisher={Wiley Online Library}
}

@article{mahaffy2013abundance,
  title={Abundance and isotopic composition of gases in the Martian atmosphere from the Curiosity rover},
  author={Mahaffy, Paul R and Webster, Christopher R and Atreya, Sushil K and Franz, Heather and Wong, Michael and Conrad, Pamela G and Harpold, Dan and Jones, John J and Leshin, Laurie A and Manning, Heidi and others},
  journal={Science},
  volume={341},
  number={6143},
  pages={263--266},
  year={2013},
  publisher={American Association for the Advancement of Science}
}

@article{richardson2002topographically,
  title={A topographically forced asymmetry in the Martian circulation and climate},
  author={Richardson, Mark I and Wilson, R John},
  journal={Nature},
  volume={416},
  number={6878},
  pages={298--301},
  year={2002},
  publisher={Nature Publishing Group UK London}
}

@article{hansen2024comparison,
  title={A comparison of CO2 seasonal activity in Mars' northern and southern hemispheres},
  author={Hansen, Candice J and Byrne, Shane and Calvin, Wendy M and Diniega, Serina and Dundas, Colin M and Hayne, Paul O and McEwen, Alfred S and McKeown, Lauren E and Piqueux, Sylvain and Portyankina, Ganna and others},
  journal={Icarus},
  volume={419},
  pages={115801},
  year={2024},
  publisher={Elsevier}
}

@article{kahre2006modeling,
  title={Modeling the Martian dust cycle and surface dust reservoirs with the NASA Ames general circulation model},
  author={Kahre, Melinda A and Murphy, James R and Haberle, Robert M},
  journal={Journal of Geophysical Research: Planets},
  volume={111},
  number={E6},
  year={2006},
  publisher={Wiley Online Library}
}

@article{kass2020mars,
  title={Mars Climate Sounder observation of Mars' 2018 global dust storm},
  author={Kass, DM and Schofield, JT and Kleinb{\"o}hl, A and McCleese, DJ and Heavens, NG and Shirley, JH and Steele, LJ},
  journal={Geophysical Research Letters},
  volume={47},
  number={23},
  pages={e2019GL083931},
  year={2020},
  publisher={Wiley Online Library}
}

@article{cantor2006mars,
  title={Mars Orbiter Camera observations of Martian dust devils and their tracks (September 1997 to January 2006) and evaluation of theoretical vortex models},
  author={Cantor, Bruce A and Kanak, Katharine M and Edgett, Kenneth S},
  journal={Journal of Geophysical Research: Planets},
  volume={111},
  number={E12},
  year={2006},
  publisher={Wiley Online Library}
}

@article{cantor2001martian,
  title={Martian dust storms: 1999 Mars orbiter camera observations},
  author={Cantor, Bruce A and James, Philip B and Caplinger, Michael and Wolff, Michael J},
  journal={Journal of Geophysical Research: Planets},
  volume={106},
  number={E10},
  pages={23653--23687},
  year={2001},
  publisher={Wiley Online Library}
}

@inproceedings{larsen2025mid,
  title={Mid Latitude Elongated Clouds on Mars},
  author={Larsen, E and Sanchez-Lavega, A and del Rio-Gaztelurrutia, T and Hernandez-Bernal, J},
  booktitle={Proceedings of the VIII Iberian Congress on Planetary Sciences and Solar System Exploration},
  pages={115},
  year={2025}
}

@article{leighton1966behavior,
  title={Behavior of Carbon Dioxide and Other Volatiles on Mars: A thermal model of the Martian surface suggests that Mars's polar caps are solid carbon dioxide.},
  author={Leighton, Robert B and Murray, Bruce C},
  journal={Science},
  volume={153},
  number={3732},
  pages={136--144},
  year={1966},
  publisher={American Association for the Advancement of Science}
}

@article{michaels2017timothy,
  title={TIMOTHY N. TITUS, SHANE BYRNE, ANTHONY COLAPRETE, FRAN{\c{C}}OIS FORGET},
  author={MICHAELS, TIMOTHY I and PRETTYMAN, THOMAS H},
  journal={The Atmosphere and Climate of Mars},
  pages={374},
  year={2017},
  publisher={Cambridge University Press}
}

@inproceedings{syed2023attribution,
  title={Attribution Patching Outperforms Automated Circuit Discovery},
  author={Aaquib Syed and Can Rager and Arthur Conmy},
  booktitle={NeurIPS Workshop on Attributing Model Behavior at Scale},
  year={2023},
  url={https://openreview.net/forum?id=tiLbFR4bJW}
}

@article{zheng2023towards,
  title={Towards Best Practices of Activation Patching in Language Models: Metrics and Methods},
  author={Zheng, Tianshu and others},
  journal={arXiv preprint arXiv:2309.16042},
  year={2023}
}

@inproceedings{huben2024sparse,
title={Sparse Autoencoders Find Highly Interpretable Features in Language Models},
author={Robert Huben and Hoagy Cunningham and Logan Riggs Smith and Aidan Ewart and Lee Sharkey},
booktitle={The Twelfth International Conference on Learning Representations},
year={2024},
url={https://openreview.net/forum?id=F76bwRSLeK}
}

@inproceedings{haon2025mechanistic,
title={Mechanistic Interpretability for Steering Vision-Language-Action Models},
author={Bear H{\"a}on and Kaylene Caswell Stocking and Ian Chuang and Claire Tomlin},
booktitle={9th Annual Conference on Robot Learning},
year={2025},
url={https://openreview.net/forum?id=YvsUD8C9QS}
}

@article{biology,
author = {Onkar Gujral  and Mihir Bafna  and Eric Alm  and Bonnie Berger },
title = {Sparse autoencoders uncover biologically interpretable features in protein language model representations},
journal = {Proceedings of the National Academy of Sciences},
volume = {122},
number = {34},
pages = {e2506316122},
year = {2025},
}

@misc{olah2022,
  title={Mechanistic interpretability, variables, and the importance of interpretable bases},
  author={Olah, Chris},
  year={2022},
  url= {https://transformer-circuits.pub/2022/mech-interp-essay/index.html}
}

@article{raissi2019pinns,
  title = {Physics-informed neural networks: A deep learning framework for solving forward and inverse problems involving nonlinear partial differential equations},
  volume = {378},
  ISSN = {0021-9991},
  url = {http://dx.doi.org/10.1016/j.jcp.2018.10.045},
  DOI = {10.1016/j.jcp.2018.10.045},
  journal = {Journal of Computational Physics},
  publisher = {Elsevier BV},
  author = {Raissi,  M. and Perdikaris,  P. and Karniadakis,  G.E.},
  year = {2019},
  month = Feb,
  pages = {686–707}
}

@article{wang2021pidon,
  title = {Learning the solution operator of parametric partial differential equations with physics-informed DeepONets},
  volume = {7},
  ISSN = {2375-2548},
  url = {http://dx.doi.org/10.1126/sciadv.abi8605},
  DOI = {10.1126/sciadv.abi8605},
  number = {40},
  journal = {Science Advances},
  publisher = {American Association for the Advancement of Science (AAAS)},
  author = {Wang,  Sifan and Wang,  Hanwen and Perdikaris,  Paris},
  year = {2021},
  month = Oct 
}

@article{Li2024pino,
  title = {Physics-Informed Neural Operator for Learning Partial Differential Equations},
  volume = {1},
  ISSN = {2831-3194},
  url = {http://dx.doi.org/10.1145/3648506},
  DOI = {10.1145/3648506},
  number = {3},
  journal = {ACM / IMS Journal of Data Science},
  publisher = {Association for Computing Machinery (ACM)},
  author = {Li,  Zongyi and Zheng,  Hongkai and Kovachki,  Nikola and Jin,  David and Chen,  Haoxuan and Liu,  Burigede and Azizzadenesheli,  Kamyar and Anandkumar,  Anima},
  year = {2024},
  month = May,
  pages = {1–27}
}

@article{yang2021bpinns,
title = {B-PINNs: Bayesian physics-informed neural networks for forward and inverse PDE problems with noisy data},
journal = {Journal of Computational Physics},
volume = {425},
pages = {109913},
year = {2021},
issn = {0021-9991},
doi = {https://doi.org/10.1016/j.jcp.2020.109913},
url = {https://www.sciencedirect.com/science/article/pii/S0021999120306872},
author = {Liu Yang and Xuhui Meng and George Em Karniadakis},
}

@article{wang2025wfpinns,
  title = {WF-PINNs: solving forward and inverse problems of burgers equation with steep gradients using weak-form physics-informed neural networks},
  volume = {15},
  ISSN = {2045-2322},
  url = {http://dx.doi.org/10.1038/s41598-025-24427-4},
  DOI = {10.1038/s41598-025-24427-4},
  number = {1},
  journal = {Scientific Reports},
  publisher = {Springer Science and Business Media LLC},
  author = {Wang,  Xianke and Yi,  Shichao and Gu,  Huangliang and Xu,  Jing and Xu,  Wenjie},
  year = {2025},
  month = Nov 
}

@article{eshaghi2025vino,
title = {Variational Physics-informed Neural Operator (VINO) for solving partial differential equations},
journal = {Computer Methods in Applied Mechanics and Engineering},
volume = {437},
pages = {117785},
year = {2025},
issn = {0045-7825},
doi = {https://doi.org/10.1016/j.cma.2025.117785},
url = {https://www.sciencedirect.com/science/article/pii/S004578252500057X},
author = {Mohammad Sadegh Eshaghi and Cosmin Anitescu and Manish Thombre and Yizheng Wang and Xiaoying Zhuang and Timon Rabczuk},
}

@article{kharazmi2019variational,
  title={Variational physics-informed neural networks for solving partial differential equations},
  author={Kharazmi, Ehsan and Zhang, Zhongqiang and Karniadakis, George Em},
  journal={arXiv preprint arXiv:1912.00873},
  year={2019}
}

@article{kharazmi2021hp,
  title={hp-VPINNs: Variational physics-informed neural networks with domain decomposition},
  author={Kharazmi, Ehsan and Zhang, Zhongqiang and Karniadakis, George Em},
  journal={Computer Methods in Applied Mechanics and Engineering},
  volume={374},
  pages={113547},
  year={2021},
  publisher={Elsevier}
}

@article{rojas2024robust,
  title={Robust variational physics-informed neural networks},
  author={Rojas, Sergio and Maczuga, Pawe{\l} and Mu{\~n}oz-Matute, Judit and Pardo, David and Paszy{\'n}ski, Maciej},
  journal={Computer Methods in Applied Mechanics and Engineering},
  volume={425},
  pages={116904},
  year={2024},
  publisher={Elsevier}
}

@article{de2024wpinns,
  title={wPINNs: Weak physics informed neural networks for approximating entropy solutions of hyperbolic conservation laws},
  author={De Ryck, Tim and Mishra, Siddhartha and Molinaro, Roberto},
  journal={SIAM Journal on Numerical Analysis},
  volume={62},
  number={2},
  pages={811--841},
  year={2024},
  publisher={SIAM}
}

@article{eusebi2024cyclone,
  title={Realistic tropical cyclone wind and pressure fields can be reconstructed from sparse data using deep learning},
  author={Eusebi, Ryan and Vecchi, Gabriel A and Lai, Ching-Yao and Tong, Mingjing},
  journal={Communications Earth \& Environment},
  volume={5},
  number={1},
  pages={8},
  year={2024},
  publisher={Nature Publishing Group UK London}
}

@article{park2025pint,
  title={Pint: Physics-informed neural time series models with applications to long-term inference on weatherbench 2m-temperature data},
  author={Park, Keonvin and Kim, Jisu and Seo, Jaemin},
  journal={arXiv preprint arXiv:2502.04018},
  year={2025}
}

@article{li2020fourier,
  title={Fourier neural operator for parametric partial differential equations},
  author={Li, Zongyi and Kovachki, Nikola and Azizzadenesheli, Kamyar and Liu, Burigede and Bhattacharya, Kaushik and Stuart, Andrew and Anandkumar, Anima},
  journal={arXiv preprint arXiv:2010.08895},
  year={2020}
}

@article{lu2021learning,
  title={Learning nonlinear operators via DeepONet based on the universal approximation theorem of operators},
  author={Lu, Lu and Jin, Pengzhan and Pang, Guofei and Zhang, Zhongqiang and Karniadakis, George Em},
  journal={Nature machine intelligence},
  volume={3},
  number={3},
  pages={218--229},
  year={2021},
  publisher={Nature Publishing Group UK London}
}

@article{wang2024cvit,
  title={Cvit: Continuous vision transformer for operator learning},
  author={Wang, Sifan and Seidman, Jacob H and Sankaran, Shyam and Wang, Hanwen and Pappas, George J and Perdikaris, Paris},
  journal={arXiv preprint arXiv:2405.13998},
  year={2024}
}

@article{cho2026mbno,
  title={MBNO: Mamba-based neural operators for solving partial differential equations},
  author={Cho, Namkyeong and Ryu, Junseung and Hwang, Hyung Ju},
  journal={Journal of Computational Physics},
  pages={114639},
  year={2026},
  publisher={Elsevier}
}

@article{rahman2024pretraining,
  title={Pretraining codomain attention neural operators for solving multiphysics pdes},
  author={Rahman, Ashiqur and George, Robert J and Elleithy, Mogab and Leibovici, Daniel and Li, Zongyi and Bonev, Boris and White, Colin and Berner, Julius and Yeh, Raymond A and Kossaifi, Jean and others},
  journal={Advances in Neural Information Processing Systems},
  volume={37},
  pages={104035--104064},
  year={2024}
}

@article{lentz2025oceantransfer,
  title={Improving ocean reanalyses of observationally sparse regions with transfer learning},
  author={Lentz, Simon and Brune, Sebastian and Kadow, Christopher and Baehr, Johanna},
  journal={Scientific Reports},
  volume={15},
  number={1},
  pages={2640},
  year={2025},
  publisher={Nature Publishing Group UK London}
}

@article{nguyen2023climax,
  title={Climax: A foundation model for weather and climate},
  author={Nguyen, Tung and Brandstetter, Johannes and Kapoor, Ashish and Gupta, Jayesh K and Grover, Aditya},
  journal={arXiv preprint arXiv:2301.10343},
  year={2023}
}

@article{subramanian2023towards,
  title={Towards foundation models for scientific machine learning: Characterizing scaling and transfer behavior},
  author={Subramanian, Shashank and Harrington, Peter and Keutzer, Kurt and Bhimji, Wahid and Morozov, Dmitriy and Mahoney, Michael W and Gholami, Amir},
  journal={Advances in Neural Information Processing Systems},
  volume={36},
  pages={71242--71262},
  year={2023}
}

@article{soares2025towards,
  title={Towards a Foundation Model for Partial Differential Equations Across Physics Domains},
  author={Soares, Eduardo and Brazil, Emilio Vital and Shirasuna, Victor and de Carvalho, Breno WSR and Malossi, Cristiano},
  journal={arXiv preprint arXiv:2511.21861},
  year={2025}
}

@article{zhao2024recfno,
  title={RecFNO: A resolution-invariant flow and heat field reconstruction method from sparse observations via Fourier neural operator},
  author={Zhao, Xiaoyu and Chen, Xiaoqian and Gong, Zhiqiang and Zhou, Weien and Yao, Wen and Zhang, Yunyang},
  journal={International Journal of Thermal Sciences},
  volume={195},
  pages={108619},
  year={2024},
  publisher={Elsevier}
}

@inproceedings{koupai2025enma,
  title={Enma: Tokenwise autoregression for generative neural pde operators},
  author={Koupa{\"\i}, Armand Kassa{\"\i} and Le Boudec, Lise and Serrano, Louis and Gallinari, Patrick},
  booktitle={Neural Information Processing Systems},
  year={2025}
}

@article{liu2023clawno,
  title={Harnessing the power of neural operators with automatically encoded conservation laws},
  author={Liu, Ning and Fan, Yiming and Zeng, Xianyi and Kl{\"o}wer, Milan and Zhang, Lu and Yu, Yue},
  journal={arXiv preprint arXiv:2312.11176},
  year={2023}
}

@article{li2023phasefielddeeponet,
  title={Phase-Field DeepONet: Physics-informed deep operator neural network for fast simulations of pattern formation governed by gradient flows of free-energy functionals},
  author={Li, Wei and Bazant, Martin Z and Zhu, Juner},
  journal={Computer Methods in Applied Mechanics and Engineering},
  volume={416},
  pages={116299},
  year={2023},
  publisher={Elsevier}
}

@article{knigge2024spacetime,
  title={Space-time continuous pde forecasting using equivariant neural fields},
  author={Knigge, David M and Wessels, David R and Valperga, Riccardo and Papa, Samuele and Sonke, Jan-Jakob and Gavves, Efstratios and Bekkers, Erik J},
  journal={Advances in Neural Information Processing Systems},
  volume={37},
  pages={76553--76577},
  year={2024}
}

@article{li2020multipole,
  title={Multipole graph neural operator for parametric partial differential equations},
  author={Li, Zongyi and Kovachki, Nikola and Azizzadenesheli, Kamyar and Liu, Burigede and Stuart, Andrew and Bhattacharya, Kaushik and Anandkumar, Anima},
  journal={Advances in Neural Information Processing Systems},
  volume={33},
  pages={6755--6766},
  year={2020}
}

@article{cao2021galerkin,
  title={Choose a transformer: Fourier or galerkin},
  author={Cao, Shuhao},
  journal={Advances in neural information processing systems},
  volume={34},
  pages={24924--24940},
  year={2021}
}

@article{wang2024lno,
  title={Latent neural operator for solving forward and inverse pde problems},
  author={Wang, Tian and Wang, Chuang},
  journal={Advances in Neural Information Processing Systems},
  volume={37},
  pages={33085--33107},
  year={2024}
}

@inproceedings{hu2026transolver,
  title={Transolver is a linear transformer: Revisiting physics-attention through the lens of linear attention},
  author={Hu, Wenjie and Liu, Sidun and Qiao, Peng and Sun, Zhenglun and Dou, Yong},
  booktitle={Proceedings of the AAAI Conference on Artificial Intelligence},
  volume={40},
  number={1},
  pages={408--416},
  year={2026}
}

@article{zhong2025linearattention,
  title={Efficient High-Accuracy PDEs Solver with the Linear Attention Neural Operator},
  author={Zhong, Ming and Yan, Zhenya},
  journal={arXiv preprint arXiv:2510.16816},
  year={2025}
}

@article{jiao2021one,
  title={One-shot learning for solution operators of partial differential equations},
  author={Jiao, Anran and He, Haiyang and Ranade, Rishikesh and Pathak, Jay and Lu, Lu},
  journal={arXiv preprint arXiv:2104.05512},
  year={2021}
}

@article{winovich2026active,
  title={Active operator learning with predictive uncertainty quantification for partial differential equations},
  author={Winovich, Nick and Daneker, Mitchell and Lu, Lu and Lin, Guang},
  journal={Journal of Computational Physics},
  pages={114791},
  year={2026},
  publisher={Elsevier}
}

@article{mialon2023liesymmetry,
  title={Self-supervised learning with lie symmetries for partial differential equations},
  author={Mialon, Gr{\'e}goire and Garrido, Quentin and Lawrence, Hannah and Rehman, Danyal and LeCun, Yann and Kiani, Bobak},
  journal={Advances in Neural Information Processing Systems},
  volume={36},
  pages={28973--29004},
  year={2023}
}

@article{Corchete2025CO2,   
AUTHOR={Corchete, Victor },
TITLE={CO2 polar mass on Mars determined from InSight pressure data},
JOURNAL={Frontiers in Astronomy and Space Sciences},
VOLUME={Volume 11 - 2024},
YEAR={2025},
URL={https://www.frontiersin.org/journals/astronomy-and-space-sciences/articles/10.3389/fspas.2024.1532334},
DOI={10.3389/fspas.2024.1532334},
ISSN={2296-987X},
  
}

@article{ruiz2025curiosity,
AUTHOR = {Ruíz, María and Sebastián-Martínez, Eduardo and Rodríguez-Manfredi, Jose Antonio and Pla-García, Jorge and de la Torre-Juarez, Manuel and Rafkin, Scot C. R.},
TITLE = {Meteorological Changes Across Curiosity Rover’s Traverse Using REMS Measurements and Comparisons Between Measurements and MRAMS Model Results},
JOURNAL = {Remote Sensing},
VOLUME = {17},
YEAR = {2025},
NUMBER = {3},
ARTICLE-NUMBER = {368},
URL = {https://www.mdpi.com/2072-4292/17/3/368},
ISSN = {2072-4292},
DOI = {10.3390/rs17030368}
}

@article{zhu2021long,
  title={Long-short transformer: Efficient transformers for language and vision},
  author={Zhu, Chen and Ping, Wei and Xiao, Chaowei and Shoeybi, Mohammad and Goldstein, Tom and Anandkumar, Anima and Catanzaro, Bryan},
  journal={Advances in neural information processing systems},
  volume={34},
  pages={17723--17736},
  year={2021}
}

\end{document}